\documentclass[a4paper,11pt]{article}
\pdfoutput=1 

\usepackage{jheppub} 
\graphicspath{{fig/}}
\usepackage[T1]{fontenc} 
\usepackage{float}
\usepackage{dcolumn}
\usepackage{bm}
\usepackage{hyperref}  
\usepackage{multirow}
\usepackage{xspace}
\usepackage{color}
\usepackage{caption}
\usepackage{amssymb}
\usepackage{amsmath}
\usepackage{amsthm}
\usepackage{placeins}

\DeclareMathOperator\erf{erf}

\newcommand{\pp}{\rm{pp}}
\newcommand{\ppbar}{\rm{p\overline{\rm p}}}

\newcommand{\qqbar}{\ensuremath{q\bar{q}}}
\newcommand{\ttbar}{t\bar{t}}

\newcommand{\sqrts}{\sqrt{s}}
\newcommand{\alphas}{\alpha_{\rm s}}
\newcommand{\alphasmZ}{\alphas(m_{\rm Z})}

\newcommand{\Wp}{\ensuremath{\mathrm{W^+}}}
\newcommand{\Wm}{\ensuremath{\mathrm{W^-}}}
\newcommand{\Wpm}{\ensuremath{\mathrm{W^{\pm}}}}
\newcommand{\Pe}{\ensuremath{\mathrm{e}}}

\newcommand{\TeV}{\ensuremath{\,\text{Te\hspace{-.08em}V}}\xspace}
\newcommand{\alpS}{\ensuremath{\alpha_S}\xspace}

\newcommand{\syst}{\ensuremath{\,\text{(syst)}}\xspace}
\newcommand{\lum}{\ensuremath{\,\text{(lumi)}}\xspace}
\newcommand{\stat}{\ensuremath{\,\text{(stat)}}\xspace}
\newcommand{\statt}{\ensuremath{\,\text{(num)}}\xspace}
\newcommand{\cmenergy}{\ensuremath{\,\text{(c.m.\,en.)}}\xspace}
\newcommand{\cmenerg}{\ensuremath{\text{c.m.\,en.}}\xspace}
\newcommand{\unit}[1]{\ensuremath{\text{\,#1}}\xspace}

\newcommand{\pT}{p_{\mathrm{T}}}
\newcommand{\mZ}{\ensuremath{m_\mathrm{Z}}}
\newcommand{\mW}{\ensuremath{m_\mathrm{W}}}
\newcommand{\mT}{\ensuremath{m_\mathrm{T}}}

\newcommand{\mcfm}{{\sc mcfm}}
\newcommand{\fewz}{{\sc fewz}}
\newcommand{\dynnlo}{{\sc dynnlo}}
\newcommand{\lhapdf}{{\sc lhapdf}}
\newcommand{\mcsanc}{{\sc mcsanc}}

\newcommand*{\eg}{e.g.\@\xspace}
\newcommand*{\ie}{i.e.\@\xspace}
\newcommand*{\cm}{c.m.\@\xspace}
\newcommand*{\vs}{vs.\@\xspace}

\renewcommand\arraystretch{1.2}

\title{Extraction of the strong coupling $\alphasmZ$ from a combined NNLO analysis of inclusive electroweak boson cross sections at hadron colliders}

\author[a]{David~d'Enterria}
\author[a,b]{\! and Andres~Poldaru}
\affiliation[a]{CERN, EP Department, 1211 Geneva, Switzerland}
\affiliation[b]{LMU, 80539 Munich, Germany}

\emailAdd{david.d'enterria@cern.ch}
\emailAdd{andres.poldaru@gmail.com}

\abstract{The inclusive cross sections of \Wp, \Wm, and Z boson production from 34 different measurements performed in proton-(anti)proton collisions at center-of-mass energies $\sqrts$~=~1.8--13~TeV, are compared to perturbative QCD calculations at next-to-next-to-leading-order (NNLO) accuracy with four sets of parton distributions functions (CT14, HERAPDF2.0, MMHT14, and NNPDF3.0 PDFs) and varying values of the strong coupling constant at the Z mass pole, $\alphasmZ$. The data-theory agreement is good within the experimental and theoretical uncertainties, with the CT14 and MMHT14 parton densities providing the most overall consistent description of all cross section data. A value of $\alphasmZ =  0.1188^{+0.0019}_{-0.0013}$ is extracted from a combined fit of the 28 experimental LHC measurements to the corresponding NNLO theoretical predictions obtained with the MMHT14 PDF set, which provides the most robust and stable QCD coupling extraction of this analysis.}

\begin{document} 

\maketitle
\flushbottom

\section{Introduction}
\label{sec:intro}

The production of electroweak W$^\pm$ and Z gauge bosons in proton-(anti)proton collisions at TeV energies provides crucial precision tests of the Standard Model (SM) of particle physics~\cite{Alioli:2016fum}. Charged- and neutral-current Drell--Yan (DY) processes, $\pp,\ppbar \to W^\pm \to \ell^\pm\nu_\ell$ and $\pp,\ppbar \to Z \to \ell^+\ell^-$ (with $\ell^\pm=e^\pm,\mu^\pm$), benefit from a very ``clean'' leptonic heavy final state and large collected data samples, leading to cross sections that are the most accurately known, both theoretically and experimentally, at hadron colliders. Experimentally, inclusive \Wp, \Wm, and Z boson production cross sections have been measured in their leptonic decay modes, over relatively broad fiducial regions, with uncertainties 4.5--6.5\% at the Tevatron~\cite{Abe:1995bm,Abbott:1999tt,Abulencia:2005ix} and 1.5--4\% at the Large Hadron Collider (LHC)~\cite{CMS:2011aa,Chatrchyan:2014mua,Aad:2011dm,Aad:2015auj,Aad:2016naf,Aaij:2015gna,Aaij:2015zlq,Aaij:2016mgv,Aaij:2016qqz}. These experimental uncertainties are clearly dominated by the absolute integrated luminosity, with statistical and systematic sources representing subleading effects, in particular at the LHC. Theoretically, the cross sections are also known at high accuracy: at next-to-next-to-leading-order (NNLO) in perturbative Quantum Chromodynamics (pQCD)~\cite{Anastasiou:2003ds,dynnlo,fewz,mcfm,mcfm8}, including next-to-leading order (NLO) electroweak (EW) corrections~\cite{Li:2012wna,mcsanc}. The theoretical uncertainties include typically 1--4\% effects from our imprecise knowledge of the (anti)proton parton distribution functions (PDFs), 1--2\% from the propagated uncertainty of the strong coupling constant at the Z mass ($\alphasmZ = 0.1181 \pm 0.0011$), and 0.4--1.2\% from missing higher-order terms (scale uncertainties)~\cite{Alioli:2016fum,Poldaru:2019dnl}.

The quality of the experimental and theoretical knowledge of the DY cross sections is such that they constitute today a key tool to accurately constrain the quark and gluon PDFs, through fits of their experimental {\it differential} distributions~\cite{CT14,HERAPDF2.0,MMHT14,NNPDF3.0,ABM12,JR09}. Recently, we proposed to exploit also the {\it absolute} inclusive W$^\pm$ and Z cross sections, $\sigma_{\rm W,Z}$, as a means to precisely derive the value of $\alphasmZ$ through detailed comparisons of the data to pQCD predictions at NNLO accuracy~\cite{Poldaru:2019dnl}. Based on this approach, the CMS collaboration has recently analyzed~\cite{Sirunyan:2019wne} twelve measurements of W$^\pm$ and Z cross sections in $\pp$ collisions at center-of-mass (\cm) energies of $\sqrts= 7$ and 8 TeV~\cite{CMS:2011aa,Chatrchyan:2014mua} to extract $\alphasmZ = 0.1175^{+0.0025}_{-0.0028}$, a result with a final propagated $\sim$2.3\% uncertainty that is comparable to that previously obtained in a similar analysis of the inclusive top-quark pair ($\ttbar$) cross sections in pp collisions at the LHC~\cite{Klijnsma:2017eqp}. Since the QCD coupling is known today through a combination of six types of observables, with a relatively poor $\sim$1\% precision~\cite{PDG},
the inclusion of new independent extraction methods is one of the suggested paths towards the reduction of the $\alphasmZ$ world average uncertainty~\cite{alphas_confs}. In this context, the purpose of our work is twofold. First, a total of 34 electroweak boson cross section measurements performed at the LHC and Tevatron are compared to the corresponding state-of-the-art NNLO pQCD theoretical predictions. Second, we combine the 12 $\alphasmZ$ extractions of the CMS analysis of Ref.~\cite{Sirunyan:2019wne} with those derived here from 16 additional W$^\pm$ and Z cross sections measured at $\sqrts$~=~7, 8, and 13~TeV by ATLAS~\cite{Aad:2011dm,Aad:2015auj,Aad:2016naf} and LHCb~\cite{Aaij:2015gna,Aaij:2015zlq,Aaij:2016mgv,Aaij:2016qqz} in order to more accurately determine $\alphasmZ$ by exploiting the maximum LHC data possible\footnote{Most W$^\pm$ and Z measurements at the LHC have been included in our analysis except those appearing only recently~\cite{Aaboud:2018nic,Aad:2019rou,Aad:2019bdc,Sirunyan:2019bzr}, when the present study was nearing completion. The Tevatron results are studied but not included in the final $\alphasmZ$ extraction as their precision is less good than that of the LHC data, as discussed below.}. A single value of the QCD coupling constant is finally extracted through a proper combination of all individual LHC $\alphasmZ$ values derived as explained below.


The paper is organized as follows. In Section~\ref{sec:2}, the theoretical tools used to compute the W$^\pm$ and Z cross sections are outlined. The list of Tevatron and LHC measurements, including fiducial criteria applied on their decay leptons, are collected in Section~\ref{sec:3}. The data are compared to the NNLO predictions in Section~\ref{sec:4}, and the preferred $\alphasmZ$ value for each measurement is derived in Section~\ref{sec:5}. The final QCD coupling constant is determined by combining all $\alphasmZ$ values through a $\chi^2$-minimization procedure that takes properly into account all propagated experimental and theoretical uncertainties and their correlations, as described in Section~\ref{sec:6}. The main conclusions of the work are summarized in Section~\ref{sec:7}. 

\section{Theoretical calculations} 
\label{sec:2}

At hadron colliders, the leading order (LO) production of W$^\pm$ and Z bosons involves the annihilation of quark-antiquark pairs of the same ($\qqbar\to \rm Z+X$) or different ($q\bar{q}'\to \rm W+X$) flavours. At NLO, the Born diagrams are supplemented with initial-state real gluon emission or virtual exchange, and gluon-quark and gluon-antiquark scatterings start to contribute. At NNLO, additional gluon radiations and/or exchanges contribute. Theoretically, the inclusive \Wpm\ and Z cross section in $\pp, \ppbar$ collisions can be computed, making use of the QCD factorization theorem~\cite{Collins:1985ue}, from the convolution of the (anti)proton PDFs ($f_i(x,\mu_F)$, evaluated at parton fractional momenta $x_i$ and factorization energy scale $\mu_F$), and the elementary subprocess partonic cross sections $\hat{\sigma}$ computed as an expansion in the QCD coupling evaluated at the renormalization scale $\mu_R$, \ie, schematically,
\begin{equation}
\!\! \sigma_{\rm pp\to W,Z+X} =\! \int\!\!\int \mathrm{d}x_1 \mathrm{d}x_2\,f_1(x_1,\mu_F)f_2(x_2,\mu_F)\, \left[\hat{\sigma}_{\textsc{lo}}+ \alphas(\mu_R)\hat{\sigma}_{\textsc{nlo}} +\alphas^2(\mu_R)\hat{\sigma}_{\textsc{nnlo}} + \cdots \right].
\end{equation}
Whereas the Born cross section, $\sigma_{\textsc{lo}}$, is a pure EW quantity, the incorporation of NLO and NNLO higher-order pQCD corrections increases the overall value of $\sigma_{\rm W,Z}$, thereby improving the data-theory agreement~\cite{Anastasiou:2003ds,dynnlo,fewz,mcfm,mcfm8}, and provides the desired dependence on $\alphasmZ$ that allows the extraction of this latter parameter from the data-theory comparisons. The size of the higher-order corrections, encoded in the so-called K-factor given by the ratio of NNLO to LO cross sections, amounts to $\rm K = \sigma_{_{\rm NNLO}}/\sigma_{_{\rm LO}}\approx$~1.35, 1.35, 1.22, 1.33, and 1.29 in the CDF, D0, ATLAS, CMS, and LHCb fiducial acceptance for \Wpm\ and Z final states, respectively, as derived here with the \mcfm~\cite{mcfm} code. Such a result indicates that indeed W and Z boson production in pp collisions is sensitive to $\alphasmZ$, through $\sim$25\% higher-order matrix-elements direct contributions to their total cross sections. In addition, already at LO, the EW boson cross sections provide sensitivity to the QCD coupling constant through the ($Q^2$-evolved) dependence on $\alphasmZ$ of the parton densities.


The electroweak gauge boson cross sections are computed at NNLO accuracy with \mcfm~v.8.0~\cite{mcfm8} implementing the fiducial acceptance criteria on the lepton $\pT$ and $\eta$ of each experimental measurement. \mcfm\ is interfaced with \lhapdf~v.6.1.6~\cite{lhapdf} to access four different PDFs: CT14~\cite{CT14}, HERAPDF2.0~\cite{HERAPDF2.0}, MMHT14~\cite{MMHT14}, and NNPDF3.0~\cite{NNPDF3.0}. All these PDF sets use the same default central value of the QCD coupling constant, $\alphasmZ = 0.118$, in their global fits of the data, and also provide various PDF fit variations over a range of $\pm 0.003$ around the default $\alphasmZ$ value\footnote{Technically, the central sets selected via \lhapdf\ in this study are: CT14nnlo\_as\_0$iii$ (for $iii$ = 115--121), HERAPDF20\_NNLO\_ALPHAS\_$iii$ (for $iii$ = 115--121), MMHT2014nnlo\_asmzlargerange (with $\alphasmZ$ = 0.115,..., 0.121 grids), and NNPDF30\_nnlo\_as\_0$iii$ (with $iii$ = 115--121).}.
However, when the  QCD coupling constant is left free in their NNLO PDF fits, the following values are preferred by the different PDF sets: $\alphasmZ = 0.1150^{+0.0036}_{-0.0024}$ (CT14)~\cite{CT14}, $0.108$ (HERAPDF2.0)~\cite{HERAPDF2.0}, and $0.1172\pm 0.0013$ (MMHT2014)~\cite{MMHT14}.
Our choice of PDF sets is based mostly on the fact that none of them used in their global fits the {\it absolute} W$^\pm$ and Z cross section data exploited here, in order to extract the parton densities themselves. This warrants that the $\alphasmZ$ values extracted at the end are truly independent of the data used to derive the PDF sets employed in the theoretical predictions. The default renormalization and factorization scales are fixed at the corresponding EW boson masses, $\mu = \mu_F = \mu_R$~=~$m_{\rm W,Z}$. 

All numerical results have been obtained using the latest values for the SM particle masses, widths, and couplings~\cite{PDG}. The so-called G$_F$ electroweak scheme, with  \mW, \mZ, and G$_F$ as input parameters, is used in all the predictions. The leptonic W and Z branching fractions are obtained in \mcfm\ from the theoretical leptonic width (computed at LO in electroweak accuracy) normalized to the total W and Z widths experimentally measured~\cite{PDG}. For simplicity, we use the default value of the charm quark mass in \mcfm\ (and in the NNPDF3.0 set), $m_c = 1.275$~GeV, rather than the slightly different values chosen by each PDF set, since the computed cross sections are found anyway to depend only at the few per mille on variations of this parameter over $m_c \approx 1.2$--1.4~GeV. For all PDF sets, we compute the NNLO cross sections at various fixed $\alphasmZ$ values over the range $[0.115\mathrm{-}0.121]$, and use the dependence of $\sigma_{\rm W,Z}$ on $\alphasmZ$ to extract the QCD coupling value preferred by each measurement as explained below. The PDF uncertainties of the theoretical predictions are obtained by taking into account the different eigenvector sets, or replicas, that come with each one of the PDFs: the 28 eigenvector pairs of CT14 (with the final cross section uncertainties divided by a factor of $\sqrt2\erf^{-1}(0.9)\approx1.645$ to convert them from 90\% to 68\% confidence level intervals as for the rest of PDF sets), the 14 eigenvector pairs and 13 variations of the HERAPDF2.0 set, the 25 eigenvector pairs of MMHT14, and the 100 replicas of NNPDF3.0. By construction, all PDF uncertainties are asymmetric except those of the NNPDF3.0 set. For the PDF uncertainties, only NNPDF3.0 provides independent replicas for each $\alphasmZ$ set, which we use in our calculations and uncertainties propagation, whereas the rest of PDFs use the same eigenvectors corresponding to the set determined with $\alphasmZ$ = 0.118. The scale uncertainties of the NNLO calculations are estimated by the usual prescription of independently varying the factorization and renormalization scales within factors of two in seven combinations: $(\mu_F,\mu_R),(\mu_F/2,\mu_R/2),(\mu_F/2,\mu_R),(\mu_F,\mu_R/2),(2\mu_F,\mu_R),(\mu_F,2\mu_R),(2\mu_F,2\mu_R)$.

The EW and photon-induced corrections to the W$^\pm$ and Z boson cross sections are evaluated at NLO accuracy with \mcsanc~v.1.01~\cite{mcsanc}. The computed EW corrections are then applied as a multiplicative correction factor, $\rm K_\mathrm{EW}=\sigma$(NLO,EW-on)/$\sigma$(NLO,EW-off), on top of the \mcfm\ cross sections. They all represent a negative correction, in the range of 0.1--4\%, of the (pure-QCD) DY cross section.
Additional higher-order terms, from other photon-induced and mixed QCD$\otimes$QED NLO processes, estimated to represent a few permille correction of the inclusive DY cross section~\cite{Bertone:2017bme,deFlorian:2018wcj}, are neglected here.

The detailed fiducial cuts on the leptonic final state of each of the 34 experimental measurements, of which 22 are listed in the next section (the same information for CMS appears in Table 1 of Ref.~\cite{Sirunyan:2019wne}), are implemented into \mcfm. All experimental and theoretical EW boson production cross sections quoted in the paper are to be understood as multiplied by their associated leptonic branching fractions, but for simplicity are referred to as ``cross section'' hereafter. A total of $\sim$20\,000 computing jobs are run, taking into account all combinations of eigenvectors/replicas of each one of the 4 PDF sets, the 5 to 7 $\alphasmZ$ values of each PDF set available among (0.115, 0.116, 0.117, 0.118, 0.119, 0.120, and 0.121), and the 34 fiducial cross sections considered. Using at the time the longest 2-week computing queue at the CERN computing center, we reach 0.2--0.6\% numerical accuracy on the theoretical predictions.

\section{Experimental data sets}
\label{sec:3}

The study presented here is based on a combined analysis of most existing measurements of W$^\pm$ and Z boson fiducial cross sections carried out by the ATLAS, CMS, and LHCb experiments at the LHC. The twelve CMS measurements, in separated electron (Z$_{\rm e}$, W$_{\rm e}^\pm$) and muon (Z$_{\mu}$, W$_{\mu}^\pm$) final states, appear compiled in Ref.~\cite{Sirunyan:2019wne}. Here, in addition we collect the seven ATLAS and nine LHCb data sets listed in Tables~\ref{tab:atlas_data} and~\ref{tab:lhcb_data}, respectively. The ATLAS results are for the combined leptonic (average of electron and muon) final states (hereafter labeled by default simply as Z and W$^\pm$, without any subindex, in all tables and plots), whereas those from LHCb are a mixture of electron, muon, and combined-lepton decays. The LHCb measurements are complementary to the ATLAS/CMS ones as they cover different rapidity phase spaces, the central region ($|\eta|\lesssim2.5$) for the latter, and the forward hemisphere ($2<\eta\lesssim 4.5$) for the former. In addition, despite not being used in the final $\alphasmZ$ extraction, but only compared to the NNLO predictions, we also list (Table~\ref{tab:tevatron_data}) the CDF and D0 measurements\footnote{The Tevatron measurements of the \Wpm\ and Z boson cross sections at $\sqrts = 630$~GeV, and similarly the older CERN S$\ppbar$S and the most recent RHIC ones, are not discussed here, as their precision is even worse than that from the 1.8 and 1.96~TeV results.}, which feature 2--3 times larger experimental uncertainties (4.5--6.5\%) than those at the LHC. All tables include the individual acceptance criteria applied on the transverse momentum ($\pT^\ell$ for charged leptons, $\pT^\nu$ for neutrinos) and rapidity $\eta^\ell$  of the decay leptons for each measurement, the mass window around the Z peak considered (or the transverse mass $\mT$ in the \Wpm\ case), as well as the breakdown of the experimental uncertainties.
The ATLAS cross sections have typical uncertainties in the range 1.8--2.8\%, smaller than the 3.4--4.6\% range of CMS~\cite{CMS:2011aa,Chatrchyan:2014mua}, whereas the LHCb data are slightly more precise (1.6--2.7\%) and include the knowledge of the \pp\ collision energy as an extra $\sim$1\% source of systematic uncertainty. 
In all cases, the main source of uncertainty is related to the integrated luminosity as shown in Table~\ref{tab:uncertainties} that summarizes the range of all individual experimental (and theoretical) uncertainties and their correlations (described in more detail in Section~\ref{sec:6}), which is an important source of information when combining all $\alphasmZ$ results as discussed in Section~\ref{sec:6}.

\begin{table}[htpb!]
\caption{Compilation of the seven ATLAS measurements of inclusive \Wpm\ and Z boson production cross sections in $\pp$ collisions at $\sqrts = 7$, 8, 13~TeV~\cite{Aad:2011dm,Aad:2015auj,Aad:2016naf}, used in this analysis. The three sources of experimental uncertainties, and the fiducial selection criteria on the lepton transverse momentum ($\pT^{\ell,\nu}$) and pseudorapidity ($\eta^\ell$), and (transverse) mass windows ($\mT$) $\mZ$ are indicated. \label{tab:atlas_data}}
\centering
\resizebox{\textwidth}{!}{
\begin{tabular}{ll}\hline
ATLAS measurement & Fiducial cross section  \\\hline
$\pp$ at $\sqrts = 7 $~TeV~\cite{Aad:2011dm} & \\
W$^+$, $\pT^\ell > 25$~GeV,  $\pT^\nu > 25$~GeV,  $| \eta^\ell| < 2.5$,  $\mT > 40$~GeV  & $2947 \pm 1_{\text{(stat)}} \pm 15_{\text{(syst)}} \pm 53_{\text{(lumi)}} \text{ pb} = 2947 \pm 55\text{ pb}$ \\
W$^-$, $\pT^\ell > 25$~GeV,  $\pT^\nu > 25$~GeV,  $| \eta^\ell| < 2.5$,  $\mT > 40$~GeV & $1964 \pm 1_{\text{(stat)}} \pm 11_{\text{(syst)}} \pm 35_{\text{(lumi)}} \text{ pb} = 1964 \pm 37\text{ pb}$ \\
Z, $\pT^\ell > 20$~GeV,  $| \eta^\ell| < 2.5$,  $\mZ = 66\mathrm{-}116$~GeV,  central & $502.2 \pm 0.3_{\text{(stat)}} \pm 1.7_{\text{(syst)}} \pm 9.0_{\text{(lumi)}} \text{ pb} = 502.2 \pm 9.2\text{ pb}$  \\\hline
$\pp$ at $\sqrts = 8 $~TeV~\cite{Aad:2015auj} & \\
Z, $\pT^\ell > 20$~GeV,  $| \eta^\ell| < 2.4$,  $\mZ = 66\mathrm{-}116$~GeV & $537.10 \pm 0.45\%_{\text{(syst)}} \pm 2.8\%_{\text{(lumi)}} \text{ pb} = 537.10 \pm 15.23 \text{ pb}$  \\ \hline

$\pp$ at $\sqrts = 13 $~TeV~\cite{Aad:2016naf} & \\	
W$^+$, $\pT^\ell > 25$~GeV,  $\pT^\nu > 25$~GeV,  $| \eta^\ell| < 2.5$,  $\mT > 50$~GeV & $4530 \pm 10_{\text{(stat)}} \pm 90_{\text{(syst)}} \pm 100_{\text{(lumi)}} \text{ pb} = 4530 \pm 130\text{ pb}$ \\
W$^-$, $\pT^\ell > 25$~GeV,  $\pT^\nu > 25$~GeV,  $| \eta^\ell| < 2.5$,  $\mT > 50$~GeV  & $3500 \pm 10_{\text{(stat)}} \pm 70_{\text{(syst)}} \pm 70_{\text{(lumi)}} \text{ pb} = 3500 \pm 100\text{ pb}$ \\
Z, $\pT^\ell > 25$~GeV,  $| \eta^\ell| < 2.5$,  $\mZ = 66\mathrm{-}116$~GeV  & $779 \pm 3_{\text{(stat)}} \pm 6_{\text{(syst)}} \pm 16_{\text{(lumi)}} \text{ pb} = 779 \pm 17\text{ pb}$  \\ \hline
\end{tabular}
}
\end{table}

\begin{table}[htpb!]
\caption{Compilation of the nine LHCb measurements of inclusive \Wpm\ and Z boson production cross sections in $\pp$ collisions at $\sqrts = 7$, 8, 13~TeV~\cite{Aaij:2015gna,Aaij:2015zlq,Aaij:2016mgv,Aaij:2016qqz}, in combined leptonic (\Wpm, Z), electron (W$^\pm_\mathrm{e}$), and muon (W$^+_{\mu}$, Z$_{\mu}$) final states, used in this analysis. The four sources of experimental uncertainties, and the fiducial selection criteria on the lepton transverse momentum ($\pT^{\ell,\mathrm{e},\mu}$) and pseudorapidity ($\eta^{\ell,\mathrm{e},\mu}$), are indicated. \label{tab:lhcb_data}}
\centering
\resizebox{\textwidth}{!}{
\begin{tabular}{ll}\hline
LHCb measurement & Fiducial cross section  \\\hline
$\pp$ at $\sqrts = 7 $~TeV~\cite{Aaij:2015gna} & \\
W$^+$, $\pT^\ell > 20$~GeV,  $2. <  \eta^\ell < 4.5$ & $878.0 \pm 2.1_{\text{(stat)}} \pm 6.7_{\text{(syst)}} \pm 9.3_{\cmenergy} \pm 15.0_{\text{(lumi)}} \text{ pb} = 878.0 \pm 19.0\text{ pb}$ \\
W$^-$, $\pT^\ell > 20$~GeV,  $2. <  \eta^\ell < 4.5$  & $689.5 \pm 2.0_{\text{(stat)}} \pm 5.3_{\text{(syst)}} \pm 6.3_{\cmenergy} \pm 11.8_{\text{(lumi)}} \text{ pb} = 689.5 \pm 14.5\text{ pb}$ \\
Z,  $\pT^\ell > 20$~GeV,  $2. <  \eta^\ell < 4.5$, $\mZ = 60\mathrm{-}120$~GeV & $76.0 \pm 0.3_{\text{(stat)}} \pm 0.5_{\text{(syst)}} \pm 1.0_{\cmenergy} \pm 1.3_{\text{(lumi)}} \text{ pb} = 76.0 \pm 1.7\text{ pb}$  \\\hline
$\pp$ at $\sqrts = 8 $~TeV~\cite{Aaij:2015zlq,Aaij:2016qqz} & \\
W$^+_e$, $\pT^\Pe > 20$~GeV,  $2. <  \eta^\Pe < 4.25$ & $1124.4 \pm 2.1_{\text{(stat)}} \pm 21.5_{\text{(syst)}} \pm 11.2_{\cmenergy} \pm 13.0_{\text{(lumi)}} \text{ pb} = 1124.4 \pm 27.6\text{ pb}$  \\
W$^-_e$, $\pT^\Pe > 20$~GeV,  $2. <  \eta^\Pe < 4.25$ & $809.0 \pm 1.9_{\text{(stat)}} \pm 18.1_{\text{(syst)}} \pm 7.0_{\cmenergy} \pm 9.4_{\text{(lumi)}} \text{ pb} = 809.0 \pm 21.6\text{ pb}$  \\
W$^+_\mu$,  $\pT^\mu > 20$~GeV,  $2. <  \eta^\mu < 4.5$ & $1093.6 \pm 2.1_{\text{(stat)}} \pm 7.2_{\text{(syst)}} \pm 10.9_{\cmenergy} \pm 12.7_{\text{(lumi)}} \text{ pb} = 1093.6 \pm 18.3\text{ pb}$ \\
W$^-_\mu$, $\pT^\mu > 20$~GeV,  $2. <  \eta^\mu < 4.5$ & $818.4 \pm 1.9_{\text{(stat)}} \pm 5.0_{\text{(syst)}} \pm 7.0_{\cmenergy} \pm 9.5_{\text{(lumi)}} \text{ pb} = 818.4 \pm 13.0\text{ pb}$ \\
Z$_\mu$, $\pT^\mu > 20$~GeV,  $2. <  \eta^\mu < 4.5$, $\mZ = 60\mathrm{-}120$~GeV & $95.0 \pm 0.3_{\text{(stat)}} \pm 0.7_{\text{(syst)}} \pm 1.1_{\cmenergy} \pm 1.1_{\text{(lumi)}} \text{ pb} = 95.0 \pm 1.7\text{ pb}$  \\ \hline
$\pp$ at $\sqrts = 13 $~TeV~\cite{Aaij:2016mgv} & \\
Z, $\pT^\ell > 20$~GeV,  $2. <  \eta^\ell < 4.5$, $\mZ = 60\mathrm{-}120$~GeV &  $194.3 \pm 0.9_{\text{(stat)}} \pm 3.3_{\text{(syst)}} \pm 7.6_{\text{(lumi)}} \text{ pb} = 194.3 \pm 8.3\text{ pb}$  \\\hline
\end{tabular}
}
\end{table}

\begin{table}[htpb!]
\caption{Compilation of the six CDF and D0 measurements of inclusive \Wpm\ and Z boson production cross sections in $\ppbar$ collisions at $\sqrts = 1.8$ and 1.96~TeV~\cite{Abe:1995bm,Abbott:1999tt,Abulencia:2005ix}, in the various leptonic final states. The three sources of experimental uncertainties, and the selection criteria on the (W) Z (transverse) masses, are indicated.\label{tab:tevatron_data}}
\centering
\resizebox{\textwidth}{!}{
\begin{tabular}{ll}\hline
Measurement & Inclusive cross section (extrapolated to full acceptance) \\\hline
$\ppbar$ at $\sqrts = 1.8$~TeV (CDF)~\cite{Abe:1995bm} & \\
W$^\pm_{\Pe}$  & $2490 \pm 20_{\text{(stat)}} \pm 80_{\text{(syst)}} \pm 90_{\text{(lumi)}} \text{ pb} = 2490 \pm 120\text{ pb}$ \\
Z$_{\Pe}$ ($66 \,\text{GeV} < \mZ < 116 \,\text{GeV}$) & $231 \pm 6_{\text{(stat)}} \pm 7_{\text{(syst)}} \pm 8_{\text{(lumi)}} \text{ pb} = 231 \pm 12 \text{ pb}$ \\\hline
$\ppbar$ at $\sqrts = 1.8$~TeV (D0)~\cite{Abbott:1999tt} & \\
W$^\pm_{\Pe}$ ($ 40 \,\text{GeV} < \mT < 120 \,\text{GeV}$) & $2310 \pm 10_{\text{(stat)}} \pm 50_{\text{(syst)}} \pm 100_{\text{(lumi)}} \text{ pb} = 2310 \pm 110\text{ pb}$\\
Z$_{\Pe}$($66 \,\text{GeV} < \mZ < 116 \,\text{GeV}$) & $221 \pm 3_{\text{(stat)}} \pm 4_{\text{(syst)}} \pm 10_{\text{(lumi)}} \text{ pb} = 221 \pm 11 \text{ pb}$ \\\hline
$\ppbar$ at $\sqrts = 1.96$~TeV (CDF)~\cite{Abulencia:2005ix} & \\
W$^\pm$ & $2749 \pm 10_{\text{(stat)}} \pm 53_{\text{(syst)}} \pm 165_{\text{(lumi)}} \text{ pb} = 2749 \pm 174\text{ pb}$ \\
Z ($ 66 \,\text{GeV} < \mZ < 116 \,\text{GeV}$) & $254.9 \pm 3.3_{\text{(stat)}} \pm 4.6_{\text{(syst)}} \pm 15.2_{\text{(lumi)}} \text{ pb} = 254.9 \pm 16.2\text{ pb}$ \\\hline
\end{tabular}
}
\end{table}

\begin{table}[htpb!]
\caption{Typical ATLAS, CMS, LHCb and Tevatron experimental and theoretical uncertainties in the W$^\pm$ and Z boson production cross sections, and their degree of correlation (described in more detail in Section~\ref{sec:6}).\label{tab:uncertainties}}
\centering
\resizebox{\textwidth}{!}{
\begin{tabular}{lcl}\\\hline
Source & Uncertainty & Degree of correlation \\\hline
Experimental: & & \\
Luminosity & 2--4\% (CMS), 2--3\% (ATLAS), 1--4\% (LHCb), 3--6\% (CDF, D0) & fully correlated per exp.\ at each $\sqrts$\\
Systematic  &  1--3\% (CMS), 0.3--2\% (ATLAS), 0.6--2\% (LHCb), 2--3\% (CDF, D0) & partially correlated  within each exp.\\
\cm\ energy  &  0.9--1.3\% (LHCb) & fully correlated at each $\sqrts$ \\
Statistical & 0.3--2.5\% (CMS), $<0.4\%$ (ATLAS), 0.2--0.5\% (LHCb), 0.4--2.6\% (CDF, D0)  & uncorrelated \\\hline
Theoretical: & &  \\
PDF & 1--4\% & partially correlated within PDF set \\
Scales & 0.4--1.2\% & partially correlated among all calculations\\
Numerical & 0.2--0.6\% & uncorrelated \\\hline
\end{tabular}
}
\end{table}

\section{Data-theory comparison}
\label{sec:4}

For each of the experimental EW gauge boson cross sections listed in Tables~\ref{tab:atlas_data}, \ref{tab:lhcb_data}, and~\ref{tab:tevatron_data}, we have computed the corresponding theoretical NNLO pQCD predictions using the four PDF sets, and 5 to 7 $\alphasmZ$ variations per PDF aforementioned. The NLO EW corrections are computed with \mcsanc\ just using one single PDF set and QCD coupling value (NNPDF3.0 and $\alphasmZ=0.118$), as those are used to adjust each one of the \mcfm\ cross sections via the multiplicative $\rm K_\mathrm{EW}$ ratio previously defined, where any PDF and $\alphasmZ$ dependencies largely cancel out. 
The numerical comparisons of all experimental and theoretical fiducial cross sections for \Wp, \Wm, and Z production in $\pp$ and $\ppbar$ collisions are listed in Tables~\ref{tab:ATLAS_7_8}--\ref{tab:Tevatron_data_vs_th2}. For each measurement, the fiducial cross section definition and the experimental result are tabulated along with their uncertainties. The \mcfm\ NNLO predictions computed with each of the four PDF sets, with the NLO EW corrections already applied, are listed including their associated PDF, $\alphas$ (obtained, as described in Section~\ref{sec:5} from the cross section change when the default $\alphasmZ = 0.118$ value is modified by $\pm 0.001$), and scale uncertainties. Similarly, the (negative) NLO EW (absolute and relative) correction factors are given. For various systems, the theoretical predictions obtained with alternative \dynnlo~\cite{dynnlo} and \fewz~\cite{fewz} NNLO pQCD calculators using various PDF sets, are also listed as provided in the original experimental references. For the ATLAS cases, the \dynnlo\ and/or \fewz\ predictions include also the NLO EW corrections as computed with \mcsanc\ or \fewz\ itself. The LHCb \fewz\ theoretical predictions, however, do not include any EW correction.
The level of agreement among the three NNLO pQCD calculators predictions for the same system(s) is relatively good, at the 1--2\% level on average, consistent with the different EW schemes and input values of key parameters used, particularly the total and leptonic W and Z widths, combined with the numerical uncertainties of the calculations\footnote{An exception are the 3--5\% differences observed between our \mcfm~8.0 and the \fewz~2.0 calculations of LHCb at $\sqrts = 13$~TeV. A more detailed study of this disagreement would require a careful comparison of the original input setups of both calculators that is beyond the scope of this paper.}. In order to estimate the impact on our $\alphasmZ$ extractions of theoretical differences of this size, we study in Section~\ref{sec:6} the effect of adding an extra 1\% uncorrelated uncertainty to all predictions (Table~\ref{tab:alphas_sensitivity}).

\begin{table}[htpb!]
\caption{Comparison of the ATLAS fiducial cross sections for \Wp, \Wm, and Z boson production in $\pp$ collisions at $\sqrts = 7$ and 8~TeV to the NNLO pQCD results obtained with \mcfm\ using the CT14, HERAPDF2.0, MMHT14, and NNPDF3.0 PDF sets, including NLO EW corrections computed with \mcsanc\ (also listed independently in the last row).\label{tab:ATLAS_7_8}}
\centering
\resizebox{\textwidth}{!}{
\begin{tabular}{ll}\hline
System & Fiducial cross section  \\\hline
$\pp\to W^+(\ell^+\nu)+X$, $\sqrts$ = 7 TeV & ATLAS ($\pT^\ell > 25 \,\text{GeV}$, $\pT^\nu > 25 \,\text{GeV}$, $|\eta| < 2.5$, $\mT > 40 \,\text{GeV}$) \\
ATLAS measurement~\cite{Aad:2011dm}     & $2947 \pm 1_{\stat} \pm 15_{\syst} \pm 53_{\lum} \unit{pb}$ \\
\mcfm\ (NNLO, CT14)                     & $2862\,^{+141}_{-74}\big|_{\,\text{(PDF)}} \pm 28_{\,(\alphas)} \pm 36_{\,(\text{scale})} \pm 18_{\stat} \unit{pb}$ \\
\mcfm\ (NNLO, HERAPDF2.0)               & $3019\,^{+41}_{-84}\big|_{\,\text{(PDF)}} \pm 11_{\,(\alphas)} \pm 36_{\,(\text{scale})} \pm 18_{\stat} \unit{pb}$ \\
\mcfm\ (NNLO, MMHT14)                   & $2880\,^{+156}_{-50}\big|_{\,\text{(PDF)}} \pm 28_{\,(\alphas)} \pm 36_{\,(\text{scale})} \pm 16_{\stat} \unit{pb}$ \\
\mcfm\ (NNLO, NNPDF3.0)                 & $2827 \pm 61_{\,\text{(PDF)}} \pm 25_{\,(\alphas)} \pm 36_{\,(\text{scale})} \pm 18_{\stat} \unit{pb}$ \\
\mcsanc\ (NLO EW, NNPDF3.0) & $-10 \unit{pb} \; (-0.4\%)$ \\\hline
$\pp\to W^-(\ell^-\bar{\nu})+X$, $\sqrts$ = 7 TeV & ATLAS ($\pT^\ell > 25 \,\text{GeV}$, $\pT^\nu > 25 \,\text{GeV}$, $|\eta| < 2.5$, $\mT > 40 \,\text{GeV}$) \\
ATLAS measurement~\cite{Aad:2011dm}     & $1964 \pm 1_{\stat} \pm 11_{\syst} \pm 35_{\lum} \unit{pb}$ \\
\mcfm\ (NNLO, CT14)                     & $1910\,^{+29}_{-88}\big|_{\,\text{(PDF)}} \pm 17_{\,(\alphas)} \pm 12_{\,(\text{scale})} \pm 8_{\stat} \unit{pb}$ \\
\mcfm\ (NNLO, HERAPDF2.0)               & $1993\,^{+13}_{-74}\big|_{\,\text{(PDF)}} \pm 7_{\,(\alphas)} \pm 12_{\,(\text{scale})} \pm 7_{\stat} \unit{pb}$ \\
\mcfm\ (NNLO, MMHT14)                   & $1930\,^{+39}_{-55}\big|_{\,\text{(PDF)}} \pm 20_{\,(\alphas)} \pm 12_{\,(\text{scale})} \pm 7_{\stat} \unit{pb}$ \\
\mcfm\ (NNLO, NNPDF3.0)                 & $1875 \pm 43_{\,\text{(PDF)}} \pm 12_{\,(\alphas)} \pm 12_{\,(\text{scale})} \pm 7_{\stat} \unit{pb}$ \\
\mcsanc\ (NLO EW, NNPDF3.0) & $-5.2 \unit{pb} \; (-0.3\%)$ \\\hline
$\pp\to Z(\ell^+\ell^-)+X$, $\sqrts$ = 7 TeV & ATLAS ($66 \,\text{GeV} < \mZ < 116 \,\text{GeV}$, $\pT^\ell > 20 \,\text{GeV}$, $|\eta| < 2.5$) \\
ATLAS measurement~\cite{Aad:2011dm}     & $502 \pm 0.3_{\stat} \pm 2_{\syst} \pm 9_{\lum} \unit{pb}$ \\
\mcfm\ (NNLO, CT14)                     & $482\,^{+10}_{-16}\big|_{\,\text{(PDF)}} \pm 5_{\,(\alphas)} \pm 2_{\,(\text{scale})} \pm 0.9_{\stat} \unit{pb}$ \\
\mcfm\ (NNLO, HERAPDF2.0)               & $499\,^{+8}_{-8}\big|_{\,\text{(PDF)}} \pm 3_{\,(\alphas)} \pm 2_{\,(\text{scale})} \pm 1_{\stat} \unit{pb}$ \\
\mcfm\ (NNLO, MMHT14)                   & $485\,^{+9}_{-8}\big|_{\,\text{(PDF)}} \pm 5_{\,(\alphas)} \pm 2_{\,(\text{scale})} \pm 1_{\stat} \unit{pb}$ \\
\mcfm\ (NNLO, NNPDF3.0)                 & $474 \pm 10_{\,\text{(PDF)}} \pm 4_{\,(\alphas)} \pm 2_{\,(\text{scale})} \pm 1_{\stat} \unit{pb}$ \\
\mcsanc\ (NLO EW, NNPDF3.0) & $-3.5 \unit{pb} \; (-0.7\%)$ \\\hline
$\pp\to Z(\ell^+\ell^-)+X$, $\sqrts$ = 8 TeV & ATLAS ($66 \,\text{GeV} < \mZ < 116 \,\text{GeV}$, $\pT^\ell > 20 \,\text{GeV}$, $|\eta| < 2.4$) \\
ATLAS measurement~\cite{Aad:2015auj}    & $537 \pm 2_{\syst} \pm 15_{\lum} \unit{pb}$ \\
\mcfm\ (NNLO, CT14)                     & $518\,^{+13}_{-16}\big|_{\,\text{(PDF)}} \pm 5_{\,(\alphas)} \pm 2_{\,(\text{scale})} \pm 1_{\stat} \unit{pb}$ \\
\mcfm\ (NNLO, HERAPDF2.0)               & $537\,^{+12}_{-8}\big|_{\,\text{(PDF)}} \pm 3_{\,(\alphas)} \pm 2_{\,(\text{scale})} \pm 1_{\stat} \unit{pb}$ \\
\mcfm\ (NNLO, MMHT14)                   & $523\,^{+8}_{-8}\big|_{\,\text{(PDF)}} \pm 6_{\,(\alphas)} \pm 2_{\,(\text{scale})} \pm 1_{\stat} \unit{pb}$ \\
\mcfm\ (NNLO, NNPDF3.0)                 & $511 \pm 11_{\,\text{(PDF)}} \pm 4_{\,(\alphas)} \pm 2_{\,(\text{scale})} \pm 1_{\stat} \unit{pb}$ \\
\mcsanc\ (NLO EW, NNPDF3.0) & $-2.5 \unit{pb} \; (-0.5\%)$ \\\hline
\end{tabular}
}
\end{table}

\begin{table}[htpb!]
\caption{Comparison of the ATLAS fiducial cross sections for \Wp, \Wm, and Z boson production in $\pp$ collisions at $\sqrts = 13$~TeV to the NNLO pQCD results obtained with \mcfm\ using the CT14, HERAPDF2.0, MMHT14, and NNPDF3.0 PDF sets, including NLO EW corrections computed with \mcsanc\ (also listed independently in the last row). Alternative \fewz$+$\dynnlo\ NNLO (plus NLO EW) results quoted in the original ATLAS work are also given for comparison.
\label{tab:ATLAS_13}}
\centering
\resizebox{\textwidth}{!}{
\begin{tabular}{ll}\hline
System & Fiducial cross section  \\\hline
$\pp\to W^+(\ell^+\nu)+X$, $\sqrts$ = 13 TeV & ATLAS ($\pT^\ell>25\,\text{GeV}$, $\pT^\nu>25\,\text{GeV}$, $|\eta^\ell|< 2.5$, $\mT > 50 \,\text{GeV}$) \\
ATLAS measurement~\cite{Aad:2016naf}    & $4530 \pm 10_{\stat} \pm 90_{\syst} \pm 100_{\lum} \unit{pb}$ \\
\fewz$+$\dynnlo\ (NNLO, CT14)~\cite{Aad:2016naf}    & $\rm 4420\,^{+130}_{-140}\big|_{(\textsc{pdf})} \pm 50_{(scale)} \pm 80_{(\alphas,\cmenerg)} \unit{pb}$ \\
\mcfm\ (NNLO, CT14)                     & $4479\,^{+342}_{-33}\big|_{\,\text{(PDF)}} \pm 37_{\,(\alphas)} \pm 50_{\,(\text{scale})} \pm 28_{\stat} \unit{pb}$ \\
\mcfm\ (NNLO, HERAPDF2.0)               & $4704\,^{+122}_{-169}\big|_{\,\text{(PDF)}} \pm 21_{\,(\alphas)} \pm 50_{\,(\text{scale})} \pm 35_{\stat} \unit{pb}$ \\
\mcfm\ (NNLO, MMHT14)                   & $4499\,^{+20}_{-338}\big|_{\,\text{(PDF)}} \pm 33_{\,(\alphas)} \pm 50_{\,(\text{scale})} \pm 36_{\stat} \unit{pb}$ \\
\mcfm\ (NNLO, NNPDF3.0)                 & $4382 \pm 104_{\,\text{(PDF)}} \pm 43_{\,(\alphas)} \pm 50_{\,(\text{scale})} \pm 38_{\stat} \unit{pb}$ \\
\mcsanc\ (NLO EW, NNPDF3.0) & $-16 \unit{pb} \; (-0.4\%)$ \\\hline
$\pp\to W^-(\ell^-\bar{\nu})+X$, $\sqrts$ = 13 TeV & ATLAS ($\pT^\ell>25\,\text{GeV}$, $\pT^\nu>25\,\text{GeV}$, $|\eta^\ell|< 2.5$, $\mT > 50 \,\text{GeV}$) \\
ATLAS measurement~\cite{Aad:2016naf}    & $3500 \pm 10_{\stat} \pm 70_{\syst} \pm 70_{\lum} \unit{pb}$ \\
\fewz$+$\dynnlo\ (NNLO, CT14)~\cite{Aad:2016naf}    & $\rm 3400\,^{+90}_{-110}\big|_{(\textsc{pdf})} \pm 40_{(scale)} \pm 60_{(\alphas,\cmenerg)} \unit{pb}$ \\
\mcfm\ (NNLO, CT14)                     & $3381\,^{+111}_{-114}\big|_{\,\text{(PDF)}} \pm 35_{\,(\alphas)} \pm 32_{\,(\text{scale})} \pm 16_{\stat} \unit{pb}$ \\
\mcfm\ (NNLO, HERAPDF2.0)               & $3552\,^{+57}_{-65}\big|_{\,\text{(PDF)}} \pm 11_{\,(\alphas)} \pm 32_{\,(\text{scale})} \pm 16_{\stat} \unit{pb}$ \\
\mcfm\ (NNLO, MMHT14)                   & $3450\,^{+19}_{-210}\big|_{\,\text{(PDF)}} \pm 35_{\,(\alphas)} \pm 32_{\,(\text{scale})} \pm 15_{\stat} \unit{pb}$ \\
\mcfm\ (NNLO, NNPDF3.0)                 & $3320 \pm 80_{\,\text{(PDF)}} \pm 26_{\,(\alphas)} \pm 32_{\,(\text{scale})} \pm 17_{\stat} \unit{pb}$ \\
\mcsanc\ (NLO EW, NNPDF3.0) & $-9.6 \unit{pb} \; (-0.3\%)$ \\\hline
$\pp\to Z(\ell^+\ell^-)+X$, $\sqrts$ = 13 TeV & ATLAS ($66 \,\text{GeV} < \mZ < 116 \,\text{GeV}$, $\pT^\ell > 25 \,\text{GeV}$, $|\eta^\ell| < 2.5$) \\
ATLAS measurement~\cite{Aad:2016naf}    & $779 \pm 3_{\stat} \pm 6_{\syst} \pm 16_{\lum} \unit{pb}$ \\
\fewz$+$\dynnlo\ (NNLO, CT14)~\cite{Aad:2016naf}    & $\rm 740\,^{+20}_{-30}\big|_{(\textsc{pdf})} \pm 10_{(scale)} \pm 10_{(\alphas,\cmenerg)} \unit{pb}$ \\
\mcfm\ (NNLO, CT14)                     & $748\,^{+27}_{-21}\big|_{\,\text{(PDF)}} \pm 8_{\,(\alphas)} \pm 5_{\,(\text{scale})} \pm 2_{\stat} \unit{pb}$ \\
\mcfm\ (NNLO, HERAPDF2.0)               & $779\,^{+14}_{-10}\big|_{\,\text{(PDF)}} \pm 4_{\,(\alphas)} \pm 5_{\,(\text{scale})} \pm 2_{\stat} \unit{pb}$ \\
\mcfm\ (NNLO, MMHT14)                   & $758\,^{+8}_{-23}\big|_{\,\text{(PDF)}} \pm 8_{\,(\alphas)} \pm 5_{\,(\text{scale})} \pm 2_{\stat} \unit{pb}$ \\
\mcfm\ (NNLO, NNPDF3.0)                 & $734 \pm 17_{\,\text{(PDF)}} \pm 6_{\,(\alphas)} \pm 5_{\,(\text{scale})} \pm 2_{\stat} \unit{pb}$ \\
\mcsanc\ (NLO EW, NNPDF3.0) & $-4.7 \unit{pb} \; (-0.6\%)$ \\\hline
\end{tabular}
}
\end{table}

\begin{table}[htpb!]
\caption{Comparison of the LHCb fiducial cross sections for \Wp, \Wm, and Z boson production in $\pp$ collisions at $\sqrts = 7$ and 13~TeV to the NNLO pQCD results obtained with \mcfm\ using the CT14, HERAPDF2.0, MMHT14, and NNPDF3.0 PDF sets, including NLO EW corrections computed with \mcsanc\ (also listed independently in the last row). Alternative \fewz\ NNLO results (without EW corrections) quoted in the original LHCb works are also given for comparison.
\label{tab:LHCb_7_13}}
\centering
\resizebox{\textwidth}{!}{
\begin{tabular}{ll}\hline
System & Fiducial cross section  \\\hline
$\pp\to W^+(\mu^+\nu)+X$, $\sqrts$ = 7 TeV & ($\pT^\ell > 20 \,\text{GeV}$, $2. < \eta^\ell < 4.5$) \\
LHCb measurement~\cite{Aaij:2015gna}    & $878 \pm 2_{\stat} \pm 7_{\syst} \pm 9_{\cmenergy} \pm 15_{\lum} \unit{pb}$ \\
\fewz\ (NNLO, NNPDF3.0; w/o EW corr.)~\cite{Aaij:2015gna}        & $867 \pm 22 \unit{pb}$ \\
\mcfm\ (NNLO, CT14)                     & $890\,^{+25}_{-29}\big|_{\,\text{(PDF)}} \pm 8_{\,(\alphas)} \pm 10_{\,(\text{scale})} \pm 5_{\stat} \unit{pb}$ \\
\mcfm\ (NNLO, HERAPDF2.0)               & $916\,^{+17}_{-44}\big|_{\,\text{(PDF)}} \pm 6_{\,(\alphas)} \pm 10_{\,(\text{scale})} \pm 5_{\stat} \unit{pb}$ \\
\mcfm\ (NNLO, MMHT14)                   & $909\,^{+29}_{-33}\big|_{\,\text{(PDF)}} \pm 8_{\,(\alphas)} \pm 10_{\,(\text{scale})} \pm 4_{\stat} \unit{pb}$ \\
\mcfm\ (NNLO, NNPDF3.0)                 & $878 \pm 20_{\,\text{(PDF)}} \pm 8_{\,(\alphas)} \pm 10_{\,(\text{scale})} \pm 5_{\stat} \unit{pb}$ \\
\mcsanc\ (NLO EW, NNPDF3.0) & $-0.91 \unit{pb} \; (-0.1\%)$ \\\hline
$\pp\to  W^-(\mu^-\bar{\nu})+X$, $\sqrts$ = 7 TeV & ( $\pT^\ell > 20 \,\text{GeV}$, $2. < \eta^\ell < 4.5$) \\
LHCb measurement~\cite{Aaij:2015gna}    & $690 \pm 2_{\stat} \pm 5_{\syst} \pm 6_{\cmenergy} \pm 12_{\lum} \unit{pb}$ \\
\fewz\ (NNLO, NNPDF3.0; w/o EW corr.)~\cite{Aaij:2015gna}  & $677 \pm 18 \unit{pb}$ \\
\mcfm\ (NNLO, CT14)                     & $704\,^{+13}_{-36}\big|_{\,\text{(PDF)}} \pm 6_{\,(\alphas)} \pm 7_{\,(\text{scale})} \pm 4_{\stat} \unit{pb}$ \\
\mcfm\ (NNLO, HERAPDF2.0)               & $738\,^{+15}_{-29}\big|_{\,\text{(PDF)}} \pm 4_{\,(\alphas)} \pm 7_{\,(\text{scale})} \pm 4_{\stat} \unit{pb}$ \\
\mcfm\ (NNLO, MMHT14)                   & $703\,^{+33}_{-17}\big|_{\,\text{(PDF)}} \pm 6_{\,(\alphas)} \pm 7_{\,(\text{scale})} \pm 4_{\stat} \unit{pb}$ \\
\mcfm\ (NNLO, NNPDF3.0)                 & $695 \pm 18_{\,\text{(PDF)}} \pm 6_{\,(\alphas)} \pm 7_{\,(\text{scale})} \pm 4_{\stat} \unit{pb}$ \\
\mcsanc\ (NLO EW, NNPDF3.0) & $-2.9 \unit{pb} \; (-0.4\%)$ \\\hline
$\pp\to Z(\mu^+\mu^-)+X$, $\sqrts$ = 7 TeV & ($60 \,\text{GeV} < \mZ < 120 \,\text{GeV}$, $\pT^\ell > 20 \,\text{GeV}$, $2. < \eta^\ell < 4.5$ ) \\
LHCb measurement~\cite{Aaij:2015gna}    & $76 \pm 0.3_{\stat} \pm 0.5_{\syst} \pm 1_{\cmenergy} \pm 1_{\lum} \unit{pb}$ \\
\fewz\ (NNLO, NNPDF3.0; w/o EW corr.)~\cite{Aaij:2015gna}        & $72.7 \pm 1.8 \unit{pb}$ \\
\mcfm\ (NNLO, CT14)                     & $76\,^{+4}_{-1}\big|_{\,\text{(PDF)}} \pm 1_{\,(\alphas)} \pm 0.9_{\,(\text{scale})} \pm 0.3_{\stat} \unit{pb}$ \\
\mcfm\ (NNLO, HERAPDF2.0)               & $81\,^{+2}_{-1}\big|_{\,\text{(PDF)}} \pm 1_{\,(\alphas)} \pm 0.9_{\,(\text{scale})} \pm 0.3_{\stat} \unit{pb}$ \\
\mcfm\ (NNLO, MMHT14)                   & $77\,^{+3}_{-1}\big|_{\,\text{(PDF)}} \pm 1_{\,(\alphas)} \pm 0.9_{\,(\text{scale})} \pm 0.3_{\stat} \unit{pb}$ \\
\mcfm\ (NNLO, NNPDF3.0)                 & $75 \pm 2_{\,\text{(PDF)}} \pm 1_{\,(\alphas)} \pm 0.9_{\,(\text{scale})} \pm 0.3_{\stat} \unit{pb}$ \\
\mcsanc\ (NLO EW, NNPDF3.0) & $-0.42 \unit{pb} \; (-0.6\%)$ \\\hline
$\pp\to Z(\ell^+\ell^-)+X$, $\sqrts$ = 13 TeV & ($60 \,\text{GeV} < \mZ < 120 \,\text{GeV}$, $\pT^\ell > 20 \,\text{GeV}$, $2. < \eta^\ell < 4.5$) \\
LHCb measurement~\cite{Aaij:2016mgv}    & $194 \pm 0.9_{\stat} \pm 3_{\syst} \pm 8_{\lum} \unit{pb}$ \\
\fewz\ (NNLO, various PDFs; w/o EW corr.)~\cite{Aaij:2016mgv}       & $193 \pm 7 \unit{pb}$ (CT14), $193 \pm 6 \unit{pb}$ (MMHT14), $188 \pm 5 \unit{pb}$ (NNPDF3.0) \\
\mcfm\ (NNLO, CT14)                     & $203\,^{+7}_{-5}\big|_{\,\text{(PDF)}} \pm 2_{\,(\alphas)} \pm 2_{\,(\text{scale})} \pm 0.9_{\stat} \unit{pb}$ \\
\mcfm\ (NNLO, HERAPDF2.0)               & $210\,^{+3}_{-9}\big|_{\,\text{(PDF)}} \pm 2_{\,(\alphas)} \pm 2_{\,(\text{scale})} \pm 0.9_{\stat} \unit{pb}$ \\
\mcfm\ (NNLO, MMHT14)                   & $205\,^{+7}_{-5}\big|_{\,\text{(PDF)}} \pm 2_{\,(\alphas)} \pm 2_{\,(\text{scale})} \pm 1_{\stat} \unit{pb}$ \\
\mcfm\ (NNLO, NNPDF3.0)                 & $195 \pm 5_{\,\text{(PDF)}} \pm 2_{\,(\alphas)} \pm 2_{\,(\text{scale})} \pm 1_{\stat} \unit{pb}$ \\
\mcsanc\ (NLO EW, NNPDF3.0) & $-1.1 \unit{pb} \; (-0.6\%)$ \\\hline
\end{tabular}
}
\end{table}

\begin{table}[htpb!]
\caption{Comparison of the LHCb fiducial cross sections for \Wp, \Wm, and Z boson production in $\pp$ collisions at $\sqrts = 8$~TeV to the NNLO pQCD results obtained with \mcfm\ using the CT14, HERAPDF2.0, MMHT14, and NNPDF3.0 PDF sets, including NLO EW corrections computed with \mcsanc\ (also listed independently in the last row). Alternative \fewz\ NNLO results (without EW corrections) quoted in the original LHCb works are also given for comparison.\label{tab:LHCb_8}}
\centering
\resizebox{\textwidth}{!}{
\begin{tabular}{ll}\hline
System & Fiducial cross section  \\\hline
$\pp\to W^+(e^+\nu)+X$, $\sqrts$ = 8 TeV & ($\pT^\ell > 20 \,\text{GeV}$, $2. < \eta^\ell < 4.25$) \\
LHCb measurement~\cite{Aaij:2016qqz}    & $1124 \pm 2_{\stat} \pm 22_{\syst} \pm 11_{\cmenergy} \pm 13_{\lum} \unit{pb}$ \\
\mcfm\ (NNLO, CT14)                     & $1095\,^{+41}_{-31}\big|_{\,\text{(PDF)}} \pm 9_{\,(\alphas)} \pm 21_{\,(\text{scale})} \pm 6_{\stat} \unit{pb}$ \\
\mcfm\ (NNLO, HERAPDF2.0)               & $1131\,^{+50}_{-42}\big|_{\,\text{(PDF)}} \pm 8_{\,(\alphas)} \pm 21_{\,(\text{scale})} \pm 7_{\stat} \unit{pb}$ \\
\mcfm\ (NNLO, MMHT14)                   & $1117\,^{+28}_{-37}\big|_{\,\text{(PDF)}} \pm 12_{\,(\alphas)} \pm 21_{\,(\text{scale})} \pm 7_{\stat} \unit{pb}$ \\
\mcfm\ (NNLO, NNPDF3.0)                 & $1081 \pm 25_{\,\text{(PDF)}} \pm 10_{\,(\alphas)} \pm 21_{\,(\text{scale})} \pm 7_{\stat} \unit{pb}$ \\
\mcsanc\ (NLO EW, NNPDF3.0) & $-3.4 \unit{pb} \; (-0.3\%)$ \\\hline
$\pp\to W^-(e^-\bar{\nu})+X$, $\sqrts$ = 8 TeV & ($\pT^\ell > 20 \,\text{GeV}$, $2. < \eta^\ell < 4.25$) \\
LHCb measurement~\cite{Aaij:2016qqz}    & $809 \pm 2_{\stat} \pm 18_{\syst} \pm 7_{\cmenergy} \pm 9_{\lum} \unit{pb}$ \\
\mcfm\ (NNLO, CT14)                     & $849\,^{+6}_{-59}\big|_{\,\text{(PDF)}} \pm 7_{\,(\alphas)} \pm 8_{\,(\text{scale})} \pm 7_{\stat} \unit{pb}$ \\
\mcfm\ (NNLO, HERAPDF2.0)               & $892\,^{+28}_{-30}\big|_{\,\text{(PDF)}} \pm 8_{\,(\alphas)} \pm 8_{\,(\text{scale})} \pm 7_{\stat} \unit{pb}$ \\
\mcfm\ (NNLO, MMHT14)                   & $853\,^{+33}_{-46}\big|_{\,\text{(PDF)}} \pm 6_{\,(\alphas)} \pm 8_{\,(\text{scale})} \pm 5_{\stat} \unit{pb}$ \\
\mcfm\ (NNLO, NNPDF3.0)                 & $840 \pm 21_{\,\text{(PDF)}} \pm 7_{\,(\alphas)} \pm 8_{\,(\text{scale})} \pm 6_{\stat} \unit{pb}$ \\
\mcsanc\ (NLO EW, NNPDF3.0) & $-2.7 \unit{pb} \; (-0.4\%)$ \\\hline
$\pp\to W^+(\mu^+\nu)+X$, $\sqrts$ = 8 TeV & ($\pT > 20 \,\text{GeV}$, $2. < \eta < 4.5$) \\
LHCb measurement~\cite{Aaij:2015zlq}    & $1094 \pm 2_{\stat} \pm 7_{\syst} \pm 11_{\cmenergy} \pm 13_{\lum} \unit{pb}$ \\
\fewz\ (NNLO, various PDFs; w/o EW corr.)~\cite{Aaij:2015zlq}       & $1078 \pm 33 \unit{pb}$ (CT14), $1093 \pm 26 \unit{pb}$ (MMHT14), $1067 \pm 28 \unit{pb}$ (NNPDF3.0) \\
\mcfm\ (NNLO, CT14)                     & $1099\,^{+32}_{-46}\big|_{\,\text{(PDF)}} \pm 9_{\,(\alphas)} \pm 17_{\,(\text{scale})} \pm 7_{\stat} \unit{pb}$ \\
\mcfm\ (NNLO, HERAPDF2.0)               & $1134\,^{+12}_{-41}\big|_{\,\text{(PDF)}} \pm 5_{\,(\alphas)} \pm 17_{\,(\text{scale})} \pm 7_{\stat} \unit{pb}$ \\
\mcfm\ (NNLO, MMHT14)                   & $1111\,^{+65}_{-26}\big|_{\,\text{(PDF)}} \pm 10_{\,(\alphas)} \pm 17_{\,(\text{scale})} \pm 8_{\stat} \unit{pb}$ \\
\mcfm\ (NNLO, NNPDF3.0)                 & $1084 \pm 25_{\,\text{(PDF)}} \pm 10_{\,(\alphas)} \pm 17_{\,(\text{scale})} \pm 7_{\stat} \unit{pb}$ \\
\mcsanc\ (NLO EW, NNPDF3.0) & $-1.2 \unit{pb} \; (-0.1\%)$ \\\hline
$\pp\to W^-(\mu^-\bar{\nu})+X$, $\sqrts$ = 8 TeV & ($\pT^\ell > 20 \,\text{GeV}$, $2. < \eta^\ell < 4.5$) \\
LHCb measurement~\cite{Aaij:2015zlq}    & $818 \pm 2_{\stat} \pm 5_{\syst} \pm 7_{\cmenergy} \pm 10_{\lum} \unit{pb}$ \\
\fewz\ (NNLO, various PDFs; w/o EW corr.)~\cite{Aaij:2015zlq}       & $817 \pm 24 \unit{pb}$ (CT14), $826 \pm 21 \unit{pb}$ (MMHT14), $805 \pm 22 \unit{pb}$ (NNPDF3.0) pb \\
\mcfm\ (NNLO, CT14)                     & $831\,^{+13}_{-59}\big|_{\,\text{(PDF)}} \pm 7_{\,(\alphas)} \pm 10_{\,(\text{scale})} \pm 5_{\stat} \unit{pb}$ \\
\mcfm\ (NNLO, HERAPDF2.0)               & $883\,^{+46}_{-8}\big|_{\,\text{(PDF)}} \pm 5_{\,(\alphas)} \pm 10_{\,(\text{scale})} \pm 6_{\stat} \unit{pb}$ \\
\mcfm\ (NNLO, MMHT14)                   & $839\,^{+13}_{-56}\big|_{\,\text{(PDF)}} \pm 8_{\,(\alphas)} \pm 10_{\,(\text{scale})} \pm 5_{\stat} \unit{pb}$ \\
\mcfm\ (NNLO, NNPDF3.0)                 & $828 \pm 21_{\,\text{(PDF)}} \pm 7_{\,(\alphas)} \pm 10_{\,(\text{scale})} \pm 5_{\stat} \unit{pb}$ \\
\mcsanc\ (NLO EW, NNPDF3.0) & $-12 \unit{pb} \; (-1.5\%)$ \\\hline
$\pp\to Z(\mu^+\mu^-)+X$, $\sqrts$ = 8 TeV & ($60 \,\text{GeV} < \mZ < 120 \,\text{GeV}$, $\pT^\ell > 20 \,\text{GeV}$, $2. < \eta^\ell < 4.5$)\\
LHCb measurement~\cite{Aaij:2015zlq}    & $95 \pm 0.3_{\stat} \pm 0.7_{\syst} \pm 1_{\cmenergy} \pm 1_{\lum} \unit{pb}$ \\
\fewz\ (NNLO, various PDFs; w/o EW corr.)~\cite{Aaij:2015zlq}       & $93.9 \pm 2.8 \unit{pb}$ (CT14), $94.8 \pm 2.3 \unit{pb}$ (MMHT14), $92.6 \pm 2.4 \unit{pb}$ (NNPDF3.0) \\
\mcfm\ (NNLO, CT14)                     & $97\,^{+3}_{-3}\big|_{\,\text{(PDF)}} \pm 1_{\,(\alphas)} \pm 1_{\,(\text{scale})} \pm 0.4_{\stat} \unit{pb}$ \\
\mcfm\ (NNLO, HERAPDF2.0)               & $103\,^{+1}_{-4}\big|_{\,\text{(PDF)}} \pm 1_{\,(\alphas)} \pm 1_{\,(\text{scale})} \pm 0.3_{\stat} \unit{pb}$ \\
\mcfm\ (NNLO, MMHT14)                   & $98\,^{+3}_{-2}\big|_{\,\text{(PDF)}} \pm 1_{\,(\alphas)} \pm 1_{\,(\text{scale})} \pm 0.4_{\stat} \unit{pb}$ \\
\mcfm\ (NNLO, NNPDF3.0)                 & $95 \pm 2_{\,\text{(PDF)}} \pm 1_{\,(\alphas)} \pm 1_{\,(\text{scale})} \pm 0.4_{\stat} \unit{pb}$ \\
\mcsanc\ (NLO EW, NNPDF3.0) & $-0.59 \unit{pb} \; (-0.7\%)$ \\\hline
\end{tabular}
}
\end{table}

\begin{table}[htpb!]
\caption{Comparison of the CDF and D0 inclusive cross sections for \Wpm\ and Z boson production in $\ppbar$ collisions at $\sqrts = 1.8$~TeV to the NNLO pQCD results obtained with \mcfm\ using the CT14, HERAPDF2.0, MMHT14, and NNPDF3.0 PDF sets, including NLO EW corrections computed with \mcsanc\ (also listed independently in the last row).\label{tab:Tevatron_data_vs_th}}
\centering
\resizebox{\textwidth}{!}{
\begin{tabular}{ll}\hline
System & Inclusive cross section (extrapolated to full acceptance)\\\hline
$\ppbar\to W(e^\pm\nu)+X$, $\sqrts$ = 1.8 TeV & (no kinematics cuts) \\
CDF measurement~\cite{Abe:1995bm}       & $2490 \pm 20_{\stat} \pm 80_{\syst} \pm 90_{\lum} \unit{pb}$ \\
\mcfm\ (NNLO, CT14)                     & $2516\,^{+59}_{-67}\big|_{\,\text{(PDF)}} \pm 11_{\,(\alphas)} \pm 13_{\,(\text{scale})} \pm 3_{\stat} \unit{pb}$ \\
\mcfm\ (NNLO, HERAPDF2.0)               & $2570\,^{+46}_{-45}\big|_{\,\text{(PDF)}} \pm 11_{\,(\alphas)} \pm 13_{\,(\text{scale})} \pm 4_{\stat} \unit{pb}$ \\
\mcfm\ (NNLO, MMHT14)                   & $2549\,^{+42}_{-72}\big|_{\,\text{(PDF)}} \pm 17_{\,(\alphas)} \pm 13_{\,(\text{scale})} \pm 3_{\stat} \unit{pb}$ \\
\mcfm\ (NNLO, NNPDF3.0)                 & $2469 \pm 56_{\,\text{(PDF)}} \pm 15_{\,(\alphas)} \pm 13_{\,(\text{scale})} \pm 3_{\stat} \unit{pb}$ \\
\mcsanc\ (NLO EW, NNPDF3.0) & $-1.2 \unit{pb} \; (-0.1\%)$ \\\hline
$\ppbar\to Z(e^+e^-)+X$, $\sqrts$ = 1.8 TeV & ($ 66 \,\text{GeV} < \mZ < 116 \,\text{GeV}$) \\
CDF measurement~\cite{Abe:1995bm}       & $231 \pm 6_{\stat} \pm 7_{\syst} \pm 8_{\lum} \unit{pb}$ \\
\mcfm\ (NNLO, CT14)                     & $228\,^{+5}_{-6}\big|_{\,\text{(PDF)}} \pm 1_{\,(\alphas)} \pm 1_{\,(\text{scale})} \pm 0.3_{\stat} \unit{pb}$ \\
\mcfm\ (NNLO, HERAPDF2.0)               & $236\,^{+3}_{-4}\big|_{\,\text{(PDF)}} \pm 1_{\,(\alphas)} \pm 1_{\,(\text{scale})} \pm 0.3_{\stat} \unit{pb}$ \\
\mcfm\ (NNLO, MMHT14)                   & $230\,^{+5}_{-5}\big|_{\,\text{(PDF)}} \pm 2_{\,(\alphas)} \pm 1_{\,(\text{scale})} \pm 0.3_{\stat} \unit{pb}$ \\
\mcfm\ (NNLO, NNPDF3.0)                 & $222 \pm 5_{\,\text{(PDF)}} \pm 2_{\,(\alphas)} \pm 1_{\,(\text{scale})} \pm 0.3_{\stat} \unit{pb}$ \\
\mcsanc\ (NLO EW, NNPDF3.0) & $-3.6 \unit{pb} \; (-2.2\%)$ \\\hline
$\ppbar\to  W(e^\pm\nu)+X$, $\sqrts$ = 1.8 TeV & ($40 \,\text{GeV} < \mW < 120 \,\text{GeV}$) \\
D0 measurement~\cite{Abbott:1999tt}     & $2310 \pm 10_{\stat} \pm 50_{\syst} \pm 100_{\lum} \unit{pb}$ \\
\mcfm\ (NNLO, CT14)                     & $2476\,^{+62}_{-63}\big|_{\,\text{(PDF)}} \pm 11_{\,(\alphas)} \pm 12_{\,(\text{scale})} \pm 3_{\stat} \unit{pb}$ \\
\mcfm\ (NNLO, HERAPDF2.0)               & $2529\,^{+44}_{-42}\big|_{\,\text{(PDF)}} \pm 11_{\,(\alphas)} \pm 12_{\,(\text{scale})} \pm 3_{\stat} \unit{pb}$ \\
\mcfm\ (NNLO, MMHT14)                   & $2506\,^{+55}_{-52}\big|_{\,\text{(PDF)}} \pm 16_{\,(\alphas)} \pm 12_{\,(\text{scale})} \pm 3_{\stat} \unit{pb}$ \\
\mcfm\ (NNLO, NNPDF3.0)                 & $2430 \pm 55_{\,\text{(PDF)}} \pm 15_{\,(\alphas)} \pm 12_{\,(\text{scale})} \pm 3_{\stat} \unit{pb}$ \\
\mcsanc\ (NLO EW, NNPDF3.0) & $-5.7 \unit{pb} \; (-0.5\%)$ \\\hline
$\ppbar\to Z(e^+e^-)+X$, $\sqrts$ = 1.8 TeV & ($66 \,\text{GeV} < \mZ < 116 \,\text{GeV}$) \\
D0 measurement~\cite{Abbott:1999tt}     & $221 \pm 3_{\stat} \pm 4_{\syst} \pm 10_{\lum} \unit{pb}$ \\
\mcfm\ (NNLO, CT14)                     & $228\,^{+6}_{-5}\big|_{\,\text{(PDF)}} \pm 1_{\,(\alphas)} \pm 1_{\,(\text{scale})} \pm 0.3_{\stat} \unit{pb}$ \\
\mcfm\ (NNLO, HERAPDF2.0)               & $236\,^{+3}_{-4}\big|_{\,\text{(PDF)}} \pm 1_{\,(\alphas)} \pm 1_{\,(\text{scale})} \pm 0.3_{\stat} \unit{pb}$ \\
\mcfm\ (NNLO, MMHT14)                   & $230\,^{+5}_{-5}\big|_{\,\text{(PDF)}} \pm 2_{\,(\alphas)} \pm 1_{\,(\text{scale})} \pm 0.3_{\stat} \unit{pb}$ \\
\mcfm\ (NNLO, NNPDF3.0)                 & $222 \pm 5_{\,\text{(PDF)}} \pm 2_{\,(\alphas)} \pm 1_{\,(\text{scale})} \pm 0.3_{\stat} \unit{pb}$ \\
\mcsanc\ (NLO EW, NNPDF3.0) & $-3.6 \unit{pb} \; (-2.2\%)$ \\\hline
\end{tabular}
}
\end{table}

\begin{table}[htpb!]
\caption{Comparison of the CDF and D0 inclusive cross sections for \Wpm\ and Z boson production in $\ppbar$ collisions at $\sqrts = 1.96$~TeV to the NNLO pQCD results obtained with \mcfm\ using the CT14, HERAPDF2.0, MMHT14, and NNPDF3.0 PDF sets, including NLO EW corrections computed with \mcsanc\ (also listed independently in the last row).\label{tab:Tevatron_data_vs_th2}}
\centering
\resizebox{\textwidth}{!}{
\begin{tabular}{ll}\hline
System & Inclusive cross section (extrapolated to full acceptance)\\\hline
$\ppbar\to W(\ell\nu)+X$, $\sqrts$ = 1.96 TeV & (no kinematics cuts) \\
CDF measurement~\cite{Abulencia:2005ix} & $2749 \pm 10_{\stat} \pm 53_{\syst} \pm 165_{\lum} \unit{pb}$ \\
\mcfm\ (NNLO, CT14)                     & $2754\,^{+68}_{-72}\big|_{\,\text{(PDF)}} \pm 13_{\,(\alphas)} \pm 15_{\,(\text{scale})} \pm 4_{\stat} \unit{pb}$ \\
\mcfm\ (NNLO, HERAPDF2.0)               & $2815\,^{+48}_{-54}\big|_{\,\text{(PDF)}} \pm 11_{\,(\alphas)} \pm 15_{\,(\text{scale})} \pm 4_{\stat} \unit{pb}$ \\
\mcfm\ (NNLO, MMHT14)                   & $2788\,^{+78}_{-42}\big|_{\,\text{(PDF)}} \pm 20_{\,(\alphas)} \pm 15_{\,(\text{scale})} \pm 4_{\stat} \unit{pb}$ \\
\mcfm\ (NNLO, NNPDF3.0)                 & $2703 \pm 60_{\,\text{(PDF)}} \pm 17_{\,(\alphas)} \pm 15_{\,(\text{scale})} \pm 4_{\stat} \unit{pb}$ \\
\mcsanc\ (NLO EW, NNPDF3.0) & $-3.6 \unit{pb} \; (-0.3\%)$ \\\hline
$\ppbar\to Z(\ell^+\ell^-)+X$, $\sqrts$ = 1.96 TeV & ($66 \,\text{GeV} < \mZ < 116 \,\text{GeV}$) \\
CDF measurement~\cite{Abulencia:2005ix} & $255 \pm 3_{\stat} \pm 5_{\syst} \pm 15_{\lum} \unit{pb}$ \\
\mcfm\ (NNLO, CT14)                     & $250\,^{+5}_{-6}\big|_{\,\text{(PDF)}} \pm 1_{\,(\alphas)} \pm 1_{\,(\text{scale})} \pm 0.3_{\stat} \unit{pb}$ \\
\mcfm\ (NNLO, HERAPDF2.0)               & $258\,^{+3}_{-5}\big|_{\,\text{(PDF)}} \pm 1_{\,(\alphas)} \pm 1_{\,(\text{scale})} \pm 0.3_{\stat} \unit{pb}$ \\
\mcfm\ (NNLO, MMHT14)                   & $252\,^{+5}_{-4}\big|_{\,\text{(PDF)}} \pm 2_{\,(\alphas)} \pm 1_{\,(\text{scale})} \pm 0.3_{\stat} \unit{pb}$ \\
\mcfm\ (NNLO, NNPDF3.0)                 & $243 \pm 5_{\,\text{(PDF)}} \pm 2_{\,(\alphas)} \pm 1_{\,(\text{scale})} \pm 0.3_{\stat} \unit{pb}$ \\
\mcsanc\ (NLO EW, NNPDF3.0) & $-4.2 \unit{pb} \; (-2.3\%)$ \\\hline

\hline
\end{tabular}
}
\end{table}
\clearpage

\begin{figure}[htbp!]
\includegraphics[width=0.49\textwidth]{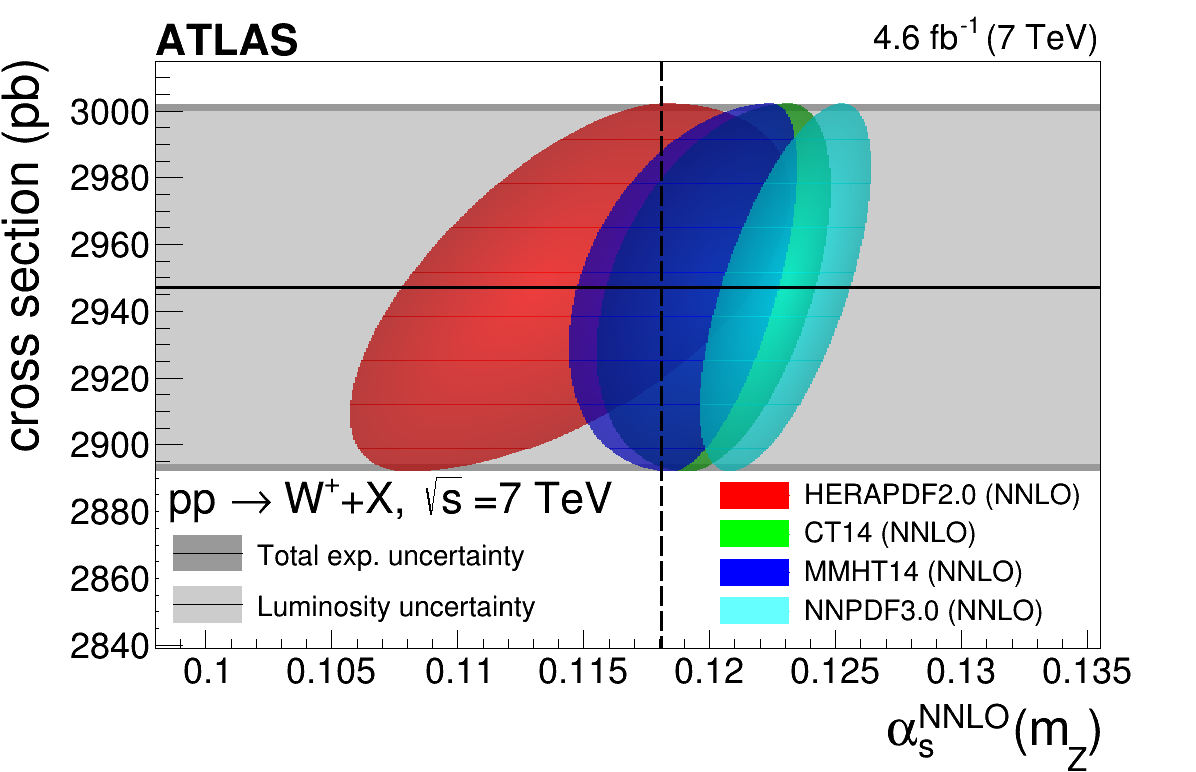}
\includegraphics[width=0.49\textwidth]{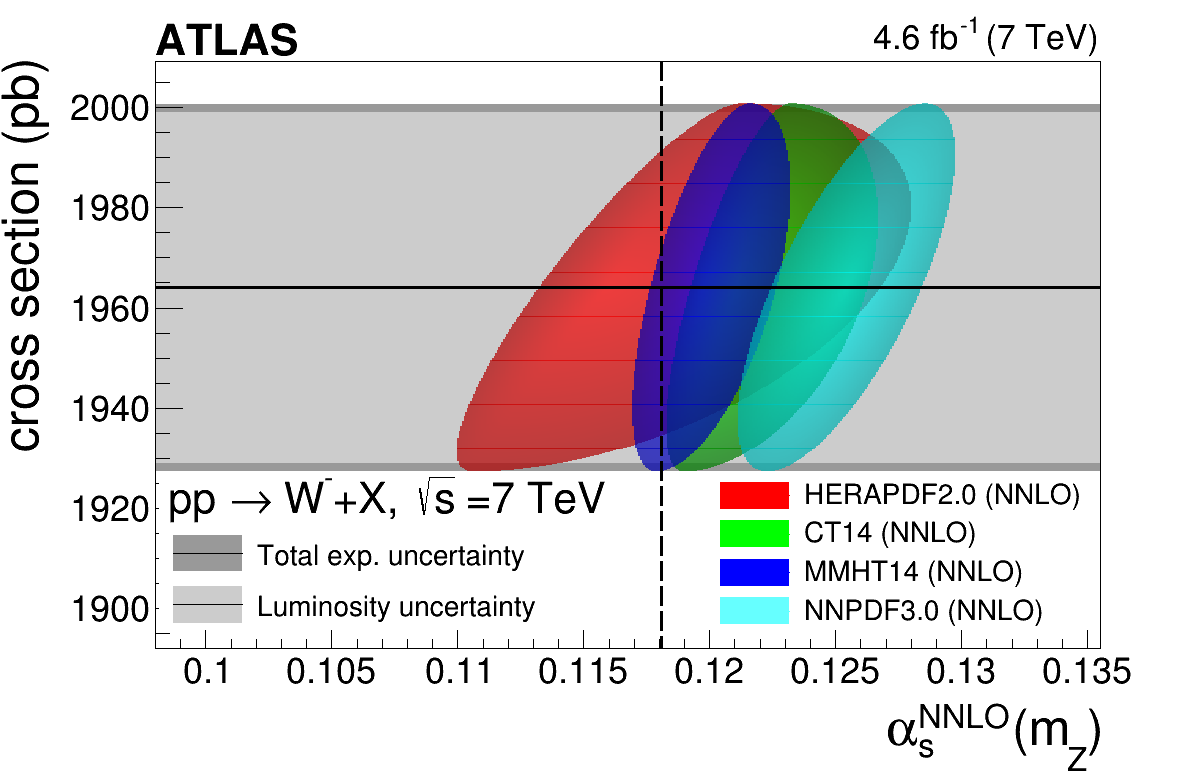}
\includegraphics[width=0.49\textwidth]{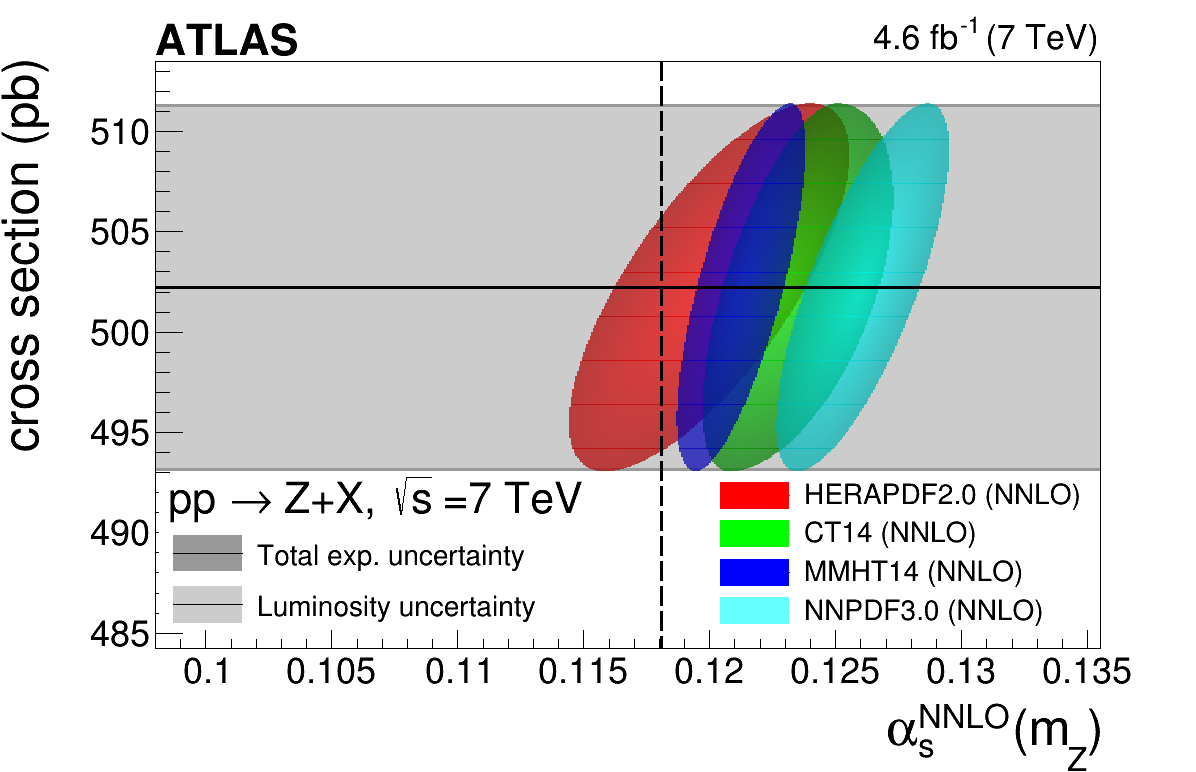}
\includegraphics[width=0.49\textwidth]{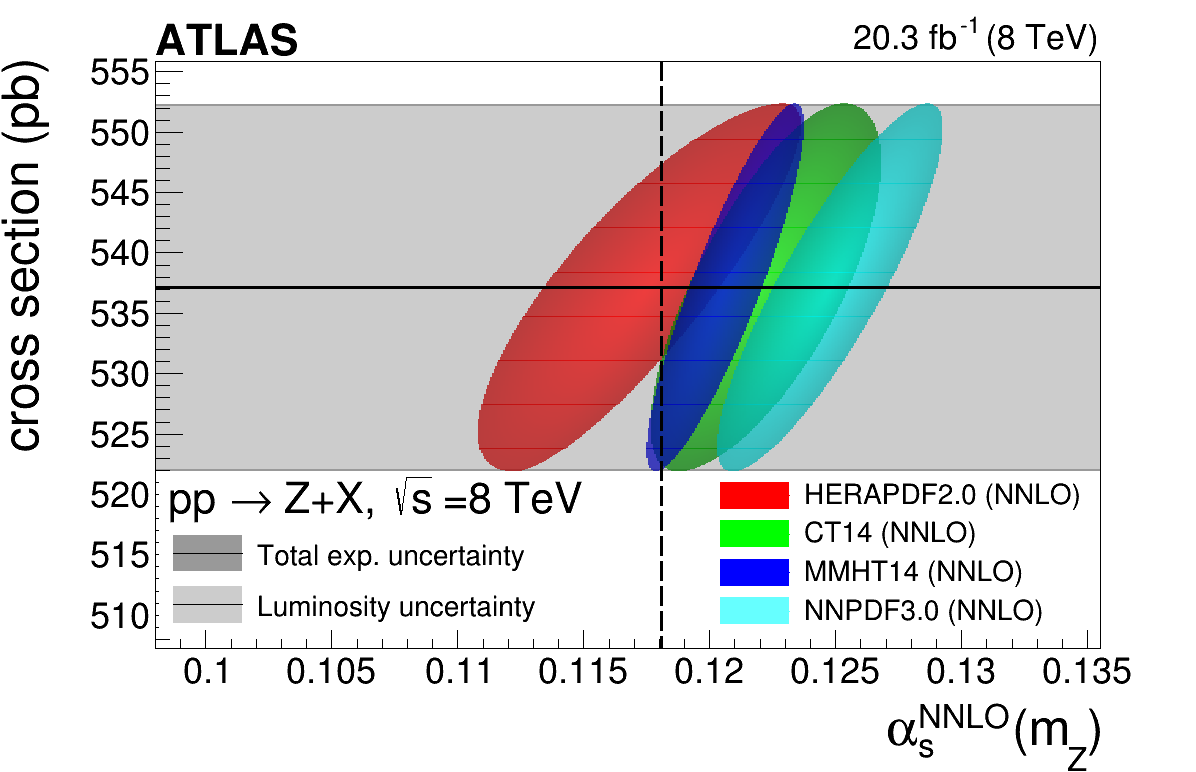}
\includegraphics[width=0.49\textwidth]{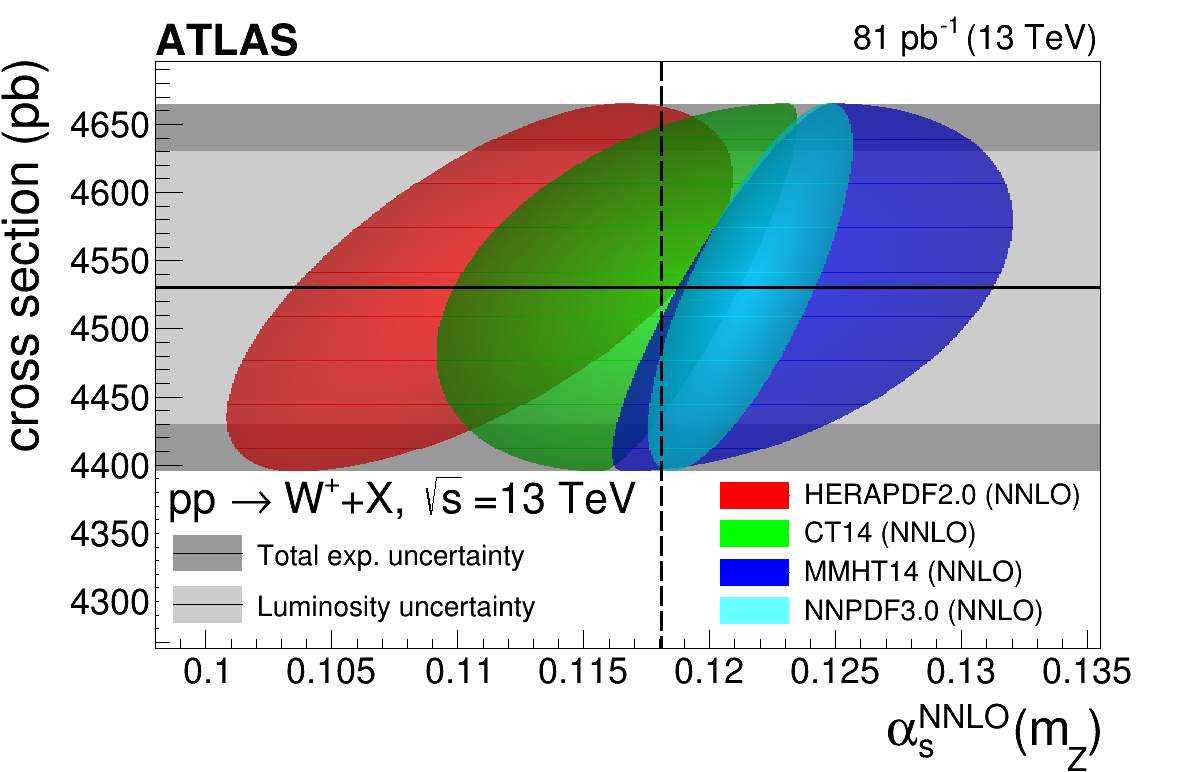}
\includegraphics[width=0.49\textwidth]{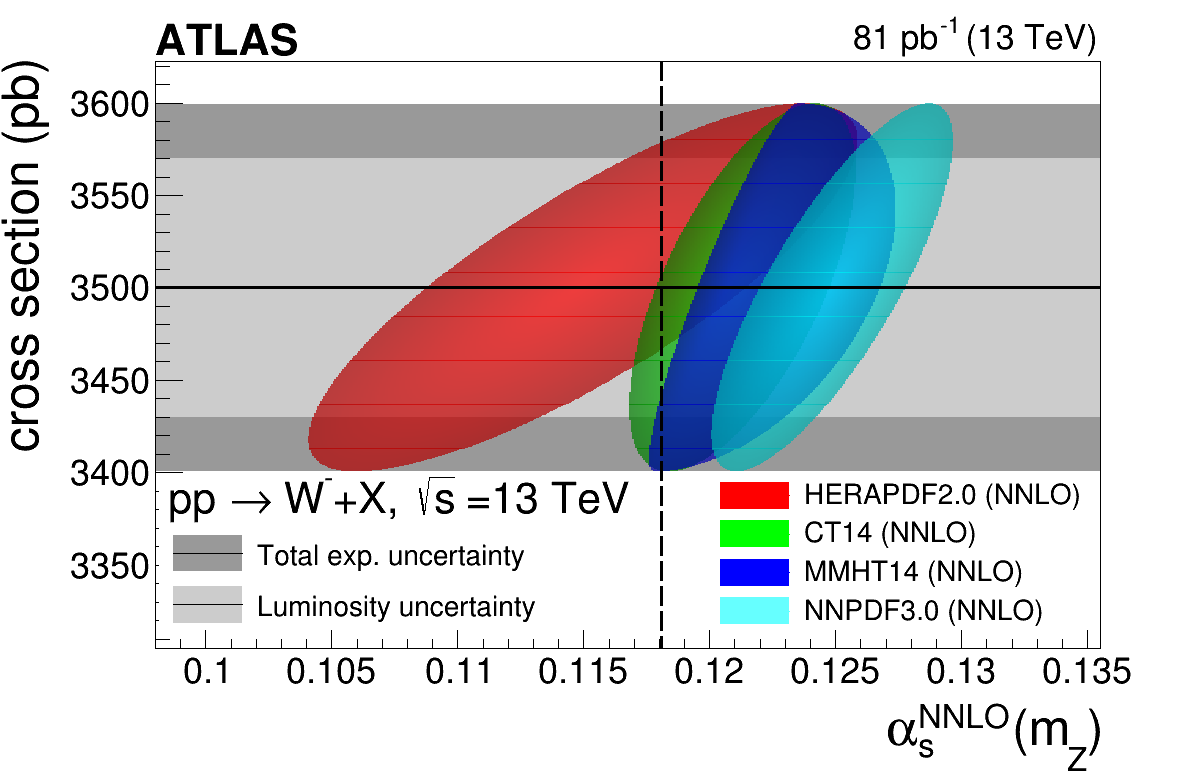}
\includegraphics[width=0.49\textwidth]{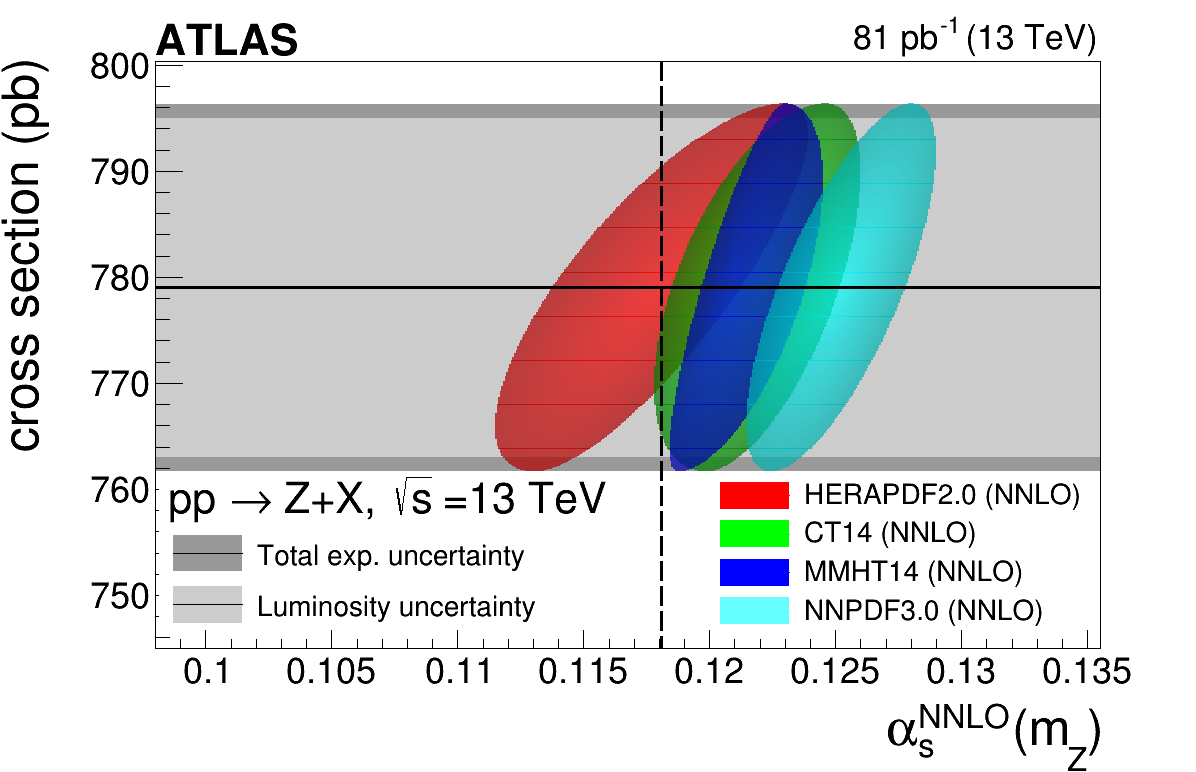}
\includegraphics[width=0.49\textwidth]{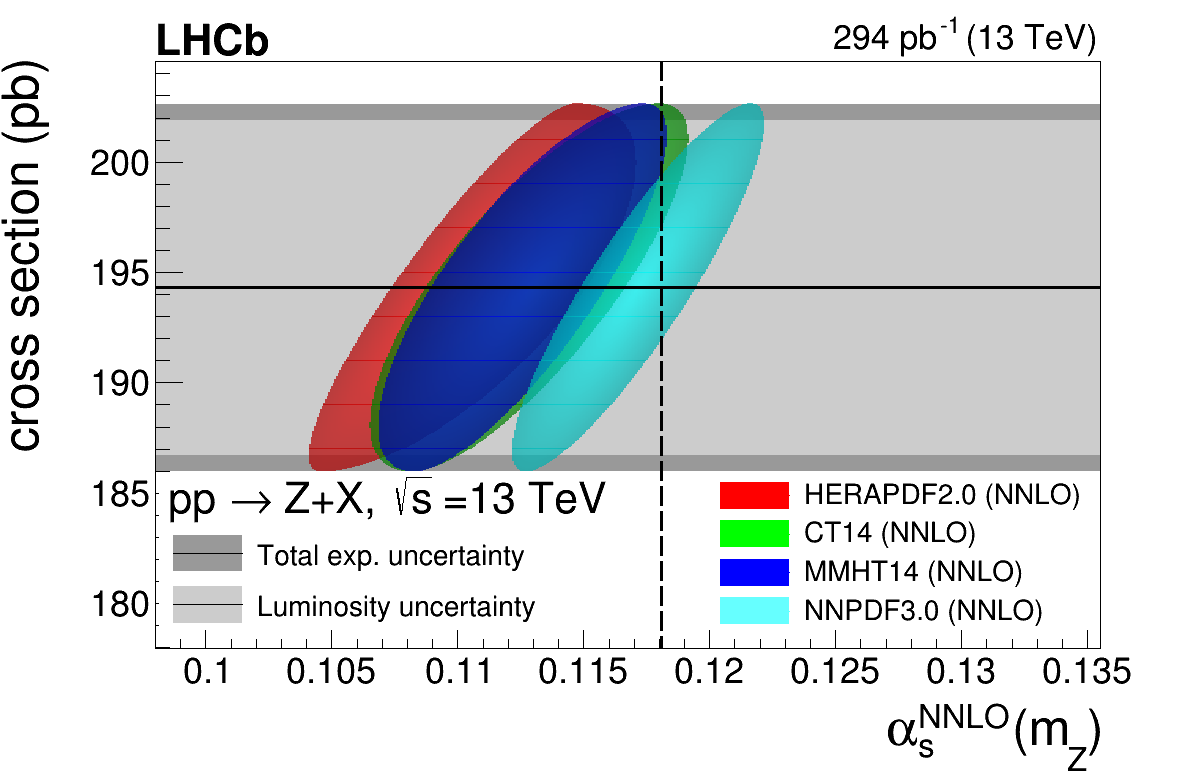}
\caption{Data-theory comparison of the ATLAS fiducial cross sections for \Wp\ (upper left), \Wm\ (upper right), and Z (second row, left) boson production in pp collisions at $\sqrts = 7$~TeV, for Z bosons at  $\sqrts = 8$~TeV (second row, right), for \Wp\ (third row, left), \Wm\ (third row, right), and Z (bottom left) bosons at $\sqrts = 13$~TeV, and for the LHCb Z boson at $\sqrts = 13$~TeV (bottom right). The experimental cross sections are plotted as horizontal black lines with outer darker (inner grey) bands indicating their total (integrated luminosity) uncertainties. The theoretical predictions are computed for each PDF set as a function of $\alphasmZ$, $x$ axis, and combined with the experimental results into Jpdfs shown as filled ellipses.
The vertical dashed line indicates the expected predictions for $\alphasmZ = 0.118$. 
\label{fig:data_th_atlas}}
\end{figure}

\begin{figure}[htbp!]
\includegraphics[width=0.49\textwidth]{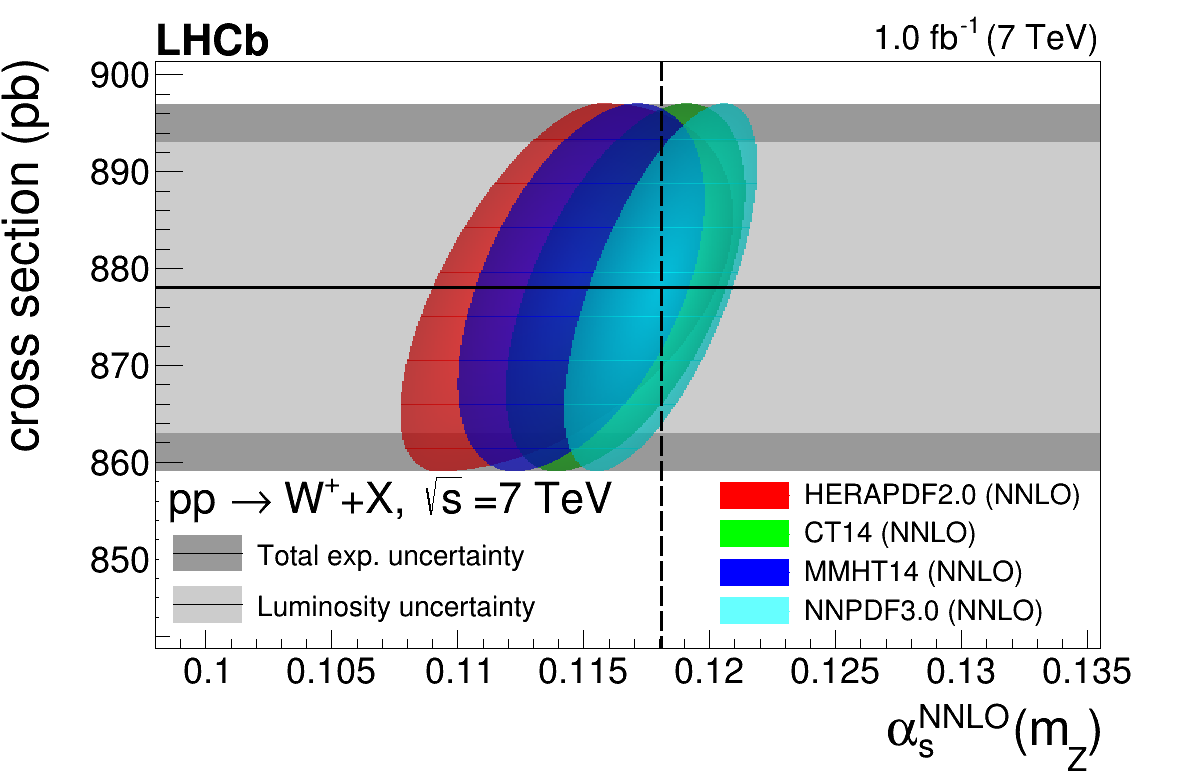}
\includegraphics[width=0.49\textwidth]{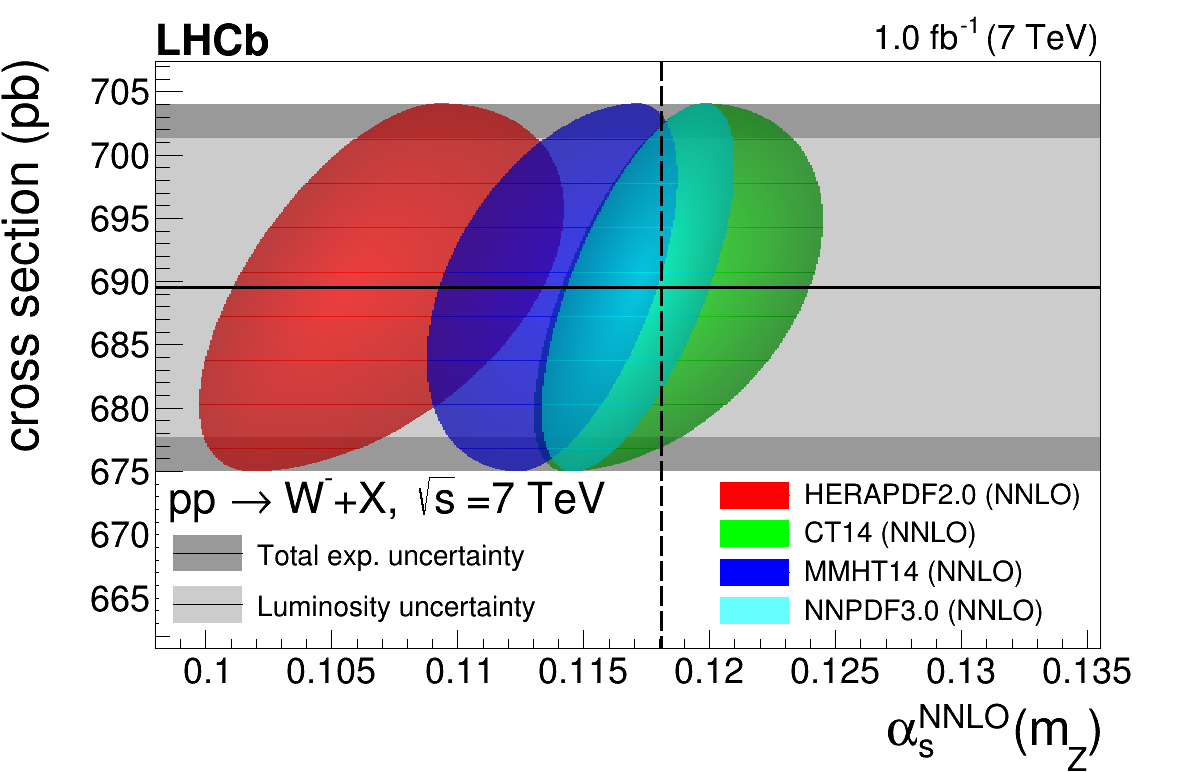}
\includegraphics[width=0.49\textwidth]{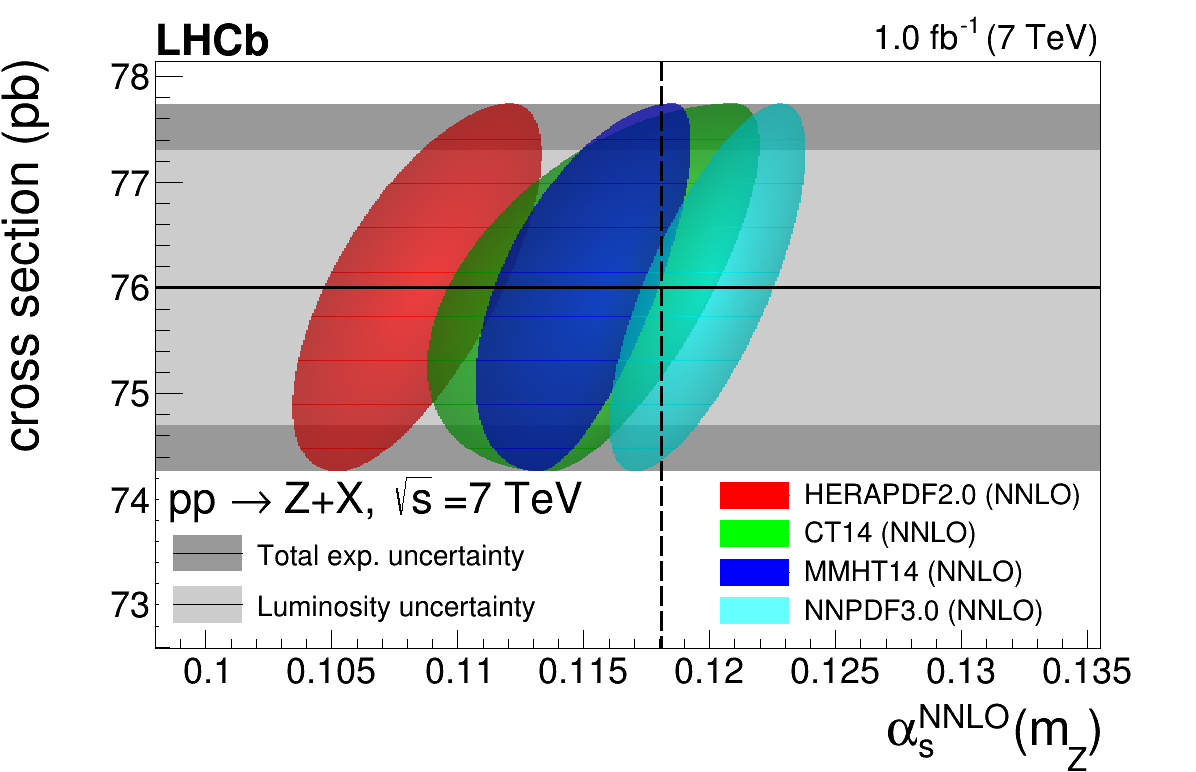}
\includegraphics[width=0.49\textwidth]{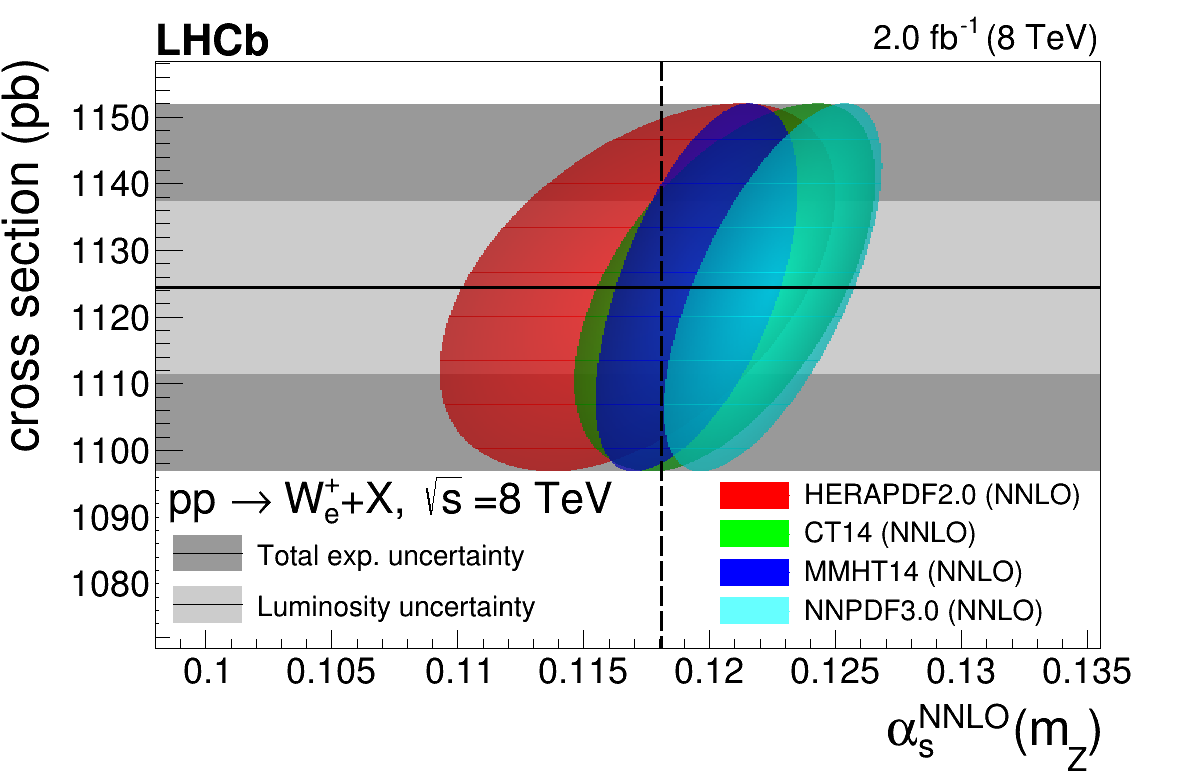}
\includegraphics[width=0.49\textwidth]{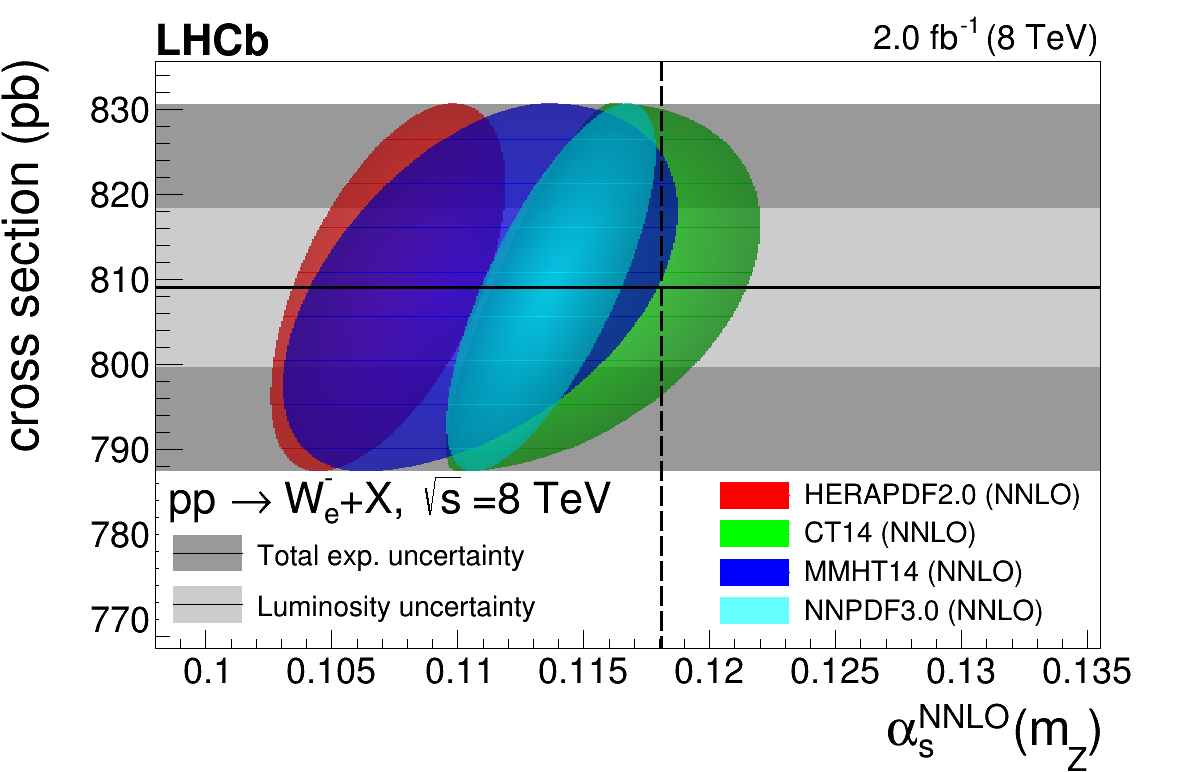}
\includegraphics[width=0.49\textwidth]{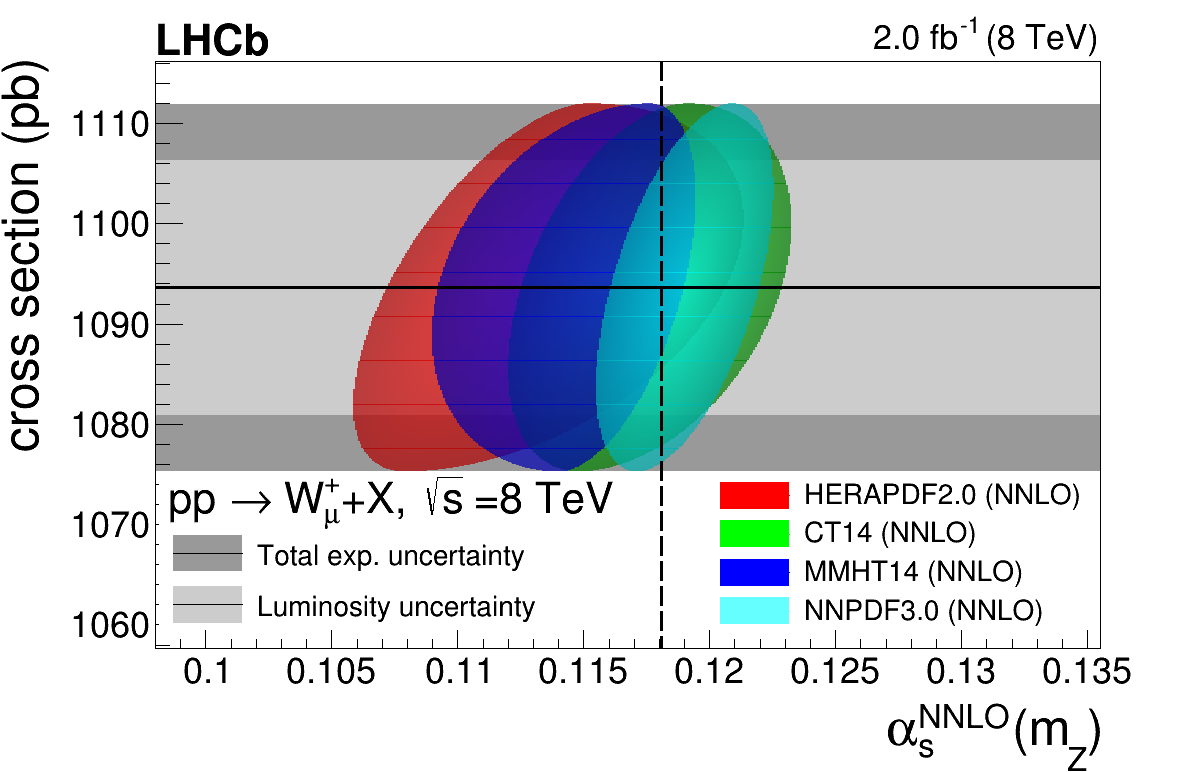}
\includegraphics[width=0.49\textwidth]{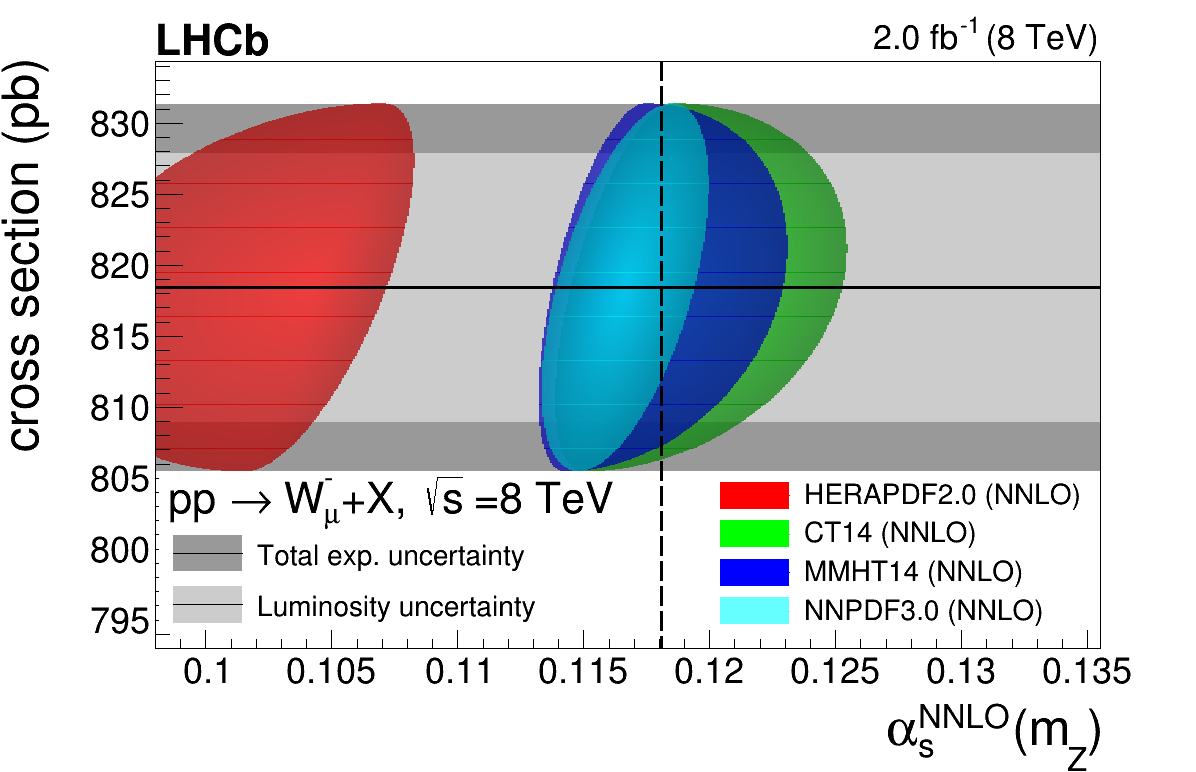}
\includegraphics[width=0.49\textwidth]{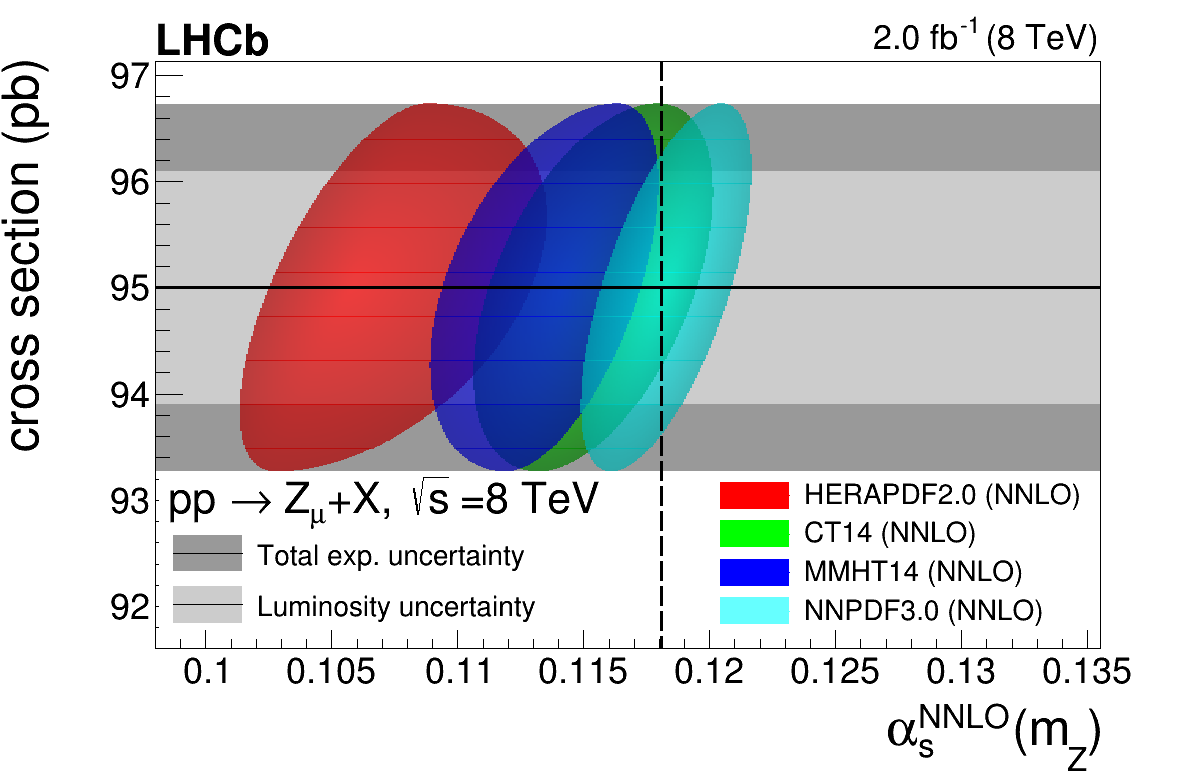}
\caption{Same as Fig.~\ref{fig:data_th_atlas} for the LHCb measurements of the cross sections for \Wp\ (upper left), \Wm\ (upper right), Z (second row, left) bosons in pp collisions at $\sqrts = 7$~TeV, and for W$^{+}_{\Pe}$ (second row, right), W$^{-}_{\Pe}$ (third row, left), W$^{+}_{\mu}$ (third row, right), W$^{-}_{\mu}$ (bottom, left) and Z$_{\mu}$ (bottom right) bosons at $\sqrts = 8$~TeV.\label{fig:data_th_lhcb}}
\end{figure}

\begin{figure}[htbp!]
\includegraphics[width=0.49\textwidth]{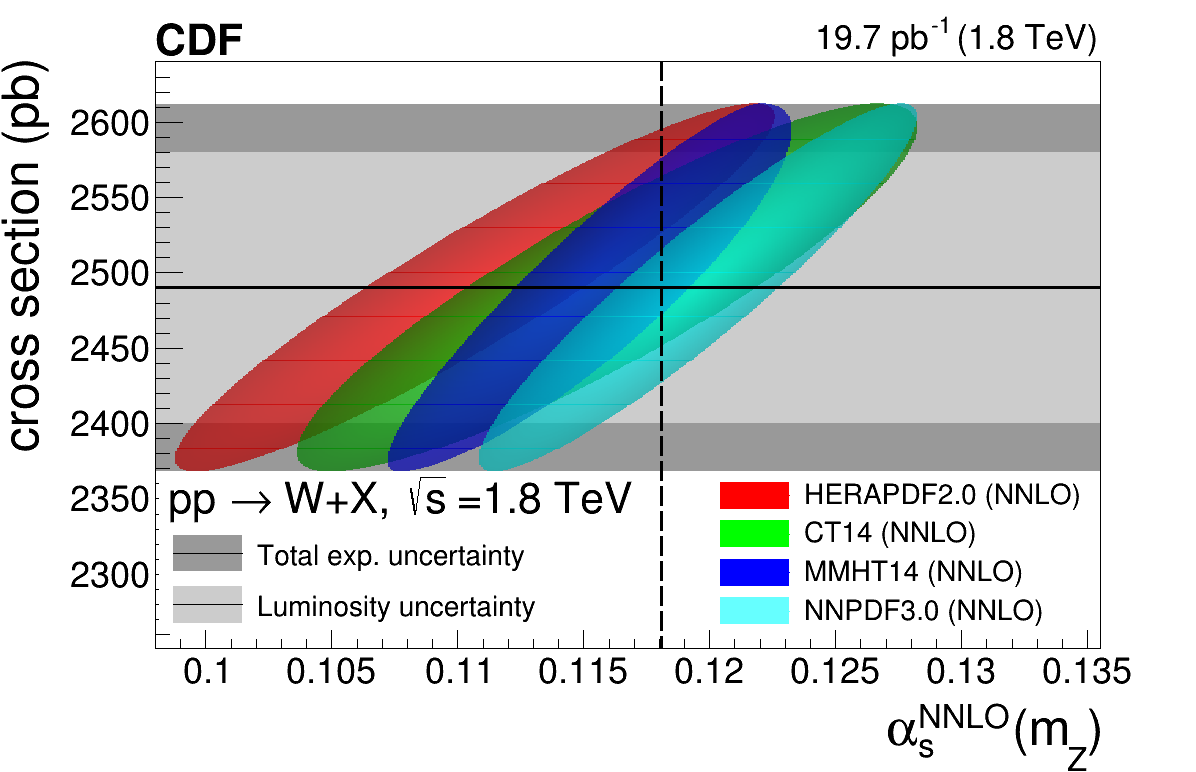}
\includegraphics[width=0.49\textwidth]{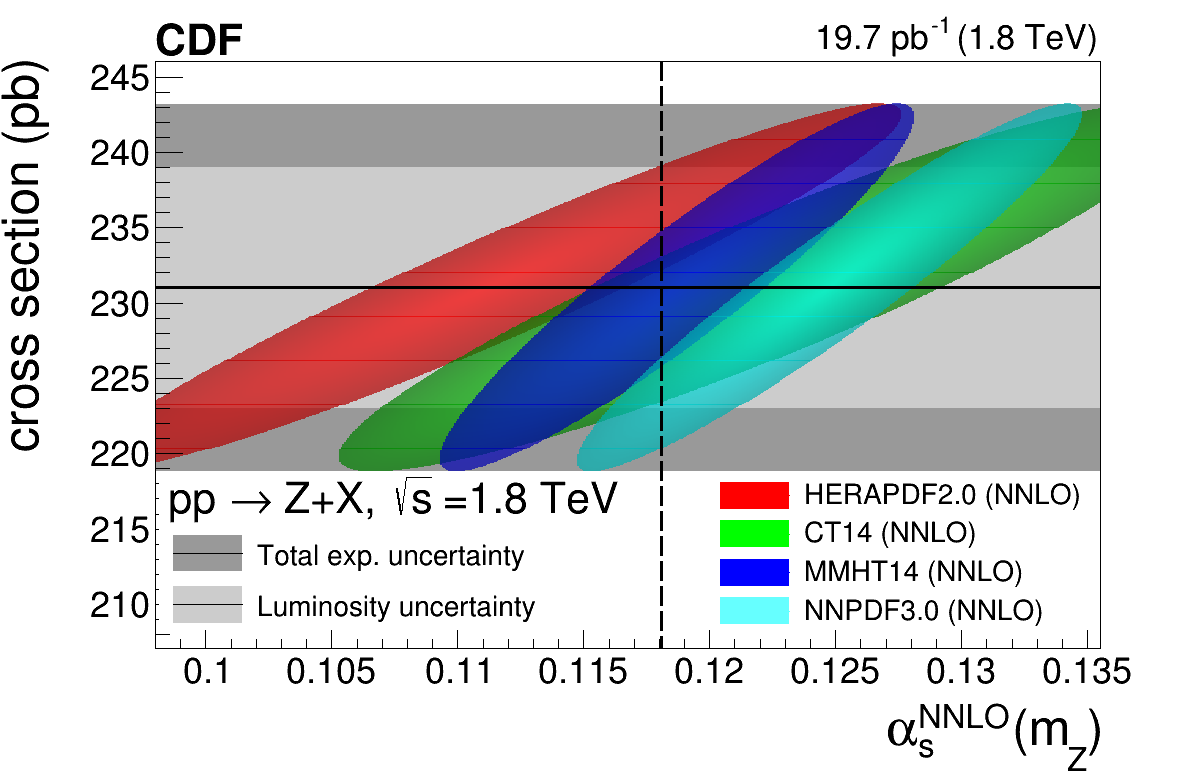}
\includegraphics[width=0.49\textwidth]{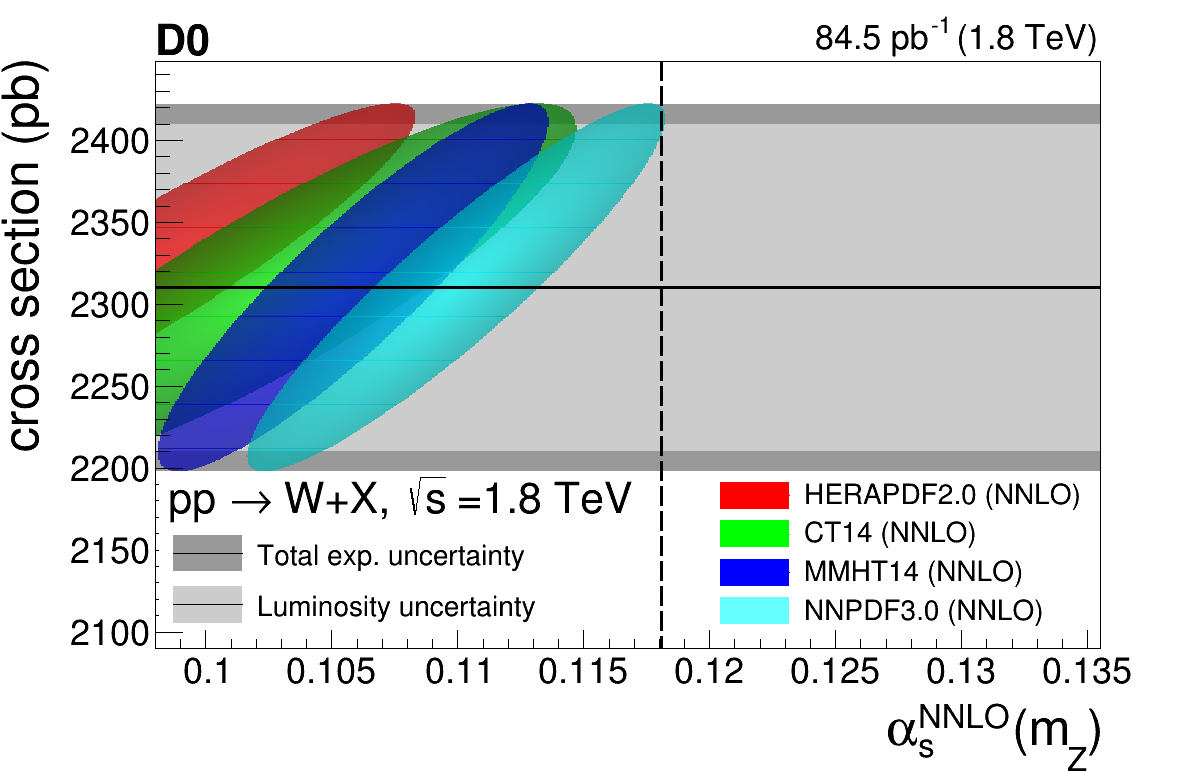}
\includegraphics[width=0.49\textwidth]{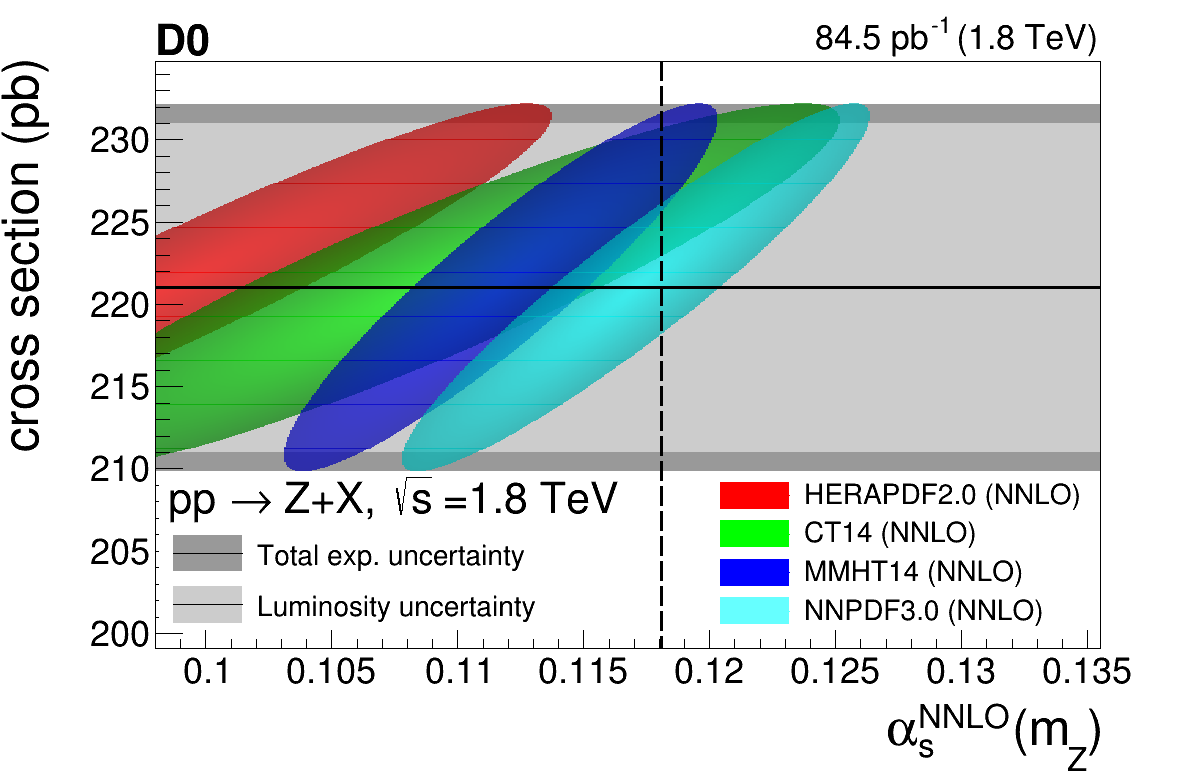}
\includegraphics[width=0.49\textwidth]{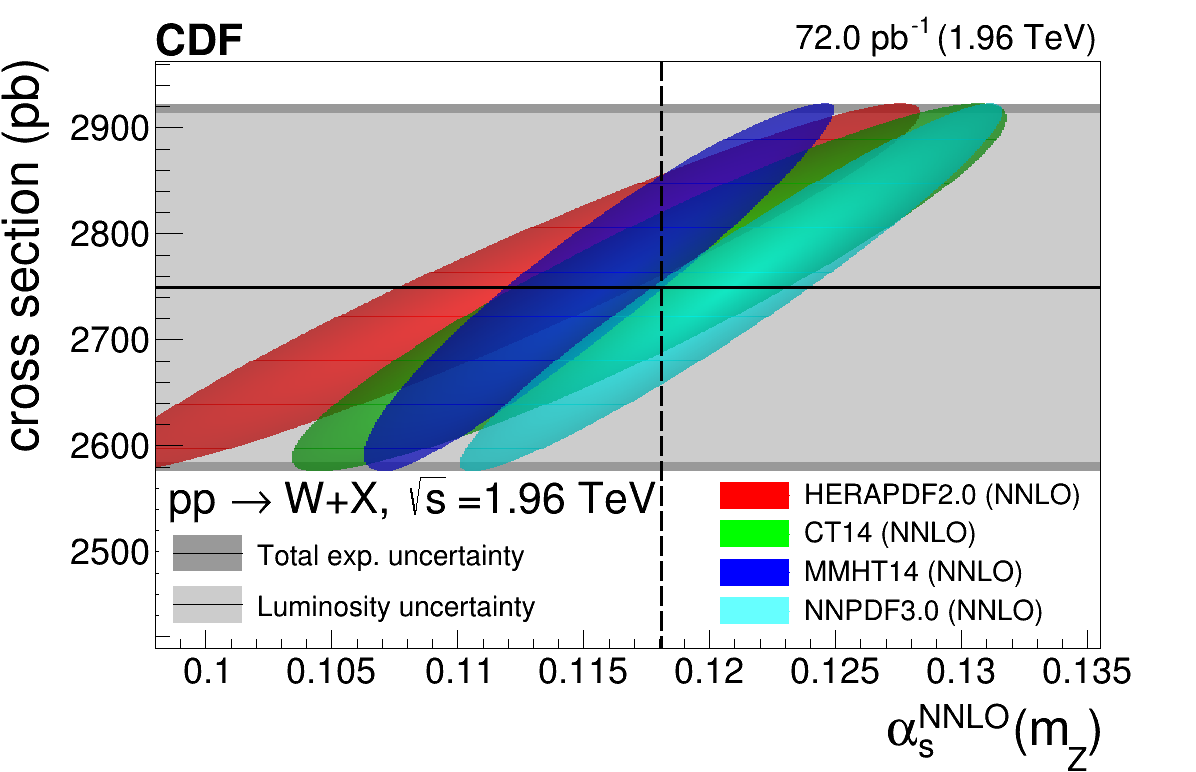}
\includegraphics[width=0.49\textwidth]{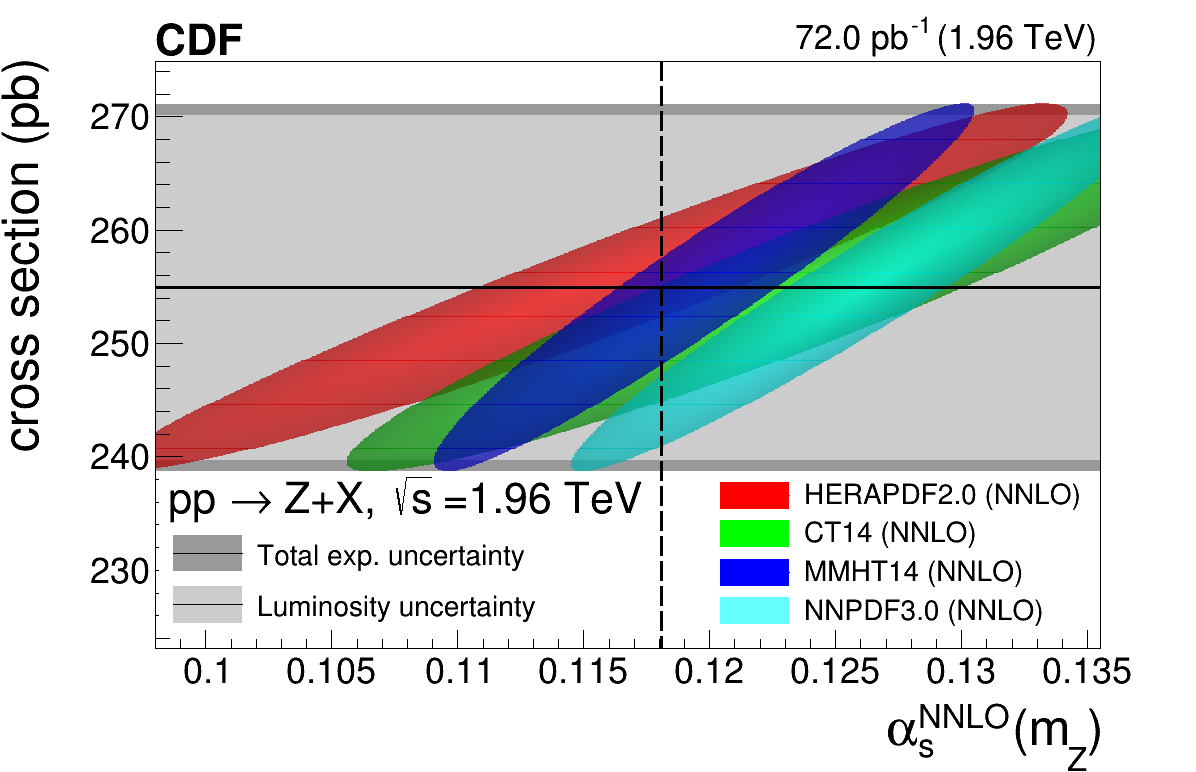}
\caption{Same as Fig.~\ref{fig:data_th_atlas} for the production cross sections in $\ppbar$ collisions at the Tevatron of \Wpm\ (CDF, upper left; D0, middle left), and Z (CDF, upper right; D0, middle right) bosons at  $\sqrts = 1.8$~TeV, and of \Wpm\ (CDF, bottom left) and Z (CDF, bottom right) bosons at  $\sqrts = 1.96$~TeV.
\label{fig:data_th_tevatron}}
\end{figure}

Figures~\ref{fig:data_th_atlas} and \ref{fig:data_th_lhcb}, and \ref{fig:data_th_tevatron} show, respectively, the ATLAS and LHCb, and Tevatron experimental cross sections (horizontal lines and bands) compared to the corresponding theoretical predictions per PDF as a function of $\alphasmZ$ (coloured ellipses). The measured cross sections are indicated by the horizontal black line with the outer dark band showing the total (quadratic sum) experimental uncertainties. The inner grey band shows the dominant integrated luminosity uncertainty. From the computed cross sections, a linear dependence of $\sigma^\mathrm{th}_\mathrm{W,Z}$ on $\alphasmZ$ is derived, and the filled ellipses are constructed to represent the contours of the joint probability density functions (Jpdfs) of the theoretical and experimental results, with a width representing a two-dimensional one standard deviation obtained from the product of both probability densities for each PDF, as explained in detail in~\cite{Poldaru:2019dnl,Sirunyan:2019wne}. The uncertainty in the theoretical cross sections is given by the quadratic sum of its associated PDF, scale, and numerical uncertainties. 
For symmetric uncertainties the Jpdfs have elliptical contours, but for asymmetric ones they correspond to two ellipses combined together. This procedure is repeated for all the 34 different measurements and for all 4 PDF sets, of which 22 are plotted as the filled ellipses shown in Figs.~\ref{fig:data_th_atlas}--\ref{fig:data_th_tevatron} (the corresponding CMS figures can be found in Ref.~\cite{Sirunyan:2019wne}).

The first observation from Figs.~\ref{fig:data_th_atlas}--\ref{fig:data_th_tevatron} is that not all theoretical predictions obtained with the default value of the QCD coupling constant consistently agree with all experimental ATLAS, LHCb, CDF, and D0 (as well as CMS, see~Ref.~\cite{Sirunyan:2019wne}) measurements. Namely, although almost all theory predictions for all PDF sets are consistent with the data within the (relatively large) experimental and theoretical uncertainties, they do not always yield the same preferred value of $\alphasmZ$. Inspecting the figures in more detail (as well as the similar figures presented in Ref.~\cite{Sirunyan:2019wne} for CMS), one can see that among the theoretical predictions, those obtained with HERAPDF2.0 (NNPDF.3.0) tend to be mostly shifted to the left (right) of the $\sigma_\mathrm{W,Z}$ \vs\ $\alphasmZ$ plots, \ie\ they prefer comparatively smaller (larger) values of the QCD coupling. Or, otherwise said, since larger $\alphasmZ$ values trivially imply larger \Wpm\ and Z cross sections, the HERAPDF2.0 (NNPDF3.0) predictions for the default $\alphasmZ = 0.118$ value tend to be mostly above (below) the data. The results computed with CT14 and MMHT14, on the other hand, appear mostly in-between those from the other two PDF sets. Thus, our first conclusions are that, in order to reproduce the experimental cross sections, HERAPDF2.0 (NNPDF3.0) tends in general to prefer a smaller (larger) value of $\alphasmZ$ than other PDFs, and that the predictions from CT14 and MMHT14 tend to be less scattered over the $\alphasmZ$ axis than those from HERAPDF2.0. A second result to point out is that, in general, the HERAPDF2.0 (MMHT14) filled ellipses have the smallest (largest) relative slope as a function of $\alphasmZ$ (a result also observed with the CMS data alone). A larger slope is advantageous for extracting the QCD coupling constant, as it indicates that the underlying $\alphasmZ$ value in the calculations has a larger impact on the computed cross sections, also leading to a lower propagated uncertainty in the $\alphasmZ$ value eventually derived by comparing the theoretical prediction to the experimental result.

The data-theory comparison plots shown in Fig.~\ref{fig:data_th_tevatron} for the Tevatron measurements feature about twice larger experimental uncertainties, and also much shallower dependence of the theoretical cross section on $\alphasmZ$, compared to their LHC counterpart measurements and calculations. The apparently smaller slopes in the Tevatron, compared to the LHC cases, is mostly due to the fact that the more imprecise experimental cross sections ``flatten out'' (via the Jpdfs convolution) the plotted ellipses.

In all plots of Figs.~\ref{fig:data_th_atlas}--\ref{fig:data_th_tevatron}, the points where the filled ellipses cross the vertical dashed line at $\alphasmZ = 0.118$ indicate the most likely cross section interval that would correspond to the baseline QCD coupling constant of all PDF sets, taking into account both the experimental and theoretical results. One can quantify the overall level of data-theory agreement 
through a goodness-of-fit test, $\chi^2 = \xi_i (M^{-1})_{ij} \xi_j$, where $M$ is the covariance matrix obtained from the sum of the covariance matrices of each single source of experimental and theoretical uncertainties (Table~\ref{tab:uncertainties}) discussed in more detail in Section~\ref{sec:6}, and $\xi_i=\sigma_{i,\text{th}}-\sigma_{i,\text{exp}}$ is the difference between the ATLAS, CMS, and LHCb experimental cross sections and their corresponding theoretical predictions for each PDF set. In the $\chi^2$ calculation, the asymmetric uncertainties of the CT14, HERAPDF2.0, and MMHT14 PDF sets are symmetrized to the largest of the two values and also separately to the smallest of the two values. Table~\ref{tab:chi2} lists the corresponding results. For the baseline QCD coupling constant value of $\alphasmZ=0.118$ of all PDF sets, the data-theory accord is better for the predictions calculated with CT14 and MMHT14 ($\rm \chi^2/ndf\lesssim 1$) compared to those obtained with the HERAPDF2.0 (when using the smaller symmetrized errors) and NNPDF3.0 sets ($\rm \chi^2/ndf\approx 2.1$).

\begin{table}[htpb!]
\caption{Overall goodness-of-fit per number of degrees of freedom, $\chi^2$/ndf, among the 28 LHC experimental measurements of \Wp, \Wm, and Z boson cross sections and the corresponding theoretical predictions obtained with the four different PDF sets for their default $\alphasmZ=0.118$ value. The first (second) row lists the results obtained symmetrizing the PDF uncertainties of the cross sections obtained with the CT14, HERAPDF2.0, and MMHT14 sets to the largest (smallest) of their respective values.\label{tab:chi2}}
\centering
\resizebox{\textwidth}{!}{
\begin{tabular}{lcccc}\\\hline
& CT14 & HERAPDF2.0 & MMHT14 & NNPDF3.0 \\
$\chi^2/$ndf (symmetrized to the largest PDF uncertainty value) & 15.8/27 & 21.8/27 & 15.7/27 & 58.8/27 \\
$\chi^2/$ndf (symmetrized to the smallest PDF uncertainty value)& 26.3/27 & 60.4/27 & 22.7/27 & 58.8/27 \\\hline
\end{tabular}
}
\end{table}

\clearpage

\section{Individual \texorpdfstring{$\alphasmZ$}{alphas(mZ)} extraction per measurement}
\label{sec:5}

The cross sections calculated with different $\alphasmZ$ values are fitted (using $\chi^2$-minimization) to a first-order polynomial, 
and the corresponding slope $k$ is extracted for each PDF and measurement. Over the considered $\alphasmZ$ range, the empirical linear fit describes well the observed $\sigma^\mathrm{th}_\mathrm{W,Z}$-versus-$\alphasmZ$ dependence for all PDF sets. The value of $\alphasmZ$ preferred by each individual measurement is determined by the crossing point of the fitted linear theoretical curve with the experimental horizontal line. It can be shown that all sources of uncertainty in the theoretical and experimental cross section, $\delta\sigma$, propagate as an $\alphasmZ$ uncertainty of $\delta\sigma/k$ size, where $k$ is the slope of the theoretical linear fit~\cite{Poldaru:2019dnl,Sirunyan:2019wne}. From this, it follows that since the luminosity (PDF) sources are the largest uncertainties in the cross section measurements (calculations), those propagate also as the dominant experimental (theoretical) uncertainties in each one of the $\alphasmZ$ values extracted.

The strong coupling values resulting from the fitting procedure described above are listed in Tables~\ref{tab:alphasmZ_atlas}, \ref{tab:alphasmZ_lhcb}, and~\ref{tab:alphasmZ_tevatron} of the Appendix for all individual ATLAS, LHCb, and Tevatron measurements, respectively, along with the uncertainty breakdowns from every source, for each PDF set. On the one hand, the LHC results feature extracted $\alphasmZ$ values with low overall uncertainty, in some cases as low as 2\%. Similar results were obtained in Ref.~\cite{Sirunyan:2019wne} from the theoretical analysis of the CMS data. On the other hand, the $\alphasmZ$ extractions based on CDF and D0 data have propagated uncertainties above 7\% in all cases, due to the less precise nature of the EW boson cross section measurements at the Tevatron. The latter extractions will, therefore, not be used in the final $\alphasmZ$ combination discussed in the next Section.


\section{Final combined \texorpdfstring{$\alphasmZ$}{alphas(mZ)} determination}
\label{sec:6}

From the 28 individual $\alphasmZ$ values extracted per PDF set from the \Wpm\ and Z cross section measurements of ATLAS (Table~\ref{tab:alphasmZ_atlas}), CMS~\cite{Sirunyan:2019wne}, and LHCb (Table~\ref{tab:alphasmZ_lhcb}), we determine a single $\alphasmZ$ per PDF by appropriately combining all results taking into account their uncorrelated, partially-, and fully-correlated experimental and theoretical uncertainties (Table~\ref{tab:uncertainties}). The following correlations among uncertainties are considered:
\begin{itemize}
\item The integrated luminosity uncertainty is fully correlated for all $\alphasmZ$ results obtained at the same $\sqrts$ for each experiment, but fully uncorrelated between different \cm\ energies or experiments. 
\item The experimental systematic uncertainties among measurements at the same $\sqrts$ are partially correlated, as quantified in Table~\ref{tab:atlas_lhcb_corrs} for the ATLAS and LHCb measurements. The correlations are taken from the corresponding experimental papers. In a few cases where only correlations were given for different bins in kinematic variables, the correlations $\rho$ of the sums of bins are computed as follows: 
\begin{equation} \displaystyle
\rho_{ij} = \frac{\sum_{m,n} \rho_{mn,ij}  \,\delta\sigma_{m,i} \,\delta\sigma_{n,j}}{\sqrt{\sum_{m,n} \rho_{mn,ii} \delta\sigma_{m,i}\delta\sigma_{n,i}} \, \sqrt{\sum_{m,n} \rho_{mn,jj} \delta\sigma_{m,j} \delta\sigma_{n,j}}}\,,
\label{eq:correlations}
\end{equation}
where the subindices $m$ and $n$ label the bins, and $i$ and $j$ label systems. The correlation matrices used for the CMS results are those described in detail in Ref.~\cite{Sirunyan:2019wne}. For LHCb, the \cm\ energy uncertainty is fully correlated at each $\sqrts$.
\item The experimental statistical uncertainty is fully uncorrelated among all $\alphasmZ$ extractions.
\item The PDF uncertainty is partially correlated for the $\alphasmZ$ values extracted with the same PDF set. A Pearson correlation coefficient is calculated as described in~\cite{Poldaru:2019dnl,Sirunyan:2019wne} by using the cross sections computed with all individual eigenvectors/replicas for each pair of W$^\pm$ and Z measurements. The correlations are found to be in the 0.4--0.8, 0.4--0.8, 0.1--0.6, and 0.8--1.0 ranges for CT14, HERAPDF2.0, MMHT14, and NNPDF3.0, respectively. 
\item The scale uncertainty is partially correlated. Similarly as for the PDF uncertainties, for each pair of measurements a Pearson correlation coefficient is calculated using the results obtained from the theoretical scale variations. The scale correlations vary over 0.0--0.8. When combining the $\alphasmZ$ estimates, each specific correlation coefficient calculated for every specific pair of estimates is used.
\item The theoretical numerical uncertainty is fully uncorrelated among all $\alphasmZ$ extractions.
\end{itemize}

\begin{table}[H]
	\caption{Correlation matrices among the systematic uncertainties of different \Wp, \Wm, and Z boson cross section measurements in ATLAS~\cite{Aad:2011dm,Aad:2015auj,Aad:2016naf} and LHCb~\cite{Aaij:2015gna,Aaij:2015zlq,Aaij:2016mgv,Aaij:2016qqz} as derived, in some cases using Eq.~(\ref{eq:correlations}), from the corresponding experimental studies.
	\label{tab:atlas_lhcb_corrs}}
	\centering
	\begin{tabular}{clll}\hline
		\multicolumn{4}{c}{ATLAS, pp at $\sqrts =7$~TeV} \\\hline
		& \Wp & \Wm & Z \\
		\Wp  & 1.00 & 0.90 & 0.36 \\
		\Wm  & 0.90 & 1.00 & 0.32 \\
		Z\;\, & 0.36 & 0.32 & 1.00 \\\hline
		\\
	\end{tabular}\hspace{1.5cm}
	\begin{tabular}{clll}\hline
		\multicolumn{4}{c}{ATLAS, pp at $\sqrts = 13$~\TeV} \\\hline
		& W$^+_\mu$ & W$^-_\mu$ & Z$_\mu$ \\
		W$^+_\mu$ & 1.00 & 0.95 & 0.21 \\
		W$^-_\mu$ & 0.95 & 1.00 & 0.20 \\
		Z$_\mu$   & 0.21 & 0.20 & 1.00 \\\hline
		\\
	\end{tabular}
\begin{tabular}{clll}\hline
	\multicolumn{4}{c}{LHCb, pp at $\sqrts =7$~TeV} \\\hline
	& \Wp & \Wm & Z \\
	\Wp  & 1.00 & 0.56 & 0.54 \\
	\Wm  & 0.56 & 1.00 & 0.48 \\
	Z\;\, & 0.54 & 0.48 & 1.00 \\\hline
\end{tabular}\hspace{0.2cm}
\begin{tabular}{clll}\hline
	\multicolumn{4}{c}{LHCb, pp at $\sqrts =8$~TeV} \\\hline
	& W$^+_\Pe$ & W$^-_\Pe$ & \\
	W$^+_\Pe$ & 1.00 & 0.89 & \\
	W$^-_\Pe$ & 0.89 & 1.00 & \\
	& & & \\\hline
\end{tabular}\hspace{0.2cm}
\begin{tabular}{clll}\hline
	\multicolumn{4}{c}{LHCb, pp at $\sqrts =8$~TeV} \\\hline
	& W$^+_\mu$ & W$^-_\mu$ & Z$_\mu$ \\
	W$^+_\mu$ & 1.00 & 0.83 & 0.61 \\
	W$^-_\mu$ & 0.83 & 1.00 & 0.61 \\
	Z$_\mu$   & 0.61 & 0.61 & 1.00 \\\hline
\end{tabular}
\end{table}

All the individual 28 $\alphasmZ$ derived per PDF set, and the correlation matrices associated with all their uncertainties are given as inputs of the \textsc{convino} v1.2~\cite{Kieseler:2017kxl} program that is employed to determine the final best estimate of all combined values. The Neyman $\chi^2$-minimization procedure is selected in the code. Identical results are obtained, when symmetrizing all uncertainties, if one uses the similar BLUE method~\cite{Nisius:2014wua} to carry out the combination. 


\begin{figure}[htbp!]
\centering
\includegraphics[width=0.9\textwidth]{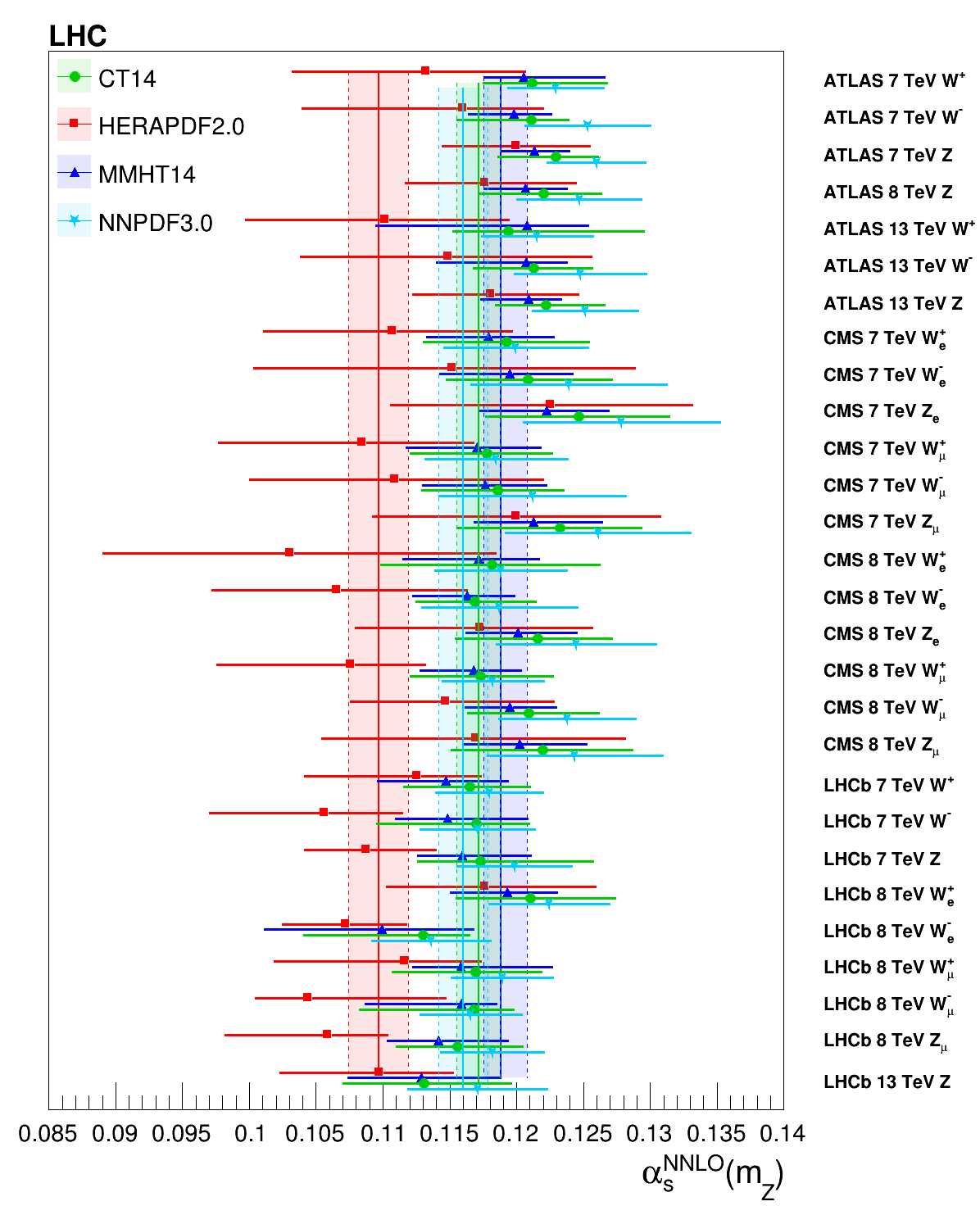}
\caption{Individual $\alphasmZ$ values extracted from each measured \Wpm\ and Z boson production cross section (symbols with error bars), and final $\alphasmZ$ values obtained combining the 28 individual determinations (vertical coloured areas), for each PDF set. The CMS $\alphasmZ$ values plotted here are those derived in Ref.~\cite{Sirunyan:2019wne}.
\label{fig:alphas_per_PDF}}
\end{figure}

\begin{table}[htbp!]
\caption{Strong coupling constant $\alphasmZ$ values extracted per PDF set by combining all the individual results obtained for each \Wpm\ and Z boson production cross section measurements (Tables~\ref{tab:alphasmZ_atlas} and \ref{tab:alphasmZ_lhcb}), listed along with their propagated total and individual uncertainties. The last column tabulates the goodness-of-fit per degree of freedom $\chi^2$/ndf of the final single combined result compared to the 28 individual $\alphasmZ$ extractions.
\label{tab:final_alphas_perPDF}}
\centering
\resizebox{\textwidth}{!}{
\renewcommand*{\arraystretch}{1.2}
\begin{tabular}{lcccccccc}\hline
PDF & $\alphasmZ$  & $\delta\stat$ & $\delta\lum$ & $\delta\syst$ & $\delta$(PDF) & $\delta$(scale) & $\delta\statt$ & $\chi^2$/ndf \\\hline
CT14       & $0.1172^{+0.0015}_{-0.0017}$ & 0.0003 & 0.0005 & 0.0006 &\,$^{+0.0011}_{-0.0013}$ & 0.0006 & 0.0003 & 23.5/27 \\
HERAPDF2.0 & $0.1097^{+0.0022}_{-0.0023}$ & 0.0004 & 0.0009 & 0.0009 & \,$^{+0.0015}_{-0.0016}$ & 0.0007 & 0.0005 & 27.0/27 \\
MMHT14     & $0.1188^{+0.0019}_{-0.0013}$ & 0.0002 & 0.0008 & 0.0003 & \,$^{+0.0015}_{-0.0007}$ & 0.0007 & 0.0002 & 19.3/27 \\
NNPDF3.0   & $0.1160 \pm 0.0018$ & 0.0006 & 0.0004 & 0.0005 & 0.0013 & 0.0006 & 0.0007 & 56.9/27 \\\hline
\end{tabular}
}
\end{table}

Figure~\ref{fig:alphas_per_PDF} shows the individual results (symbols with horizontal error bars) and the final combined $\alphasmZ$ value (vertical coloured areas) extracted for each PDF set. The width of the vertical coloured areas in the plot indicates the size of the total propagated uncertainty in the final QCD coupling value derived for each PDF set. Table~\ref{tab:final_alphas_perPDF} lists the obtained $\alphasmZ$ values, along with the uncertainty breakdowns from every source, determined for each PDF set through the combination of the 28 individual extractions. The total $\alphasmZ$ uncertainties derived for NNPDF3.0 are symmetric by construction, and a small asymmetry propagates into the final extractions for the other PDF sets. The total (symmetrized) uncertainties amount to $\sim$1.4\% for CT14, $\sim$2.1\% for HERAPDF2.0, $\sim$1.3\% for MMHT14 and $\sim$1.6\% for NNPDF3.0. 

The last column of Table~\ref{tab:final_alphas_perPDF} lists the goodness-of-fit per degree of freedom ($\chi^2$/ndf) of the final single combined result compared to the 28 individual $\alphasmZ$ extractions. The results obtained with CT14, HERAPDF2.0, and MMHT14 feature all an overall good agreement among the final combined $\alphasmZ$ and the individual extractions from each measurement, as indicated by the $\chi^2$/ndf\,$\lesssim 1$ value. The NNPDF3.0 extraction, however, shows a bad accord between the final and individual $\alphasmZ$ values obtained. Indeed, although the 28 individual extractions appear to have QCD coupling values on the ``high'' side, the final combined value is shifted down to $\alphasmZ \approx 0.116$ outside of the region around $\alphasmZ\approx 0.120$ defined by most of the individual estimates. Such a seemingly counterintuitive behaviour, also observed in the analysis of the CMS data alone~\cite{Sirunyan:2019wne}, often found in the literature under the name ``Peelle's pertinent puzzle''~\cite{Neudecker:2014}, is due to the presence of strong correlations among individual extractions, with the lowest $\alphasmZ$ values derived having smaller uncertainties than the others. The underlying tension apparent between NNPDF3.0 and the weak boson measurements at the LHC studied here, has been solved in the latest NNPDF3.1 global fit~\cite{Ball:2017nwa}.

\begin{figure}[htbp!]
\centering
\includegraphics[width=0.58\textwidth]{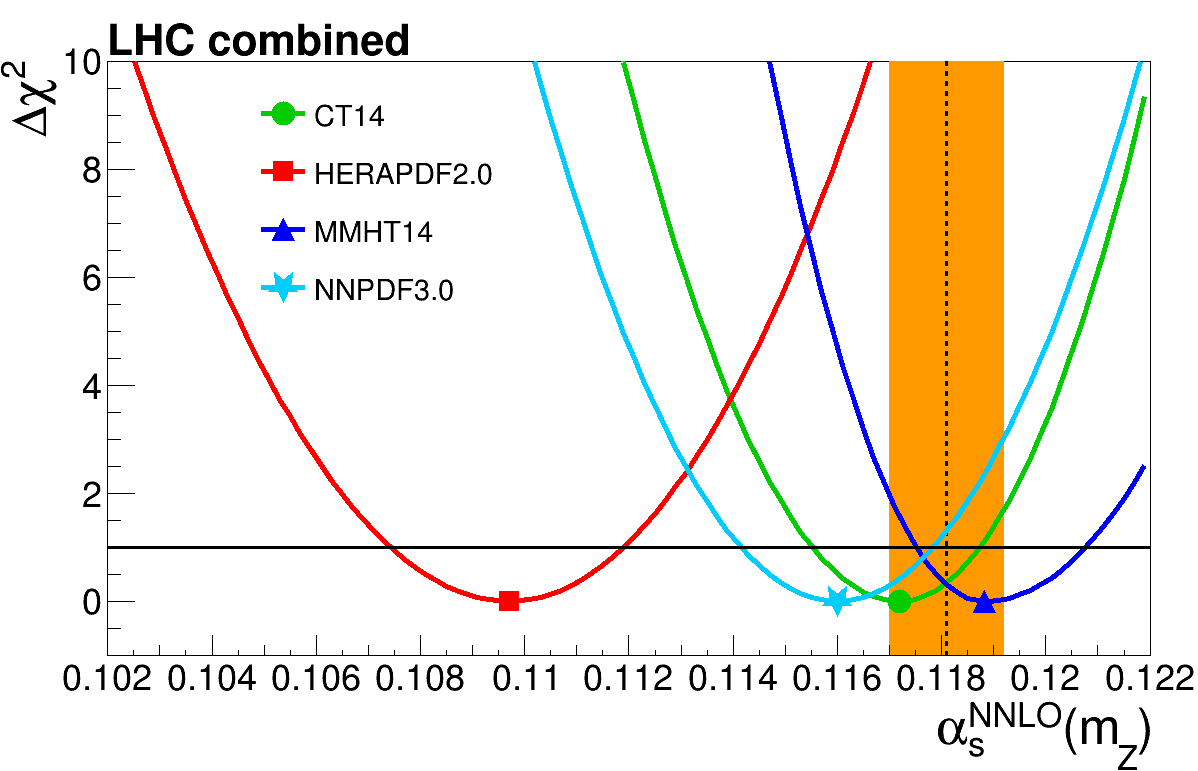}
\includegraphics[width=0.41\textwidth]{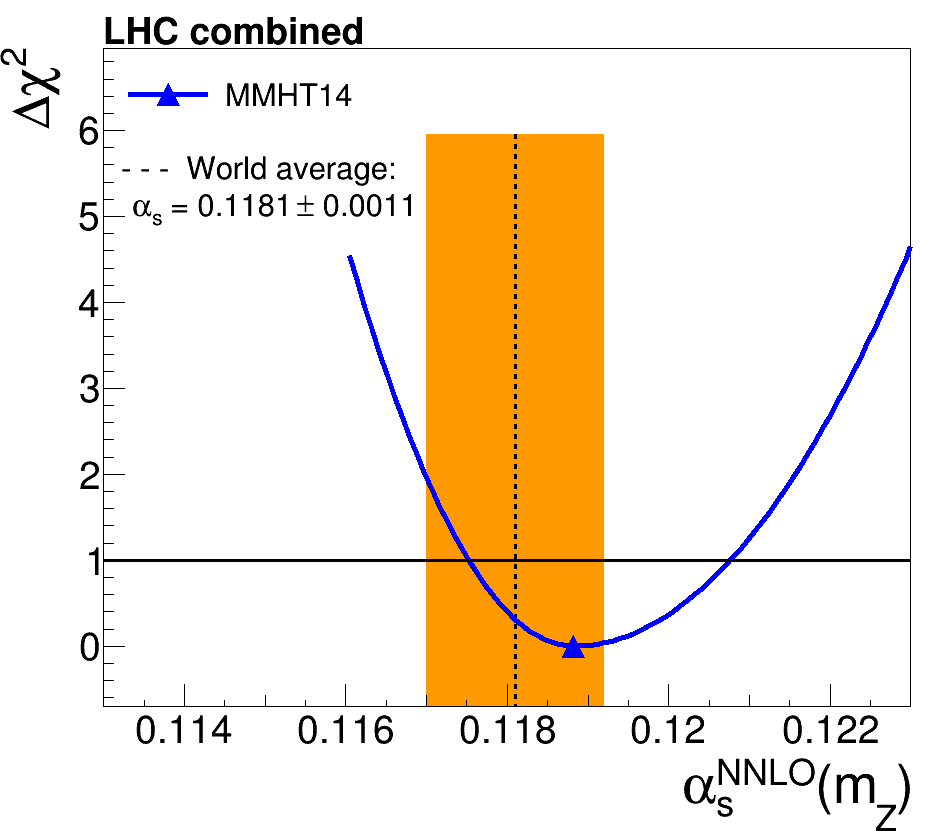}
\caption{Final $\alphasmZ$ values extracted from the analysis of the electroweak boson inclusive cross sections at the LHC using the CT14, HERAPDF2.0, MMHT14, and NNPDF3.0 PDF sets (left), and $\alphasmZ$ extraction from the MMHT14 PDF alone (right), compared to the current world average (vertical orange band).
\label{fig:final_alphas_per_PDF}}
\end{figure}

The final $\alphasmZ$ extractions are plotted in Fig.~\ref{fig:final_alphas_per_PDF} (left) compared with the current world average of $\alphasmZ = 0.1181 \pm 0.0011$ (orange band) for each individual PDF. The (asymmetric) parabolas are constructed to have a minimum at the combined value and are fitted to go through $\Delta\chi^2=1$ (horizontal black lines) at the one std.\,deviation  uncertainties quoted in Table~\ref{tab:final_alphas_perPDF}.

\begin{table}[!htbp]
	\caption{Sensitivity of the final $\alphasmZ$ extractions per PDF set to various experimental and theoretical ingredients. Top row: When using only the 7, 8, and 13~TeV measurements separately. Second row: When combining only the data of each experiment.
	Third row: When considering the extreme case of all the PDF eigenvectors/replicas or/and the scale uncertainties as fully correlated.
	Fourth row: When removing all theoretical uncertainties in the combination (center), and also first subtracting from (left) or adding to (right) each of the 28 $\alphasmZ$ central values their corresponding quadratic sum of scale and PDF uncertainties.
	Bottom row: When symmetrizing the PDF uncertainties by taking the maximum of the negative and positive values (left), and when adding a 1\% uncorrelated uncertainty to all cross sections (center), and largest differences observed in the $\alphasmZ$ values per PDF set, compared to the original results in Table~\ref{tab:final_alphas_perPDF}, after all the above cross-checks (right).
\label{tab:alphas_sensitivity}}
	\centering
	\resizebox{\textwidth}{!}{
	\renewcommand*{\arraystretch}{1.2}
	\begin{tabular}{lccc}\hline
		PDF & $\alphasmZ$  &  $\alphasmZ$  &  $\alphasmZ$  \\\hline
		    & [7\TeV data] & [8\TeV data] & [13\TeV data] \\
		CT14       & $0.1188^{+0.0022}_{-0.0024}$ & $0.1152^{+0.0024}_{-0.0023}$ & $0.1210^{+0.0031}_{-0.0033}$  \\
		HERAPDF2.0 & $0.1105^{+0.0029}_{-0.0028}$ &  $0.1085^{+0.0027}_{-0.0025}$ & $0.1134^{+0.0042}_{-0.0041}$   \\
		MMHT14     & $0.1197^{+0.0022}_{-0.0020}$ & $0.1186 \pm 0.0017$ & $0.1187^{+0.0029}_{-0.0030}$  \\
		NNPDF3.0   & $0.1171 \pm 0.0023$ &  $0.1147 \pm 0.025$ & $0.1207 \pm 0.0033$  \\\hline
		     & [ATLAS data] & [LHCb data] & [CMS data]~\cite{Sirunyan:2019wne} \\
		CT14 & $0.1214 \pm 0.0017$ & $0.1131 \pm 0.0024$ & $0.1163^{+0.0024}_{-0.0031}$  \\
		HERAPDF2.0 & $0.1170^{+0.0038}_{-0.0037}$ & $0.1071 \pm 0.0021$ & $0.1072^{+0.0043}_{-0.0040}$ \\
		MMHT14 & $0.1208^{+0.0016}_{-0.0020}$ &  $0.1166^{+0.0023}_{-0.0024}$ & $0.1186 \pm 0.0025$ \\
		NNPDF3.0 & $0.1222 \pm 0.0026$  & $0.1169 \pm 0.0032$ & $0.1147 \pm 0.0023$ \\\hline
		    & [PDF corr.\ = 1] & [scale corr.\ = 1] & [PDF corr.\ = scale corr.\ = 1] \\
		CT14 & $0.1181 \pm 0.0010$ & $0.1176^{+0.0017}_{-0.0016}$ & $0.1186 \pm 0.0009$  \\
		HERAPDF2.0 & $0.1071 \pm 0.0014$ & $0.1100^{+0.0023}_{-0.0022}$ & $0.1098 \pm 0.0013$\\
		MMHT14 & $0.1192 \pm 0.0009$ &  $0.1192^{+0.0016}_{-0.0014}$ & $0.1191 \pm 0.0009$ \\
		NNPDF3.0 & $0.1144 \pm 0.0017$  & $0.1157 \pm 0.0018$ & $0.1148 \pm 0.0016$ \\\hline
           & [$-$ th.\ shift, comb.\ w/o th.\ unc.]  & [comb.\ w/o th.\ unc.] & [$+$ th.\ shift, comb.\ w/o th.\ unc.] \\
        CT14       & $0.1142 \pm 0.0009$ & $0.1184 \pm 0.0009$ & $0.1203 \pm 0.0009$ \\
		HERAPDF2.0 & $0.1042 \pm 0.0012$ & $0.1090 \pm 0.0012$ & $0.1123 \pm 0.00012$ \\
		MMHT14     & $0.1163 \pm 0.0008$ & $0.1188 \pm 0.0008$ & $0.1209 \pm 0.0008$ \\
		NNPDF3.0   & $0.1160 \pm 0.0009$ & $0.1183 \pm 0.0009$ & $0.1205 \pm 0.0009$ \\\hline
		     & [symm.\ PDF uncert.] & [$+1\%$ uncorr.\,uncert.] & Largest differences \\
		CT14 & $0.1176 \pm 0.0023$ & $0.1171^{+0.0016}_{-0.0018}$ & $(+0.0042,-0.0041)$ \\
		HERAPDF2.0 & $0.1121 \pm 0.0027$ & $0.1101 \pm 0.0023$ & $(+0.0073,-0.0026)$ \\
		MMHT14 & $0.1200 \pm 0.0016$ &  $0.1191^{+0.0016}_{-0.0015}$ & $(+0.0020,-0.0022)$ \\
		NNPDF3.0 & $0.1160 \pm 0.0018$  & $0.1170 \pm 0.0022$ & $(+0.0062,-0.0016)$ \\\hline
	\end{tabular}
}
\end{table}

\begin{figure}[htbp!]
\centering
\includegraphics[width=0.97\textwidth]{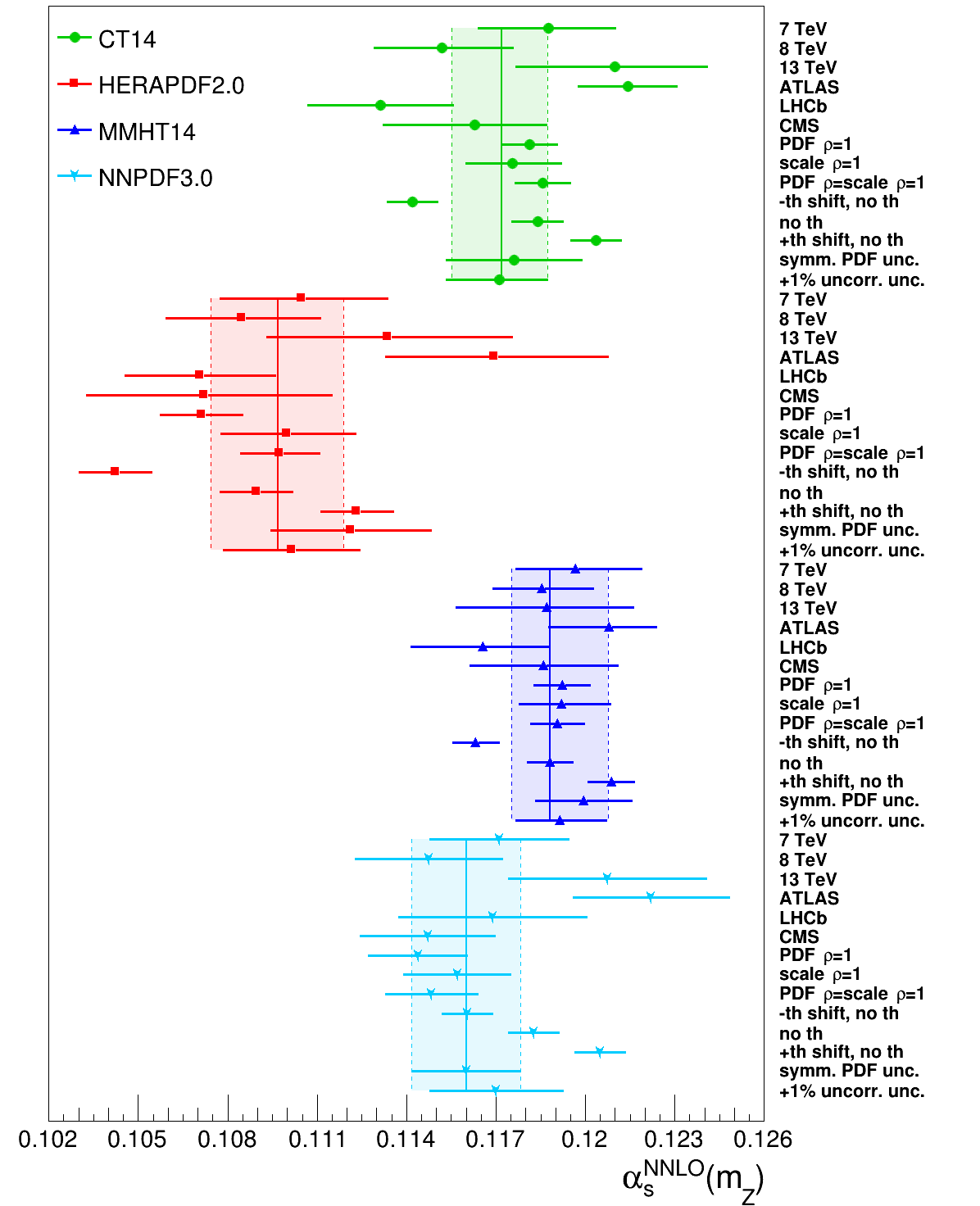}
\caption{Sensitivity of the $\alphasmZ$ values derived per PDF set to the various cross-checks reported in Table~\ref{tab:alphas_sensitivity}. The different symbols with error bars indicate groups of $\alphasmZ$ values obtained using different subsets of the data or varying different underlying uncertainties or correlations, individually listed on the right. The shaded areas and vertical lines show the final results of the default $\alphasmZ$ extraction per PDF (Table~\ref{tab:final_alphas_perPDF}). The CMS $\alphasmZ$ values plotted are those derived in~\cite{Sirunyan:2019wne}.
\label{fig:alphas_sensitivity}}
\end{figure}

To analyse the robustness and stability of the final $\alphasmZ$ extractions to the underlying data sets, their uncertainties and correlations, we repeat the {\sc convino} combination varying several ingredients as explained next. First, the $\alphasmZ$ are extracted by combining separately only those  measurements at the same \cm\ energies of $\sqrts = 7$, 8, and 13~TeV, as shown in the top rows of Table~\ref{tab:alphas_sensitivity} and in the three top data points of each PDF grouping of Fig.~\ref{fig:alphas_sensitivity}. This separation of data sets yields final $\alphasmZ$ values consistent with those derived from the combined ones listed in Table~\ref{tab:final_alphas_perPDF}, but with the 8~(13)~TeV extractions often systematically preferring lower (higher) values of $\alphasmZ$ compared to those derived at 7~TeV (except for MMHT14, where all three extractions yield values very close to each other). A second cross-check is carried out by using only data from each experiment separately. The ATLAS and LHCb data tend to prefer slightly higher and lower, respectively, values of the QCD coupling, with the CMS results appearing mostly in between. 
The presence of seemingly different $\alphasmZ$ values extracted from subsets of the experimental data is not of a concern, because the default extraction using all data shows a  good agreement among the individual extractions and the final $\alphasmZ$ value derived from their combination, as indicated by the $\chi^2/\mathrm{ndf}\approx1$ (except for NNDPF3.0) listed in Table~\ref{tab:final_alphas_perPDF}.

A third test is performed to cross-check the Pearson's coefficients method used to estimate the underlying correlations among PDF replicas/eigenvectors and scale uncertainties. We have studied the $\alphasmZ$ and $\chi^2$ values obtained when the correlations among the PDF or scale uncertainties in the 28 different systems are consistently varied over $\rho = 0$--1. Our final $\alphasmZ$ values are robust with respect to all such systematic variations, except for the PDF correlation scan of the NNPDF3.0 and HERAPDF2.0 sets for extreme correlation coefficients (approaching zero for the former, and unity for latter), where the result is slightly outside their one std.\ deviation default uncertainty. As an example, in the third row of Table~\ref{tab:alphas_sensitivity}, we list the QCD couplings extracted when we set the PDF and/or the scale uncertainty correlations all to unity. For these cases, 
except for HERAPDF2.0 for which the results go slightly outside the original uncertainty region,
the $\alphasmZ$ values stay well within the uncertainties of the original results.
A fourth robustness test of our $\alphasmZ$ extraction procedure is shown in the fourth row of Table~\ref{tab:alphas_sensitivity}, where we perform an alternative $\alphasmZ$ combination of the 28 individual estimates without scale and PDF uncertainties, 
as done \eg\ in~\cite{Klijnsma:2017eqp} for the $\alphasmZ$ averaging from the data-theory comparison of $\ttbar$ cross sections. The results of ignoring the theoretical uncertainties during the combination are listed in the middle column of the fourth row, as well as also repeating this operation but shifting first all 28 $\alphasmZ$ extractions, to the minus (left column) and plus (right column) direction with respect to the central value, by one std.\ deviation corresponding to the PDF and scale uncertainties added in quadrature. Whereas the results for all PDFs show variations consistent with the other tests reported in the table, in the NNPDF3.0 case we see that the removal of theoretical uncertainties in the averaging of individual extractions yields a final central $\alphasmZ$ value that is consistent with the individual estimates themselves (\ie\ it does not suffer from the ``Peelle's pertinent puzzle'' effect mentioned above).
The last tests include repeating the full $\alphasmZ$ combination after symmetrizing the PDF uncertainties, or after adding an uncorrelated 1\% numerical uncertainty to all theoretical cross sections. The latter choice aims at evaluating the impact of small extra uncertainties coming \eg\ from the use of different calculators for the theoretical pQCD cross sections (see Tables~\ref{tab:ATLAS_13}, \ref{tab:LHCb_7_13} and \ref{tab:LHCb_8}), or from potentially overlooked experimental uncertainties (LHCb quotes a $\sim$1\% uncertainty from the knowledge of the \cm\ energy, which is not explicitly considered by ATLAS and CMS).
As one can see in Fig.~\ref{fig:alphas_sensitivity}, the results of MMHT14 are the most stable against any variations in the analysis, whereas a few larger-than-1-standard-deviation changes appear for the rest of the PDF sets. 
In the bottom-right row of Table~\ref{tab:alphas_sensitivity}, we quantify the maximum size of the $\alphasmZ$ variations induced by all the aforementioned cross-checks. As can be seen also in Fig.~\ref{fig:alphas_sensitivity}, some CT14 tests can change $\alphasmZ$ by up to $0.0042$ (over 2.5 times larger than the uncertainty of the default result for this PDF set, $\alphasmZ = 0.1172^{+0.0015}_{-0.0017}$), whereas those of HERAPDF2.0 and NNPDF3.0 are quite asymmetric towards the positive side of the QCD coupling value, \ie\ towards values more consistent with the other two PDFs: maximum variations of $^{+0.0073}_{-0.0026}$ and $^{+0.0062}_{-0.0016}$, respectively, are observed that are more than three standard-deviations away from the default extraction for each one of these PDF sets (Table~\ref{tab:final_alphas_perPDF}). 

The HERAPDF2.0 result is $\sim$2--3 standard deviations away from the other $\alphasmZ$ extractions. Since there exists a generic anticorrelation between the values of $\alphas(Q^2)$ and the parton densities evaluated at $(x,Q^2)$ themselves, particularly for the gluon and in turn (through perturbative evolution) for the sea quarks, this result is indicative of intrinsic underlying differences between HERAPDF2.0 and the rest of PDFs. A study carried out with \textsc{apfel} v2.7.1~\cite{Carrazza:2014gfa} in Ref.~\cite{Sirunyan:2019wne} shows that the HERAPDF2.0 $u$-quark densities (and the overall quark-antiquark luminosities) are enhanced by about 5\% compared to the rest of the PDFs in the $(x,Q^2)$ region of relevance for EW boson production at the LHC. This fact increases the relative weight of the LO contributions to the  theoretical total W$^\pm$ and Z boson production cross sections, thereby relatively pushing down the cross section contributions from higher-order pQCD diagrams that are more sensitive to $\alphasmZ$. The effective result is thereby a comparatively reduced $\alphasmZ$ value. The fact that HERAPDF2.0 uses deep-inelastic scattering data alone in its fit and does not include LHC results, which chiefly constrain the gluon density, leads to a very weak $\alphas$ sensitivity for this PDF set, as indicated by the shallower slope of the $\sigma_{\rm{th}}$--$\alphasmZ$ dependence seen in Figs.~\ref{fig:data_th_atlas}--\ref{fig:data_th_tevatron}.
In agreement with the discussions of~\cite{Sirunyan:2019wne}, 
we conclude that one would need an updated refit of the HERAPDF set to an extended set of experimental data, including LHC results, before relying on the QCD coupling constant derived following the procedure described here.

Among all $\alphasmZ$ extractions, we consider the value obtained with MMHT14
as the most reliable in this analysis for several reasons, partially concomitant with those outlined in the CMS-only study~\cite{Sirunyan:2019wne}. First, the cross sections computed with the MMHT14 PDF for many measurements feature the largest sensitivity to $\alphasmZ$ variations, \ie\ the $\sigma^\mathrm{th}_\mathrm{W,Z}$ \vs\ $\alphasmZ$ dependencies observed for this PDF set have a larger $k$ slope than for the others 
(Figs.~\ref{fig:data_th_atlas} and \ref{fig:data_th_lhcb}). 
Second,
the level of agreement among the combined and individual $\alphasmZ$ extractions is the best of all PDFs (as indicated by the $\chi^2$/ndf~=~19.3/27 value listed in Table~\ref{tab:final_alphas_perPDF}). 
Third, MMHT14 
features the lowest relative (symmetrized) propagated uncertainty of all final $\alphasmZ$ results (Table~\ref{tab:final_alphas_perPDF}). 
Last but not least, the QCD coupling values extracted with MMHT14 show the largest stability and robustness of all PDFs with respect to variations in the data sets and in the assumptions of underlying uncertainties (Fig.~\ref{fig:alphas_sensitivity}).
Using the MMHT14 extraction as our baseline result, we obtain a final value of the QCD coupling constant at the Z pole mass of 
$\alphasmZ = 0.1188^{+0.0019}_{-0.0013}$ (symmetrized to $\alphasmZ = 0.1188 \pm 0.0016$),
with a total uncertainty of $\sim$1.3\%. The right plot of Fig.~\ref{fig:final_alphas_per_PDF} shows the asymmetric $\alphasmZ$ parabola extracted from the MMHT14 results alone compared to the current world average (orange band). 
This final extraction is fully consistent with the PDG world average, and has an overall uncertainty that is better than that of other recent determinations at this level of (NNLO) theoretical accuracy, such as those from EW precision fits~\cite{Haller:2018nnx}, and $\ttbar$ cross sections~\cite{Klijnsma:2017eqp,Sirunyan:2018goh}.
In terms of precision, our determination compares well with the $\alphasmZ = 0.1191 \pm 0.0018$ value extracted from the theoretical analysis of $\tau$ lepton hadronic decays, which has an uncertainty of $\sim$1.5\%~\cite{PDG,Boito:2014sta,Pich:2016mgv}. 

The alternative of providing a single final $\alphasmZ$ value combining the results derived from different PDF sets is not obvious because there are complicated correlations among all the parton densities fits, as they all use many identical data sets. If, nonetheless, one prefers to combine the results from the CT14, MMHT14, and NNPDF3.0 analyses (discarding HERAPDF2.0 for the reasons mentioned above), an unbiased approach to do so, in line with the PDG practice~\cite{PDG} as well as with the procedure employed to produce the PDF4LHC combined PDF set~\cite{Butterworth:2015oua} and with other $\alphasmZ$ extractions from collider data~\cite{Klijnsma:2017eqp,Sirunyan:2019wne}, is to average them without applying any weighting. By taking the straight average of the mean values and of the uncertainties of the results from these three PDFs, we would obtain a final value of the QCD coupling constant of $\alphasmZ = 0.1173 \pm 0.0017$, with a total (symmetrized) uncertainty of 1.4\%.


\section{Summary}
\label{sec:7}

We have presented a study of the production cross sections of electroweak gauge (\Wpm\ and Z) bosons in proton-(anti)proton ($\pp$, $\ppbar$) collisions at the LHC and Tevatron colliders at center-of-mass energies of $\sqrts = 1.8, 1.96, 7, 8$, and 13~TeV, aiming at the extraction of the QCD coupling constant at the Z mass scale. Thirty-four different experimental data sets available, corresponding to different fiducial criteria on the electron and/or muon decay final states at the LHC and Tevatron, have been individually compared to theoretical pQCD predictions computed at next-to-next-to-leading-order (NNLO) accuracy with the CT14, HERAPDF2.0, MMHT14, and NNPDF3.0 parton distribution functions (PDFs). An overall good data-theory agreement, within the experimental and theoretical uncertainties, is found for all measurements. 
A more detailed analysis of the 7 and 9 measurements from ATLAS and LHCb, respectively, has been carried out with the aim of extracting a precise value of $\alphasmZ$ via detailed data-theory comparisons. A total of 28 ``preferred'' values of the QCD coupling per PDF set are thereby extracted by combining the ATLAS and LHCb results with those from the CMS data analyzed in Ref.~\cite{Sirunyan:2019wne}.
The largest experimental (theoretical) propagated uncertainties are associated with the integrated luminosity (intra-PDF) uncertainties. A $\chi^2$-minimization procedure is employed to combine all 28 individual $\alphasmZ$ extractions per PDF set, properly taking into account all individual sources of experimental and theoretical uncertainties, and their correlations. The following combined values are extracted for the four different PDFs: $\alphasmZ = 0.1172^{+0.0015}_{-0.0017}$ (CT14), $0.1097^{+0.0022}_{-0.0023}$ (HERAPDF2.0), $0.1188^{+0.0019}_{-0.0013}$ (MMHT14), and $0.1160\pm 0.0018$ (NNPDF3.0). Among the four extractions, those based on the MMHT14 PDF appear as the most reliable in this analysis given that their cross sections are the most sensitive to the underlying $\alphasmZ$ value, the level of accord between individual and combined $\alphasmZ$ extractions is the best among PDF sets, and the derived $\alphasmZ$ values are the most robust and stable with respect to variations in the data sets and uncertainties. The final result derived 
from the MMHT14 calculations, 
$\alphasmZ = 0.1188^{+0.0019}_{-0.0013}$, has a $\sim$1.3\% uncertainty that is better than any other observable currently measured at hadron colliders, and comparable to that of the analysis of hadronic $\tau$ lepton decay data.

This work confirms that the total inclusive \Wpm\ and Z boson cross sections at hadron colliders are new promising observables that can provide useful constraints on the value of the QCD coupling constant, and that can eventually help to reduce the current relatively large uncertainty of the $\alphasmZ$ world average. The future availability of N$^3$LO codes, with one additional higher degree of theoretical accuracy than the current state-of-the-art, for the calculation of inclusive \Wpm\ and Z boson production cross sections will allow for further reductions of the propagated scale uncertainties. Such a result, combined with upcoming electroweak boson measurements at the LHC with $\sim$1\% experimental uncertainties, thanks to further reduced integrated luminosity uncertainties, will enable future $\alphasmZ$ extractions with propagated uncertainties below the 1\% level.\\


\hspace{-0.85cm}
{\bf Acknowledgments--}
We are indebted to Weichen~Xiao for his valuable help in an early version of this study. We thank Stefano Camarda and Steve Farry for useful feedback on the ATLAS and LHCb, respectively, experimental measurements and correlations.

\appendix
\section{Independent \texorpdfstring{$\alphasmZ$}{alphaS(mZ)} extractions per EW boson cross section measurement}

Tables~\ref{tab:alphasmZ_atlas}, \ref{tab:alphasmZ_lhcb} and~\ref{tab:alphasmZ_tevatron} list the $\alphasmZ$ values obtained (along with the propagated uncertainty breakdowns from every source) from each individual ATLAS, LHCb, and Tevatron electroweak cross section measurement, respectively, for each PDF set. 

\begin{table}[htpb!]
\renewcommand\arraystretch{1.3}
\caption{Extracted $\alphasmZ$ values from the comparison of the \Wp, \Wm, and Z boson production cross sections measured by ATLAS to NNLO pQCD predictions. For each measurement, four results are extracted, one per PDF set. The breakdown of the propagated uncertainties from different experimental (statistical, integrated luminosity, and systematic) and theoretical (PDF, scale, and numerical) sources is provided for each $\alphasmZ$ estimate.\label{tab:alphasmZ_atlas}}
\centering
\resizebox{\textwidth}{!}{
\begin{tabular}{llccccccc}\hline
  Cross section & PDF & $\alphasmZ$
  & $\delta_{\alpS}\stat$ & $\delta_{\alpS}\lum$ & $\delta_{\alpS}\syst$ & $\delta_{\alpS}$(PDF) & $\delta_{\alpS}$(scale) & $\delta_{\alpS}\,\statt$ \\\hline
  ATLAS & CT14 & 0.1211 $^{+0.0056}_{-0.0036}$ & 0.0000 & 0.0019 & 0.0005 & $^{+0.0051}_{-0.0027}$ & 0.0013 & 0.0007 \\
  W$^+$  (7~TeV) & HERAPDF2.0 & 0.1132 $^{+0.0074}_{-0.0101}$ & 0.0001 & 0.0049 & 0.0014 & $^{+0.0038}_{-0.0078}$ & 0.0034 & 0.0017 \\
  & MMHT14 & 0.1205 $^{+0.0061}_{-0.0030}$ & 0.0000 & 0.0019 & 0.0005 & $^{+0.0056}_{-0.0018}$ & 0.0013 & 0.0006 \\
  & NNPDF3.0 & 0.1229 $^{+0.0036}_{-0.0036}$ & 0.0000 & 0.0021 & 0.0006 & $^{+0.0024}_{-0.0024}$ & 0.0015 & 0.0007 \\
  
  ATLAS & CT14 & 0.1211 $^{+0.0028}_{-0.0056}$ & 0.0001 & 0.0020 & 0.0006 & $^{+0.0017}_{-0.0051}$ & 0.0007 & 0.0005 \\
  W$^-$  (7~TeV) & HERAPDF2.0 & 0.1160 $^{+0.0060}_{-0.0121}$ & 0.0001 & 0.0051 & 0.0016 & $^{+0.0018}_{-0.0107}$ & 0.0017 & 0.0010 \\
  & MMHT14 & 0.1198 $^{+0.0028}_{-0.0035}$ & 0.0001 & 0.0018 & 0.0006 & $^{+0.0020}_{-0.0028}$ & 0.0006 & 0.0004 \\
  & NNPDF3.0 & 0.1254 $^{+0.0047}_{-0.0047}$ & 0.0001 & 0.0030 & 0.0009 & $^{+0.0033}_{-0.0033}$ & 0.0010 & 0.0006 \\
  
  ATLAS & CT14 & 0.1229 $^{+0.0032}_{-0.0044}$ & 0.0001 & 0.0021 & 0.0004 & $^{+0.0024}_{-0.0038}$ & 0.0005 & 0.0002 \\
  Z  (7~TeV) & HERAPDF2.0 & 0.1200 $^{+0.0055}_{-0.0056}$ & 0.0001 & 0.0040 & 0.0008 & $^{+0.0036}_{-0.0038}$ & 0.0009 & 0.0005 \\
  & MMHT14 & 0.1214 $^{+0.0026}_{-0.0025}$ & 0.0001 & 0.0018 & 0.0003 & $^{+0.0018}_{-0.0015}$ & 0.0004 & 0.0002 \\
  & NNPDF3.0 & 0.1260 $^{+0.0037}_{-0.0037}$ & 0.0001 & 0.0025 & 0.0005 & $^{+0.0026}_{-0.0026}$ & 0.0006 & 0.0003 \\
  
  ATLAS & CT14 & 0.1220 $^{+0.0044}_{-0.0048}$ & 0.0000 & 0.0032 & 0.0005 & $^{+0.0028}_{-0.0035}$ & 0.0005 & 0.0002 \\
  Z  (8~TeV) & HERAPDF2.0 & 0.1177 $^{+0.0068}_{-0.0060}$ & 0.0000 & 0.0053 & 0.0008 & $^{+0.0041}_{-0.0027}$ & 0.0008 & 0.0004 \\
  & MMHT14 & 0.1207 $^{+0.0031}_{-0.0031}$ & 0.0000 & 0.0027 & 0.0004 & $^{+0.0014}_{-0.0014}$ & 0.0004 & 0.0002 \\
  & NNPDF3.0 & 0.1247 $^{+0.0047}_{-0.0047}$ & 0.0000 & 0.0038 & 0.0006 & $^{+0.0026}_{-0.0026}$ & 0.0006 & 0.0003 \\ \hline
  
  ATLAS & CT14 & 0.1194 $^{+0.0102}_{-0.0041}$ & 0.0003 & 0.0027 & 0.0025 & $^{+0.0094}_{-0.0009}$ & 0.0014 & 0.0008 \\
  W$^+$  (13~TeV) & HERAPDF2.0 & 0.1102 $^{+0.0093}_{-0.0108}$ & 0.0005 & 0.0048 & 0.0043 & $^{+0.0059}_{-0.0082}$ & 0.0024 & 0.0017 \\
  & MMHT14 & 0.1208 $^{+0.0046}_{-0.0113}$ & 0.0003 & 0.0031 & 0.0028 & $^{+0.0006}_{-0.0104}$ & 0.0015 & 0.0011 \\
  & NNPDF3.0 & 0.1215 $^{+0.0042}_{-0.0042}$ & 0.0002 & 0.0023 & 0.0021 & $^{+0.0024}_{-0.0024}$ & 0.0012 & 0.0009 \\
  
  ATLAS & CT14 & 0.1213 $^{+0.0045}_{-0.0045}$ & 0.0003 & 0.0020 & 0.0020 & $^{+0.0032}_{-0.0033}$ & 0.0009 & 0.0005 \\
  W$^-$  (13~TeV) & HERAPDF2.0 & 0.1149 $^{+0.0107}_{-0.0111}$ & 0.0009 & 0.0062 & 0.0062 & $^{+0.0051}_{-0.0058}$ & 0.0029 & 0.0014 \\
  & MMHT14 & 0.1207 $^{+0.0031}_{-0.0067}$ & 0.0003 & 0.0020 & 0.0020 & $^{+0.0005}_{-0.0060}$ & 0.0009 & 0.0004 \\
  & NNPDF3.0 & 0.1248 $^{+0.0050}_{-0.0050}$ & 0.0004 & 0.0027 & 0.0027 & $^{+0.0029}_{-0.0029}$ & 0.0012 & 0.0007 \\
  
  ATLAS & CT14 & 0.1222 $^{+0.0044}_{-0.0038}$ & 0.0004 & 0.0022 & 0.0008 & $^{+0.0037}_{-0.0028}$ & 0.0007 & 0.0003 \\
  Z (13~TeV) & HERAPDF2.0 & 0.1181 $^{+0.0065}_{-0.0059}$ & 0.0009 & 0.0046 & 0.0017 & $^{+0.0040}_{-0.0028}$ & 0.0014 & 0.0006 \\
  & MMHT14 & 0.1209 $^{+0.0024}_{-0.0036}$ & 0.0004 & 0.0020 & 0.0007 & $^{+0.0010}_{-0.0029}$ & 0.0006 & 0.0002 \\
  & NNPDF3.0 & 0.1251 $^{+0.0040}_{-0.0040}$ & 0.0005 & 0.0026 & 0.0010 & $^{+0.0027}_{-0.0027}$ & 0.0008 & 0.0003 \\ \hline
\end{tabular}
}
\end{table}
\begin{table}[htpb!]
\renewcommand\arraystretch{1.3}
\caption{Extracted $\alphasmZ$ values from the comparison of the \Wp, \Wm, and Z boson production cross sections measured by LHCb to NNLO pQCD predictions. For each measurement, four results are extracted, one per PDF set. The breakdown of the propagated uncertainties from different experimental (statistical, integrated luminosity, and systematic including \cm\ energy) and theoretical (PDF, scale, and numerical) sources is provided for each $\alphasmZ$ estimate.\label{tab:alphasmZ_lhcb}}
\centering
\resizebox{\textwidth}{!}{
\begin{tabular}{llccccccc}\hline
  Cross section & PDF & $\alphasmZ$
   & $\delta_{\alpS}\stat$ & $\delta_{\alpS}\lum$ & $\delta_{\alpS}\syst$ & $\delta_{\alpS}$(PDF) & $\delta_{\alpS}$(scale) & $\delta_{\alpS}\,\statt$ \\\hline
  LHCb & CT14 & 0.1165 $^{+0.0046}_{-0.0049}$ & 0.0003 & 0.0020 & 0.0016 & $^{+0.0035}_{-0.0039}$ & 0.0014 & 0.0006 \\
  W$^+$  (7~TeV) & HERAPDF2.0 & 0.1126 $^{+0.0048}_{-0.0085}$ & 0.0004 & 0.0026 & 0.0020 & $^{+0.0029}_{-0.0076}$ & 0.0017 & 0.0008 \\
  & MMHT14 & 0.1147 $^{+0.0047}_{-0.0051}$ & 0.0003 & 0.0019 & 0.0015 & $^{+0.0037}_{-0.0043}$ & 0.0013 & 0.0005 \\
  & NNPDF3.0 & 0.1180 $^{+0.0040}_{-0.0040}$ & 0.0003 & 0.0020 & 0.0015 & $^{+0.0028}_{-0.0028}$ & 0.0013 & 0.0007 \\
  
  LHCb & CT14 & 0.1170 $^{+0.0040}_{-0.0075}$ & 0.0004 & 0.0022 & 0.0015 & $^{+0.0025}_{-0.0068}$ & 0.0014 & 0.0007 \\
  W$^-$  (7~TeV) & HERAPDF2.0 & 0.1057 $^{+0.0058}_{-0.0087}$ & 0.0005 & 0.0030 & 0.0021 & $^{+0.0039}_{-0.0075}$ & 0.0018 & 0.0011 \\
  & MMHT14 & 0.1148 $^{+0.0060}_{-0.0039}$ & 0.0003 & 0.0019 & 0.0013 & $^{+0.0053}_{-0.0028}$ & 0.0012 & 0.0007 \\
  & NNPDF3.0 & 0.1171 $^{+0.0043}_{-0.0043}$ & 0.0004 & 0.0021 & 0.0015 & $^{+0.0031}_{-0.0031}$ & 0.0013 & 0.0008 \\
  
  LHCb & CT14 & 0.1173 $^{+0.0085}_{-0.0047}$ & 0.0006 & 0.0027 & 0.0023 & $^{+0.0074}_{-0.0022}$ & 0.0019 & 0.0006 \\
  Z  (7~TeV) & HERAPDF2.0 & 0.1088 $^{+0.0052}_{-0.0047}$ & 0.0006 & 0.0025 & 0.0022 & $^{+0.0035}_{-0.0026}$ & 0.0018 & 0.0006 \\
  & MMHT14 & 0.1159 $^{+0.0052}_{-0.0034}$ & 0.0005 & 0.0020 & 0.0017 & $^{+0.0042}_{-0.0015}$ & 0.0014 & 0.0004 \\
  & NNPDF3.0 & 0.1198 $^{+0.0043}_{-0.0043}$ & 0.0005 & 0.0022 & 0.0019 & $^{+0.0028}_{-0.0028}$ & 0.0015 & 0.0005 \\ \hline
  
  LHCb & CT14 & 0.1210 $^{+0.0064}_{-0.0055}$ & 0.0002 & 0.0015 & 0.0029 & $^{+0.0049}_{-0.0036}$ & 0.0025 & 0.0007 \\
  W$^+_\text{e}$  (8~TeV) & HERAPDF2.0 & 0.1177 $^{+0.0083}_{-0.0074}$ & 0.0003 & 0.0018 & 0.0033 & $^{+0.0068}_{-0.0057}$ & 0.0028 & 0.0009 \\
  & MMHT14 & 0.1193 $^{+0.0038}_{-0.0043}$ & 0.0002 & 0.0011 & 0.0020 & $^{+0.0023}_{-0.0031}$ & 0.0017 & 0.0006 \\
  & NNPDF3.0 & 0.1224 $^{+0.0045}_{-0.0045}$ & 0.0002 & 0.0014 & 0.0025 & $^{+0.0027}_{-0.0027}$ & 0.0022 & 0.0008 \\
  
  LHCb & CT14 & 0.1130 $^{+0.0035}_{-0.0090}$ & 0.0003 & 0.0013 & 0.0027 & $^{+0.0009}_{-0.0083}$ & 0.0011 & 0.0010 \\
  W$^-_\text{e}$  (8~TeV) & HERAPDF2.0 & 0.1072 $^{+0.0046}_{-0.0048}$ & 0.0002 & 0.0012 & 0.0024 & $^{+0.0034}_{-0.0037}$ & 0.0009 & 0.0009 \\
  & MMHT14 & 0.1099 $^{+0.0068}_{-0.0088}$ & 0.0003 & 0.0016 & 0.0033 & $^{+0.0056}_{-0.0079}$ & 0.0013 & 0.0009 \\
  & NNPDF3.0 & 0.1136 $^{+0.0045}_{-0.0045}$ & 0.0003 & 0.0013 & 0.0027 & $^{+0.0030}_{-0.0030}$ & 0.0011 & 0.0009 \\
  
  LHCb & CT14 & 0.1169 $^{+0.0050}_{-0.0063}$ & 0.0003 & 0.0015 & 0.0016 & $^{+0.0039}_{-0.0055}$ & 0.0020 & 0.0008 \\
  W$^+_\mu$  (8~TeV) & HERAPDF2.0 & 0.1117 $^{+0.0057}_{-0.0098}$ & 0.0004 & 0.0025 & 0.0026 & $^{+0.0023}_{-0.0083}$ & 0.0034 & 0.0014 \\
  & MMHT14 & 0.1158 $^{+0.0068}_{-0.0036}$ & 0.0002 & 0.0012 & 0.0013 & $^{+0.0063}_{-0.0026}$ & 0.0017 & 0.0008 \\
  & NNPDF3.0 & 0.1189 $^{+0.0038}_{-0.0038}$ & 0.0002 & 0.0013 & 0.0014 & $^{+0.0027}_{-0.0027}$ & 0.0018 & 0.0007 \\
  
  LHCb & CT14 & 0.1168 $^{+0.0030}_{-0.0086}$ & 0.0003 & 0.0013 & 0.0012 & $^{+0.0018}_{-0.0083}$ & 0.0014 & 0.0008 \\
  W$^-_\mu$  (8~TeV) & HERAPDF2.0 & 0.1044 $^{+0.0103}_{-0.0040}$ & 0.0004 & 0.0020 & 0.0018 & $^{+0.0097}_{-0.0016}$ & 0.0021 & 0.0012 \\
  & MMHT14 & 0.1159 $^{+0.0026}_{-0.0072}$ & 0.0002 & 0.0012 & 0.0011 & $^{+0.0016}_{-0.0069}$ & 0.0012 & 0.0007 \\
  & NNPDF3.0 & 0.1166 $^{+0.0038}_{-0.0038}$ & 0.0003 & 0.0014 & 0.0012 & $^{+0.0030}_{-0.0030}$ & 0.0014 & 0.0008 \\
  
  LHCb & CT14 & 0.1156 $^{+0.0050}_{-0.0046}$ & 0.0004 & 0.0015 & 0.0018 & $^{+0.0039}_{-0.0034}$ & 0.0018 & 0.0005 \\
  Z$_\mu$  (8~TeV) & HERAPDF2.0 & 0.1059 $^{+0.0045}_{-0.0077}$ & 0.0005 & 0.0019 & 0.0023 & $^{+0.0023}_{-0.0067}$ & 0.0023 & 0.0005 \\
  & MMHT14 & 0.1142 $^{+0.0052}_{-0.0038}$ & 0.0004 & 0.0014 & 0.0017 & $^{+0.0044}_{-0.0026}$ & 0.0017 & 0.0005 \\
  & NNPDF3.0 & 0.1182 $^{+0.0039}_{-0.0039}$ & 0.0004 & 0.0014 & 0.0016 & $^{+0.0027}_{-0.0027}$ & 0.0016 & 0.0005 \\ \hline
  
  LHCb & CT14 & 0.1131 $^{+0.0066}_{-0.0061}$ & 0.0005 & 0.0045 & 0.0019 & $^{+0.0041}_{-0.0032}$ & 0.0014 & 0.0005 \\
  Z (13~TeV) & HERAPDF2.0 & 0.1097 $^{+0.0055}_{-0.0074}$ & 0.0005 & 0.0046 & 0.0020 & $^{+0.0017}_{-0.0053}$ & 0.0014 & 0.0006 \\
  & MMHT14 & 0.1129 $^{+0.0060}_{-0.0055}$ & 0.0005 & 0.0041 & 0.0018 & $^{+0.0036}_{-0.0027}$ & 0.0013 & 0.0005 \\
  & NNPDF3.0 & 0.1171 $^{+0.0052}_{-0.0052}$ & 0.0005 & 0.0041 & 0.0018 & $^{+0.0024}_{-0.0024}$ & 0.0012 & 0.0005 \\ \hline
\end{tabular}
}
\end{table}

\begin{table}[htpb!]
\renewcommand\arraystretch{1.3}
\caption{Extracted $\alphasmZ$ values from the comparison of the \Wp, \Wm, and Z boson production cross sections measured by CDF and D0 to NNLO pQCD predictions. For each measurement, four results are extracted, one per PDF set. The breakdown of the propagated uncertainties from different experimental (statistical, integrated luminosity, and systematic) and theoretical (PDF, scale, and numerical) sources is provided for each $\alphasmZ$ estimate.\label{tab:alphasmZ_tevatron}}
\centering
\resizebox{\textwidth}{!}{
\begin{tabular}{llccccccc}\hline
Cross section & PDF & $\alphasmZ$
 & $\delta_{\alpS}\stat$ & $\delta_{\alpS}\lum$ & $\delta_{\alpS}\syst$ & $\delta_{\alpS}$(PDF) & $\delta_{\alpS}$(scale) & $\delta_{\alpS}\,\statt$ \\\hline
CDF & CT14 & 0.1158 $^{+0.0122}_{-0.0125}$ & 0.0018 & 0.0080 & 0.0071 & $^{+0.0053}_{-0.0060}$ & 0.0011 & 0.0003 \\
\Wpm\ (1.8~TeV) & HERAPDF2.0 & 0.1108 $^{+0.0119}_{-0.0119}$ & 0.0018 & 0.0082 & 0.0073 & $^{+0.0042}_{-0.0041}$ & 0.0012 & 0.0003 \\
& MMHT14 & 0.1149 $^{+0.0076}_{-0.0084}$ & 0.0012 & 0.0053 & 0.0047 & $^{+0.0025}_{-0.0043}$ & 0.0008 & 0.0002 \\
& NNPDF3.0 & 0.1195 $^{+0.0089}_{-0.0089}$ & 0.0013 & 0.0060 & 0.0053 & $^{+0.0037}_{-0.0037}$ & 0.0009 & 0.0002 \\

CDF & CT14 & 0.1220 $^{+0.0167}_{-0.0170}$ & 0.0076 & 0.0101 & 0.0088 & $^{+0.0064}_{-0.0070}$ & 0.0017 & 0.0003 \\
Z (1.8~TeV) & HERAPDF2.0 & 0.1112 $^{+0.0162}_{-0.0165}$ & 0.0076 & 0.0102 & 0.0089 & $^{+0.0042}_{-0.0052}$ & 0.0017 & 0.0004 \\
& MMHT14 & 0.1187 $^{+0.0094}_{-0.0094}$ & 0.0043 & 0.0057 & 0.0050 & $^{+0.0033}_{-0.0034}$ & 0.0009 & 0.0002 \\
& NNPDF3.0 & 0.1247 $^{+0.0103}_{-0.0103}$ & 0.0047 & 0.0062 & 0.0054 & $^{+0.0039}_{-0.0039}$ & 0.0010 & 0.0002 \\ \hline

CDF & CT14 & 0.1175 $^{+0.0141}_{-0.0142}$ & 0.0008 & 0.0125 & 0.0040 & $^{+0.0052}_{-0.0054}$ & 0.0011 & 0.0003 \\
\Wpm\ (1.96~TeV) & HERAPDF2.0 & 0.1122 $^{+0.0161}_{-0.0162}$ & 0.0009 & 0.0147 & 0.0047 & $^{+0.0043}_{-0.0048}$ & 0.0013 & 0.0004 \\
& MMHT14 & 0.1159 $^{+0.0096}_{-0.0090}$ & 0.0005 & 0.0083 & 0.0027 & $^{+0.0039}_{-0.0021}$ & 0.0008 & 0.0002 \\
& NNPDF3.0 & 0.1208 $^{+0.0109}_{-0.0109}$ & 0.0006 & 0.0098 & 0.0031 & $^{+0.0036}_{-0.0036}$ & 0.0009 & 0.0002 \\

CDF & CT14 & 0.1236 $^{+0.0180}_{-0.0183}$ & 0.0035 & 0.0160 & 0.0048 & $^{+0.0056}_{-0.0063}$ & 0.0013 & 0.0003 \\
Z (1.96~TeV) & HERAPDF2.0 & 0.1150 $^{+0.0187}_{-0.0193}$ & 0.0037 & 0.0171 & 0.0052 & $^{+0.0038}_{-0.0061}$ & 0.0014 & 0.0004 \\
& MMHT14 & 0.1199 $^{+0.0108}_{-0.0106}$ & 0.0021 & 0.0096 & 0.0029 & $^{+0.0034}_{-0.0026}$ & 0.0008 & 0.0002 \\
& NNPDF3.0 & 0.1261 $^{+0.0119}_{-0.0119}$ & 0.0023 & 0.0106 & 0.0032 & $^{+0.0037}_{-0.0037}$ & 0.0009 & 0.0002 \\ \hline

D0 & CT14 & 0.1028 $^{+0.0118}_{-0.0119}$ & 0.0009 & 0.0092 & 0.0046 & $^{+0.0057}_{-0.0058}$ & 0.0011 & 0.0003 \\
\Wpm\ (1.8~TeV) & HERAPDF2.0 & 0.0968 $^{+0.0117}_{-0.0117}$ & 0.0010 & 0.0097 & 0.0048 & $^{+0.0042}_{-0.0041}$ & 0.0012 & 0.0003 \\
& MMHT14 & 0.1059 $^{+0.0078}_{-0.0077}$ & 0.0006 & 0.0062 & 0.0031 & $^{+0.0034}_{-0.0032}$ & 0.0008 & 0.0002 \\
& NNPDF3.0 & 0.1099 $^{+0.0085}_{-0.0085}$ & 0.0007 & 0.0068 & 0.0034 & $^{+0.0038}_{-0.0038}$ & 0.0008 & 0.0002 \\

D0 & CT14 & 0.1090 $^{+0.0166}_{-0.0161}$ & 0.0039 & 0.0131 & 0.0053 & $^{+0.0074}_{-0.0064}$ & 0.0017 & 0.0003 \\
Z (1.8~TeV) & HERAPDF2.0 & 0.0985 $^{+0.0150}_{-0.0153}$ & 0.0038 & 0.0128 & 0.0051 & $^{+0.0043}_{-0.0052}$ & 0.0017 & 0.0004 \\
& MMHT14 & 0.1118 $^{+0.0086}_{-0.0086}$ & 0.0021 & 0.0070 & 0.0028 & $^{+0.0034}_{-0.0033}$ & 0.0009 & 0.0002 \\
& NNPDF3.0 & 0.1170 $^{+0.0096}_{-0.0096}$ & 0.0023 & 0.0078 & 0.0031 & $^{+0.0039}_{-0.0039}$ & 0.0010 & 0.0002 \\ \hline
\end{tabular}
}
\end{table}

\clearpage


\end{document}